\RequirePackage{fix-cm}
\documentclass{svjour3} 
\usepackage[margin=1.0in]{geometry}
\smartqed  
\usepackage{graphicx}
\usepackage{psfrag}
\usepackage{pstricks}
\usepackage{epsfig}
\usepackage{hyperref}
\graphicspath{{Figures/}}
\usepackage{subcaption}
\usepackage{amsmath,amssymb,amsfonts}
\usepackage[maxfloats=100]{morefloats}
\usepackage{color}
\usepackage{tabularx}
\usepackage{tabulary}
\usepackage{multirow}
\usepackage{array,mathtools,booktabs,siunitx}
\usepackage{bm}
\newcommand{\ra}[1]{\renewcommand{\arraystretch}{#1}}
\usepackage{tikz} 

\usepackage{comment}
\usepackage{float}

\newcommand*{\xdash}[1][3em]{\rule[0.5ex]{#1}{0.55pt}}

\newcommand{\cmnt}[1]{\ignorespaces}

\begin{document}


%
\title{Assessment of End-to-End and Sequential Data-driven Learning of Fluid Flows}

\author{Shivakanth Chary Puligilla \and Balaji Jayaraman}
\institute{Shivakanth chary Puligilla \at
         School of Mechanical and Aerospace Engineering, Oklahoma State University, Stillwater, OK, USA\\
              \email{shivakanthchary.puligilla@okstate.edu}           
           \and
           Balaji Jayaraman (Corresponding Author) \at
         School of Mechanical and Aerospace Engineering, Oklahoma State University, Stillwater, OK, USA\\
         \email{balaji.jayaraman@okstate.edu}     
}

\date{Received: date / Accepted: date}
\maketitle


%

\abstract{
In this work we explore the advantages of end-to-end learning  of multilayer maps offered by feed forward neural-networks (FFNN) for learning and predicting dynamics from transient fluid flow data. While machine learning in general depends on data quality and quantity relative to the underlying dynamics of the system, it is important for a given \cmnt{data-driven} learning architecture to make the most of this available information. To this end, we focus on data-driven problems where there is a need to predict over reasonable time into the future with limited data availability. Such function approximation or time series prediction is in contrast to many applications of machine learning such as pattern recognition and parameter estimation that leverage vast datasets. 
In this study, we interpret the suite of recently popular data-driven learning approaches that approximate the dynamics as Markov linear model in a higher-dimensional feature space as a multilayer architecture similar to neural networks. However, there exists a couple of key differences:  (i) Markov linear models employ layer-wise learning in the sense of linear regression whereas neural networks represent end-to-end learning in the sense of nonlinear regression. We show through examples of data-driven modeling of canonical fluid flows that FFNN-like methods owe their success to leveraging the extended learning parameter space available in end-to-end learning without overfitting the data. In this sense, the Markov linear models behave as shallow neural networks.  (ii) The second major difference is that while the FFNN is by design a forward architecture, the class of Markov linear methods that approximate the Koopman operator are bi-directional, i.e., they incorporate both forward and backward maps in order to learn a linear map that can provide insight into spectral characteristics. In this study, we assess both reconstruction as well as predictive performance of temporally evolving dynamic using limited snapshots of data  for canonical nonlinear fluid flows including the transient limit-cycle attractor in a cylinder wake and the instability-driven dynamics of buoyant Boussinesq flow. 
}


\maketitle


\section{Introduction}\label{s:intro}

Fluid flows are predominantly multiscale phenomena occurring over a wide range  of length and time scales such as transition \cite{Edstrand:18parallel}, turbulence \cite{Wu:17transitional} and flow separation \cite{Deem:18Exp}. Direct numerical simulation (DNS) of such realistic high Reynolds number flows even in their canonical forms is a challenge even with current computing capacity. On the other hand, advances in experimental techniques for visualization and data acquisition have led to an abundance of fluid flow measurement data, but these measurements are often sparse and in many cases the underlying phenomenology or governing model is not known. In both these cases, there is a need for efficient data-driven models to serve the twin goals of (i) system modulation to achieve desired effects, i.e. flow control \cite{Kim:07ARev,Brunton:15CLT} or (ii) forecasting for informed decision making\cite{Cao:07Reduced,Fang:09POD,Benner:15survey} or both. 
Additionally, data-driven models also allow for extraction of dynamical and physical characteristics to generate insight~\cite{Bagheri:13,Rowley:09} into the system behavior. In flow control applications linear operator based control is often preferred so that one can leverage the expertise accumulated from the past~\cite{Rowley:17ARev}. Consequently, learning a linear system model is attractive as evidenced by voluminous recent literature in this area~\cite{SchmidDMD:10,Williams:15} including that from our team \cite{Lu:18sparse}. However, such methods have their inherent limitations and perform inadequately with small amounts of data. In this paper, we explore the potential of machine learning frameworks for nonlinear function representations  to extend the horizon of prediction for canonical fluid flows. Particularly for this article, we explore  bluff body wake flows and buoyancy-driven mixing. 

A good data-driven model should perform well in both system identification and prediction using limited amounts of data. In addition, these models need to be computationally tractable which makes dimensionality reduction essential. System identification enables learning of stability and physical characteristics such as unstable modes and coherent structures. For example, proper orthogonal decomposition (POD) \cite{Lumley:70POD} via singular value decomposition (SVD) \cite{Trefethen:97} and its close cousin, the Dynamic mode decomposition (DMD) \cite{SchmidDMD:10} are well known methods to extract such relevant spectral information. However, the capacity of DMD for long-term prediction is underwhelming~\cite{Lu:17,Lu:18sparse}. POD-based methods that use Galerkin projection onto the flow governing equations are more successful as long as the basis remains relevant to the flow evolution, but require knowledge of the system.  In this study, we focus on purely data-driven scenarios without knowledge of governing equations.  By long-time predictions, we imply evolving the system model over multiple characteristic time-scales beyond the training regime. The other prominent use of such models is to forecast the system evolution along different trajectories. Obviously, the precise definitions of 'long-time' prediction or forecasting is physics dependent. For example, a limit-cycle system evolving on a stable attractor will be more amenable to prediction from limited data as compared to more complex nonlinear mixing dynamics.  In the case of cylinder wake flow explored in this study, forecasting represents predicting the limit cycle \cite{Berkooz:93POD,Noack:03hierarchy} dynamics using limited data in the transient unstable wake region. We explore such cases as they are sensitive to error growth and hence, used to evaluate a given model.  Errors in model learning can be attributed to limited training data, measurement noise, model over fitting and insufficient validation~\cite{bishop1995neural,christopher2016pattern}. The contribution from this paper is a systematic exploration and assessment of how nonlinear regression-based data-driven models perform relative to commonly used linear regression models for dynamically evolving fluid flows.

There are two classes of approaches for modeling dynamical systems from limited data, namely Markov and non-Markov models. For a given current state $\pmb{x}_t$ and future state $\pmb{x}_{t+T}$ of a dynamical system, a Markov model \cite{Wu:17var}, under some transformation $g,\ h$, evolves the system state as $ g(\pmb{x}_{t+T}) = \mathcal{K} h(\pmb{x}_t)$. Learning such an operator $\mathcal{K}$ is the key to building such models. Markovian processes are minimally memory dependent and popular approaches for modeling such systems include dynamic mode decomposition (DMD) \cite{SchmidDMD:10,Rowley:09} and Feed forward neural networks (FFNNs). Recently, linear operator \cite{Rowley:17ARev,Mezic:05,Williams:15,Lu:18sparse} methods for modeling nonlinear dynamics have been related to the Koopman operator \cite{Koopman:31} theoretic framework. 
The Koopman approximation-based methods are a special case of Markov models that employ symmetric transformations of the input and out to the same feature space (i.e. $g = h$). On the other hand, if the model incorporates copious amounts of memory of the state variables to predict a future state, then it is considered non-Markovian. Recurrent neural networks(RNN) are good examples of non-Markovian models and have been employed for learning dynamical systems both in the past~\cite{Hopfield:82PNAS,Hochreiter:97LSTM} and in recent times~\cite{Soltani:16,yu:learning}.  Although these have shown success, they are very hard to build and train~\cite{Bengio:94IEEE} as compared to standard feed forward neural networks (FFNNs)~\cite{bengio2015deep}. This is because, the standard backpropagation-based algorithms can lead to exploding or vanishing gradient problems.  

\par While Markov models are popular, especially the linear variants, their success is often tied to two aspects: (i) the ability of the projection or maps to the feature space \cite{Rowley:17ARev,Taira:17aiaa,Lu:18sparse} to accurately map data without loss of information while incorporating the appropriate degree of nonlinearity and (ii) their ability to capture the evolution of the dynamics in the feature space \cite{Lu:18sparse}. This renders many such learning methodologies into an exercise in identifying the optimal `magic' feature maps. A common approach to building such nonlinear map operators is to layer multiple `elementary' maps \cite{Williams:14arXiv,Williams:15,Lu:18sparse}.  While DMD \cite{SchmidDMD:10,Rowley:09} employs a single-layer map operator based on singular value decomposition (SVD) of the training data, its multilayer variant EDMD \cite{Williams:15} layers a second nonlinear functional map over the SVD. This approach is effective if one knows the nature of the nonlinearity {\it a priori}, but often results in a high-dimensional feature space. The kernel variant of this method, KDMD \cite{Williams:14arXiv} helps reduce dimension, but limited by the approximation capabilities of the kernel function.  

A major limitation of all such multilayer methods is related to the sequential learning of the feature maps independent of each other, i.e. learning occurs through local features as a one time-measure in a specified direction and the upstream map isnot adjusted for a downstream map. This `one-way and one-time' learning process limits the representational capacity of the model for handling nonlinear fluid flows.  Deep neural networks (DNNs) have been employed to identify multilayer maps without such limitations. Particularly, such DNN-based multilayer maps have been used to embed the nonlinear dynamical system into a Koopman function space~\cite{Shiva:18AIAA,Otto:17KDN,Lusch:17KDN} governed by linear dynamics \cite{Mezic:05}. They are expected to provide improved performance due to the `two-way and iterative' process of learning the model so that the optimal nonlinear multilayer map can be discovered instead of the assumed structure.   In this paper we term the former as \emph{Multilayer Sequential Maps} or (MSMs) and the latter as \emph{Multilayer End-to-end Maps} (MEMs). 

In this work, we carefully and systematically asses the predictive performance of both the sequential (MSM) and end-to-end (MEM) learning of Markov models of complex nonlinear dynamics using limited data. In particular, for the sequential maps (MSMs) we restrict ourselves to the popular class of Koopman approximation methods such as Dynamic Mode Decomposition~\cite{SchmidDMD:10,Rowley:09} and its extensions~\cite{Williams:15}. For the end-to-end learning architectures (MEMs) we focus on different types of feed forward neural networks (FFNN), a robust approach for learning the embedded nonlinearity in the dynamics from data.  In all the case studies considered, the maps are carefully chosen so as to minimize variability so that we can focus purely on the effect of the \emph{sequential versus end-to-end optimization} on the learning of the dynamics. Proper orthogonal decomposition (POD) is used as the first layer in all the above architectures in order to operate in a low-dimensional feature space. 

The outcomes of our study indicate that MSMs in spite of including multiple layers behave more like shallow neural networks while MEMs carry the advanced function approximation capabilities of deep learning tools. It is well known that while shallow NN are known to possess universal function approximation properties \cite{Hornik:89}, it usually requires exponentially more neurons (features) for accurate prediction as compared to deeper architectures. Deep neural networks (DNNs) offer a low-dimensional (short) and layered (deep) alternative for high (almost exponential) representational capacity of complex data.  This low-dimensional feature space also helps limit overfitting in a relative sense, i.e. as compared to MSMs. In particular, we observed that for a similar architecture, i.e same number of layers and feature dimension, MEMs offer robust and accurate learning performance using the same training data as compared to MSMs by leveraging an extended learning parameter space with elements estimated concurrently using  nonlinear regression techniques. Similar performance from MSMs require very `tall' layers that cause overfitting. These ideas are illustrated using different flow case studies including  transient dynamical evolution of a cylinder wake towards a limit-cycle attractor and a transient buoyancy-driven mixing layer. The organization of this paper is as follows.
In section \ref{s:methods} we present an overview of data-driven Markov models for transient dynamical systems and their connections to neural networks (section~\ref{ss:markovNN}) and Koopman theoretic methods (section  \ref{ss:markovkoopman}). In section \ref{ss:markovLOC}, we describe multilayer sequential maps (MSM) for Markov modeling and its two variants in subsections \ref{sss:DMDLOC} and  \ref{sss:EDMDLOC}. In section~\ref{ss:markovGOC} we introduce feed forward neural network based Markov representations. The numerical examples and discussion of the modeling performance is presented in section \ref{s:results} and the various outcomes are summarized with discussion in section \ref{s:conclusions}.

\section{Data-driven Markov Models for Transient Dynamical Systems} \label{s:methods}
\par Extraction of high-fidelity Markov models from snapshot (time) data of nonlinear dynamical systems is a major need in science and engineering, where measurement data can be the only available piece of information. It is advantageous to learn the model in a low-dimensional feature space to both simplify the learning process and also improve efficiency. Most Markov models are built as linear operators in the feature space to take advantage of the powerful linear systems machinery for control \cite{Kim:07ARev}, optimization and spectral analysis \cite{Rowley:09} although this is not necessary for the following formulation.
Given a discrete-time dynamical fluid system that evolves as below:
\begin{equation}
\pmb y = \pmb x_{t+T}= \pmb{F}(\pmb x_t)= \pmb{F}(\pmb x)
\label{e:nonlinear}
\end{equation}
where ${\pmb x},{\pmb y} \in \mathcal{M}$ are $N$-dimensional state vectors ($\mathbb{R}^N$), e.g., velocity components at discrete locations in a flow field at a current instant $t$, and separated by an appropriate unit of time $T$. To be explicit, ${\bm x} \triangleq {\bm x_t}$ and ${\pmb y} \triangleq {\pmb x}_{t+T}$. Operator $\pmb {F}$ evolves the dynamical system nonlinearly from $\pmb x$ to $\pmb y$, i.e. $\pmb {F}: \mathcal{M}\to \mathcal{M}$. This representation can easily be made relevant to continuous time systems as well in the limit $t\rightarrow0$. A general (linear) Markov description of such a dynamical system is given in eqn.~\eqref{e:genMarkov} : 
\begin{equation}
{\bm g}(\pmb y)={\bm g}(\pmb x_{t+T})=\mathcal{D}{\bm h}(\pmb x_t)=\mathcal{D}{\bm h}(\pmb x).
\label{e:genMarkov}
\end{equation}
Here,  ${\bm g}(\pmb y)$ and ${\bm h}(\pmb y)$  are vector-valued transformations (components of $\bm {g,h}$ are scalar-valued) to a feature space.
In general, $\bm {g,h} \in \mathcal{F}$ (where $\mathcal{F}$ is a function space) are infinite-dimensional, but approximated into a finite-dimensional vector in practice and $\mathcal{D}: \mathcal{F} \to \mathcal{F}$.  
Without loss of generality we use the first order Markov process approximation of the dynamical system, i.e. $\bm g(\bm x_{t+T})=\mathcal{D}{\bm h}(\bm x_{t})$ in the above discussion.
That said, the algorithms presented here can easily be generalized to $n^{th}$ order processes. Mathematically, we can represent such as system as $\bm g(\bm x_{t+T})=\mathcal{D}{\bm h}({\bm x_t},{\bm x_{t-T}},{\bm x_{t-2T}},{\bm x_{t-3T}}...{\bm x_{t-(n-1)T}})$. 
In subsection \ref{ss:markovNN} we explore how feedforward neural networks and the popular Koopman approximation-based methods build such Markov representations. 

\subsection{ Markov Model Approximation using Feed-forward Neural Networks (FFNNs)}\label{ss:markovNN}

\begin{figure}
\begin{center}
\includegraphics[width=0.81\columnwidth]{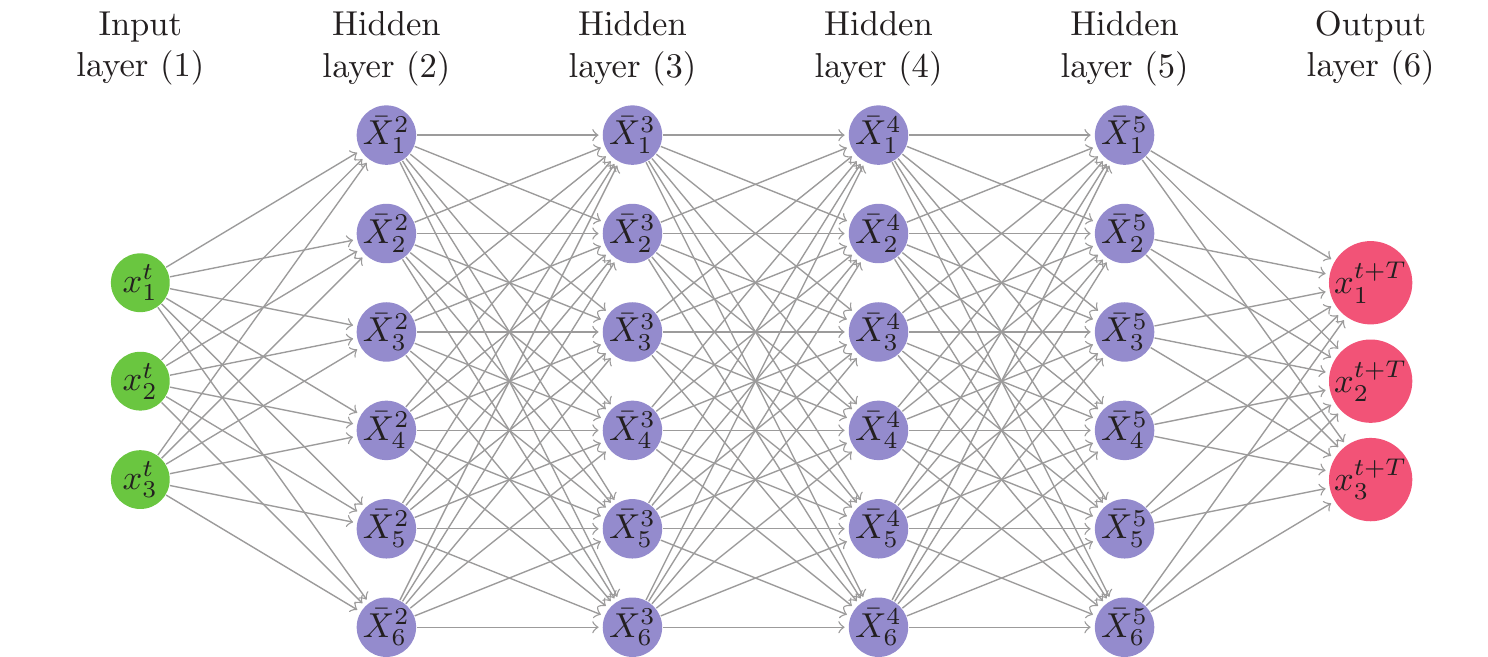}
\caption{A six-level Feed forward Neural Network architecture for building a Markov model. \label{f:6levelFFNN}}
\end{center}
\end{figure}

The key to developing a model for the Markovian dynamics is to learn the transformations $\bm g,\bm h$ and the operator $\mathcal{D}$. As mentioned earlier, each of ${\bm g,\bm h}$ and $\mathcal{D}$ can be either linear or nonlinear.  FFNNs are excellent function approximators~\cite{Hornik:89} that one can use to learn these maps ${\bm g,\bm h} \textnormal{ or }\mathcal{D}$ for a given training data. A standard FFNN architecture as shown in fig.~\ref{f:6levelFFNN} involves a linear map, $\Theta_l$ applied to the features ($\bar{X}^l$) at any given layer followed by the application of a nonlinear activation function, $\mathcal{N}_l$. Usually, $l=1\dots L$, indicating a $L$-layered network governed by the recursive relationship as shown in eqn.~\eqref{eq:DNN1}. $\bar{X}^1$ and  $\bar{X}^L$ represent the input ($\pmb{x}^t$) and output ($\pmb x^{t+T}$) features respectively. It is common to include a bias term inside the parentheses in eqn.~\eqref{eq:DNN1}, i.e. $\bar{X}^{l+1}=\mathcal{N}_l\left(\Theta_l \bar{X}^l +  b^l \right)$ for improved approximation properties. 
\begin{equation}
\bar{X}^{l+1}=\mathcal{N}_l\left(\Theta_l \bar{X}^l \right)
\label{eq:DNN1}
\end{equation}

\begin{equation}
{\pmb x}^{t+T}=\bar{X}^{L}=\mathcal{N}_{L-1}\left(\Theta_{L-1}\mathcal{N}_{L-2}\left(\Theta_{L-2}
\mathcal{N}_{L-3} \left(\cdots\left(\Theta_1 {\pmb x}^t \right)\right)\right)\right)=FFNN({\pmb x}^t)
 \label{eq:DNN2}
\end{equation}

Using this framework, we can develop an evolutionary model that approximates $\pmb g,\ \pmb h \textnormal{ and } \mathcal{D}$ as shown below in eqn.~\eqref{eq:DNN2}.
The above represents a nonlinear regression model of the data which can be interpreted in many ways.  
A convenient interpretation adopted in this article is that $\bm{g}$ and $\mathcal{D}$ are identity maps, i.e. $\bm g(a)=\mathcal{I}a=a$ and $\mathcal{D}a=Ia=a$ for any $a\in \mathbb{R}^N$ and $\bm{h}$ is given by the `layering' of $\Theta_l,\ \mathcal{N}_l$, i.e. $\bm h(a)={\it FFNN}(a)$. For this interpretation, the feature space coincides with the input state space. Other interpretations are possible by splitting the FFNN into a combination of $\bm h$ and $\mathcal{D}$. A key takeaway here is the use of a layered map to approximate $\bm g$ or $\bm h$  or both, which is nonlinear and can be asymmetric, i.e. $\bm h \neq \bm g$. Typically, it is not possible to have $\bm g=\bm h$ using a standard FFNN, but one can use machine learning tools such as deep autoencoder networks~\cite{hinton2006reducing} to accomplish this as reported in~\cite{Shiva:18AIAA}. Irrespective of the chosen interpretation, another key takeaway is that the learning process is end-to-end, i.e., it includes estimating the entire set of $\Theta_l$ (with $\mathcal{N}_l$ specified) by inverting the nonlinear system in eqn.~\eqref{eq:DNN2} requiring significant training cost.  For this reason, it is not uncommon to reduce of the dimension of  input and output features using projections onto some sparse basis such as the singular vectors of the data matrix as used in this study. 
\subsection{Markov Model Approximation using Koopman Framework}\label{ss:markovkoopman}
In the earlier section, we interpreted FFNNs as an asymmetric Markov model with a linear transition operator, identity map, $\pmb g$ and a nonlinear map $\pmb h$.
In this section, we explore how first order Markov models can be represented through the class of Koopman operator-theoretic frameworks\cite{Mezic:05,Koopman:31} for modeling nonlinear dynamics.  A Markov process is approximated by the Koopman operator \cite{Mezic:05,Koopman:31,Rowley:09} under conditions of $\bm g=\bm h$ and $\mathcal{D}=\mathcal{K}$ with $\mathcal{K}$ being linear. 
In the Koopman framework, the feature space is the space characterized by a vector of observables $\pmb g(\pmb x),\pmb h (\pmb x)$ and the feature maps, $\pmb g,\pmb h$ represent a vector of observable functions. The operator theoretic view\cite{Mezic:05,Williams:15}  interprets $\mathcal{K}$ as operating on the space of functions $\mathcal{K}: \mathcal{F} \to \mathcal{F}$. 
When $\pmb g$ and $\pmb h$ are identical, then the linear operator $\mathcal{K}$ evolves the Markovian dynamics in Eq. (\ref{e:genMarkov}) as a Koopman evolutionary model given by:
\begin{equation}
	\mathcal{K}\bm g(\pmb x)=\bm g(\pmb y)=g({\pmb F}(\pmb  x)). 
	\label{e:koopman1}
\end{equation}
This representation is exact when $\bm {g,h} \in \mathcal{F}$ span the infinite-dimensional function space, $\mathcal{F}$ in which the koopman operator, $\mathcal{K}$ acts. However, one often uses a finite-dimensional approximation in practice.  
Since, the Koopman operator has the effect of operating on the functions of state space as shown in eqn. (\ref{e:koopman1}), it is commonly referred to as a composition operator where $\circ$ represents the composition between $\bm g$ and the exact model describing the dynamical system, $\pmb F$. 
\begin{equation}
	\mathcal{K}{\bm g}={\bm g}{\circ}{\pmb F}.
	\label{e:koopman2}
\end{equation}
Being a linear operator, the products of Koopman spectral analysis  such as the eigenfunctions ($\phi_{j}$), eigenmodes ({$v_{j}$) and eigenvalues ({$\mu_{j}$) can be leveraged to reconstruct the transformation $\bm g(\pmb x)$ as shown in eqn. \eqref{e:evolution1} provided the elements of $\pmb g$ lie in the span of $\phi$. 
If this is true, then the evolutionary model can be represented as in eqn.~\eqref{e:evolution2}.
\begin{equation}
\bm g(\pmb x)=\sum_{j=1}^{\infty} \phi_{j}{\pmb v}_{j}
\label{e:evolution1}
\end{equation}
\begin{equation}
\bm g(\pmb y)=\mathcal{K}{\bm g}(\pmb x)=\sum_{j=1}^{\infty} \phi_{j}{\pmb v}_{j}\mu_{j}
\label{e:evolution2}
\end{equation}
In practice, a temporal sequence of data, $(\pmb x^{T}, \pmb x^{t+T}\dots)$ generated by a nonlinear dynamical system as in eqn.~\eqref{e:nonlinear} needs to be represented using a Koopman Markov framework as in eqn.~\eqref{e:koopman_sys} where, $\pmb g,\pmb h$ is yet to be identified (or modeled). 
\begin{equation}\label{e:koopman_sys}
\bm g(\pmb{x}^{t+T})=\bm g(\pmb{F}(\pmb{x}^{t}))=\mathcal{K} \bm g(\pmb{x}^{t}) 
\end{equation} 
Arranging the data into snapshot pairs as $X=(\pmb x^{t} \dots\ \pmb x^{t+(M-2)T}, \pmb x^{t+(M-1)T})$ and $Y=(\pmb x^{t+T}\dots\ \pmb x^{t+(M-1)T}, \pmb x^{t+MT})$ such that $(X,Y) \in \mathbb{R}^{N \times M}$, eqn.~\eqref{e:nonlinear} can be recast as $Y=\pmb{F}{X}$ with a corresponding quasi-linear form given by ${Y} \approx \pmb A(X)X$ with $N, M$ representing the dimensions of instantaneous system state and data snapshots respectively.
The observable function $\bm g$ is unknown and modeled as a finite-dimensional map $C \in \mathcal{R}^{N \times K}$ that can either be functional or data-driven. It is common to treat the map as a projection of the input state onto an appropriate basis such that the dynamics evolve in a feature space that is  low-dimensional ($K\ll N$). This would require $\pmb x^t$ be spanned accurately by the basis forming the columns of $C$.
 
 A finite-dimensional approximation of $\mathcal{K}$ (in eqn.~\eqref{e:koopman_sys}) corresponding to the choice of $C \in \mathbb{R}^{N \times K}$  can be obtained using the following method. The approximation of $\pmb g$ is given by the relationship $C \pmb g(\pmb x^t)\approx \pmb x^t$ for a single snapshot and $C \pmb g(X)=X=C\bar{X}$ for a collection of snapshots. Substituting $X = C\bar{X}=C \pmb g(X)$ and $Y = C\bar{Y}=C \pmb g(Y)$ in the linearized model for the dynamical system $Y \approx A(X)X$, we get
 \begin{equation}
 C^{+} \pmb A(X)C\bar{X} = \bar{Y}, 
 \label{e:conv4}
 \end{equation}
 where $C^{+}$, is the left Moore-Penrose pseudo-inverse.  $C^{+}$ should be computable as $C^TC$ is not likely to be rank deficient for $K\ll N$. Relating eqn.~\eqref{e:koopman1} to eqn.~\eqref{e:conv4} for $\pmb g(X)=\bar{X}$ and $\pmb g(Y)=\bar{Y}$, we get a linear finite-dimensional approximation for the Koopman operator, $\mathcal{K}$ given by  $\mathcal{K}\approx {\Theta} \triangleq C^{+} \pmb A(X)C$ that governs the evolution of the dynamics in the feature space. $\bar{X} \in \mathbb{R}^{K \times M}$ and $\bar{Y} \in \mathbb{R}^{K \times M}$ are the representations of the state in the feature space. 
Naturally, the fidelity of the above approximation to the dynamical system in eqn.~\eqref{e:nonlinear} depends on the choice of $C$ as an approximation to $\bm g$.  Further, for ${\Theta}$ to be truly linear, it is easy to infer that $C$ will have to evolve with the state $\pmb x^t$ as $C(\pmb x^t)$. For detailed discussion on the architecture and choice of map maps we refer to Lu and Jayaraman \cite{Lu:18sparse}. For a chosen $C$ and given $X,Y$, we can learn ${\Theta}$ by minimizing the frobenius norm of $\|{{\bar Y}} - {\Theta}{\bar X}\|_{F}$ via $\Theta = \bar Y {\bar X}^+$. In principle one could minimize the $2$-norm $\|{{\bar Y}} - {\Theta}{\bar X}\|_{2}$ at the risk of added complexity, but the Frobenius norm serves an efficient alternative.

\section{Modeling the Feature Maps $g,h$}\label{ss:markovLOC}\label{s:featuremaps}
In the previous section, we interpreted FFNNs as learning one side of a feature map, i.e., $\pmb h$ in a Markov model with $\pmb g,\ \mathcal{D}$ being modeled as identity maps. Contrastingly, the Koopman theoretic methods assume a model for the feature maps, i.e. $\pmb g=\pmb h = C$ and learns the linear transition or Koopman operator $\mathcal{D}=\mathcal{K}\approx \Theta$ using data. As the Koopman framework is symmetric it allows one to learn a low-dimensional linear transition operator that can be used for spectral analysis. The FFNN's offer no such luxury \cmnt{as the data-driven learning is focused on identifying the feature map}. The key to the success of both approaches for extended predictions relies on the accuracy of the feature map approximations either using data (FFNNs) or through models (Koopman). A prominent approach to improving the fidelity in Koopman approximation methods is to sequentially layer elementary maps (both functions and basis projections) in a supervised fashion and then approximate the Koopman operator in the resulting feature space. While this layering approach is similar to FFNNs, there exist key differences, We will explore these in the following subsections. 
 
\subsection{Multilayer Sequential Maps (MSMs) for Koopman Approximation}\label{ss:markovLOC}

For Koopman approximation methods, the basis space onto which the input state is mapped should evolve with the state itself, i.e., $C$ should be $C(X)$ so that $\Theta \approx \mathcal{K}$ can be linear. However, this often leads to a futile search for `magic' basis \cmnt{when the nature of the dynamical system is unknown}. Alternative approaches such as Extended Dynamic Mode Decomposition or EDMD~\cite{Williams:15} include approximating the functional form of the map from data using a dictionary of basis functions. However, the dependence of $C$ on the choice of functions populating the dictionary and the relative ease with which the feature dimension grows, limits these approaches. In \cite{Lu:18sparse}, Jayaraman and collaborators propose an alternate approach to building complex and efficient maps through layering of elementary operators based on the hypothesis that \emph{deeper and shorter is better than taller and shallow} operators. This approach is similar to the kernel DMD framework~\cite{Williams:14arXiv} that combined EDMD with a kernel principal component analysis (KPCA) to improve efficiency. This evolution of methods align with the recent successes in \emph{deep learning} ideas for artificial intelligence~\cite{bengio2007challenge} in spite of the need to tune many hyperparameters. It is worth noting that both strategies increase the number of model parameters to be learned, but layering offers a systematic way to build model complexity for data-driven learning as compared to shallow architectures~\cite{Hornik:89}.  A generalized way of building $C$ is to layer recursively multiple mapping operators such as: 
\begin{equation}
X = C_{1}C_{2}{\bar{X}^2} = {\mathcal{C}}_{ML}{\bar{X}}^2,
\label{e:mconv1}
\end{equation}	
\begin{equation}
Y = C_{1}C_{2}{\bar{Y}^2} = {\mathcal{C}}_{ML}{\bar{Y}^2},
\label{e:mconv2}
\end{equation}
where ${\bar{X}}^2$ and ${\bar{Y}}^2$ represent the features at the $2^{nd}$ layer and ${\mathcal{C}}_{ML}$ represents the multilayer sequential map. A schematic of such a model is presented in fig~\ref{f:MLCF} where the state $X$ is operated by a two-layer map $C^{+}_{1}C^{+}_{2}$ to yield ${\bar{X}}^2$. Similarly,  $C^{+}_{1}C^{+}_{2}$ operate on $Y$, $\bar{Y}^1$ respectively to  yield ${\bar{Y}}^2$ . An approximate linear Koopman operator $\Theta$ is then learned from ${\bar{X}}^2$  and ${\bar{Y}}^2$. 
\begin{figure}
\begin{center}
\includegraphics[width=1.0\columnwidth]{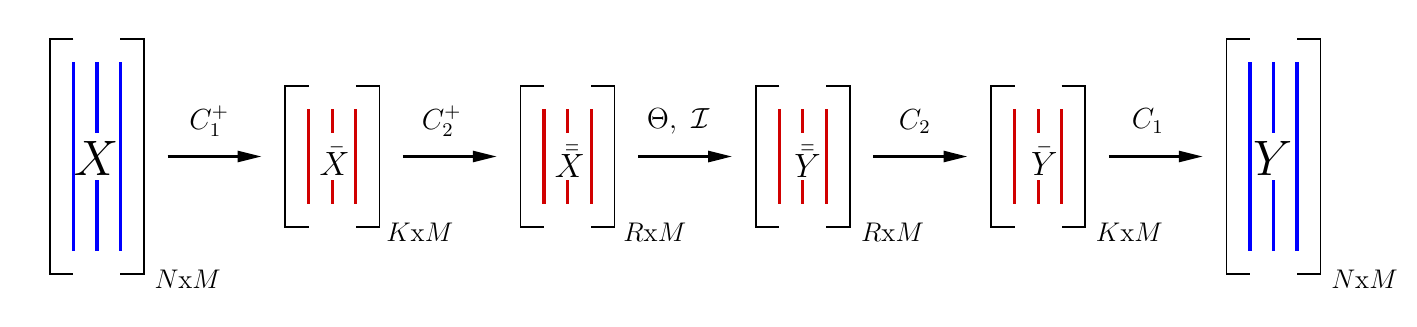}
\caption{Schematic of a six-layer representation of the multilayer sequential map (MSM) framework to approximate the Koopman operator. $X,\ Y$ represent state space matrices and $C^{+}_{i}$, $C_{i}$, represent the elemental maps and its inverse operation respectively. The arrows indicate direction of the maps, i.e., $C^{+}_{1}$ acts on $X$ to yield $\bar{X}^1$ and $C_{1}$ acts on $\bar{Y}^1$ to yield $Y$.  $\Theta$ represents the approximation Koopman operator shown in Eq.(\ref{e:koopman_sys}). The size of data matrices in the high ($X$) and low dimensional($\bar{X}$ or $\bar{\bar{X}}$ ) space is also shown.}
\label{f:MLCF}
\end{center}
\end{figure}
Substituting eqns.~\eqref{e:mconv1} and \eqref{e:mconv2} into eqn.\eqref{e:conv4}, we have:
\begin{equation}
{{\mathcal{C}}_{ML}}^{+}A{{\mathcal{C}}_{ML}}{\bar {X}^2} = \Theta{\bar {X}^2}= {\bar {Y}^2},
\label{e:mconv4}
\end{equation}
with $\Theta \triangleq {{\mathcal{C}}_{ML}}^{+}A{{\mathcal{C}}_{ML}}$. Instead of the two-layer map, we can have a deep architecture with ${{\mathcal{C}}_{ML}}^{+}=C_{L}^{+}...C_{1}^{+}C_{2}^{+}C_{3}^{+}$ and ${{\mathcal{C}}_{ML}}=C_{3}C_{2}C_{1}...C_{L}$ where $2(L+1)$ represent the total number of layers in the design. The encoder map ${{\mathcal{C}}_{ML}}^{+}$ can be computed as long as the elemental maps, $C_i$, are invertible in a generalized sense. Although this MSM formulation is designed for Koopman approximations, i.e. $\bm g = \bm h = \mathcal{C}_{ML}$, one can build generalized Markov versions of this model  i.e. $\bm g =  \mathcal{C}_{ML1}$   \&  $ \bm h = \mathcal{C}_{ML2}$. A key limitation of such MSM frameworks is that  $C_i$ and consequently $\mathcal{C}_{ML}$ are usually predetermined maps (or functions) and the Koopman approximation relies only on the local features. In the following subsections \ref{sss:DMDLOC} and \ref{sss:EDMDLOC} we present some of the Koopman approximation methods in the MSM context.

\subsubsection{DMD as a Four-level Multilayer Sequential Map (MSM) based Markov Model}\label{sss:DMDLOC}
\begin{figure}
\begin{center}
\includegraphics[width=0.9\columnwidth]{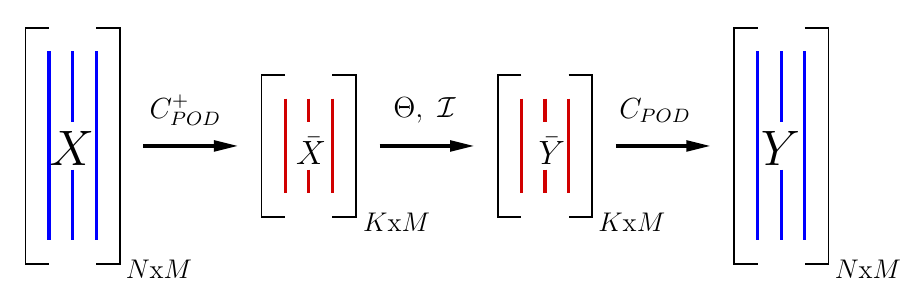}
\end{center}
\caption{Dynamic Mode Decomposition (DMD):  four-level MSM framework with linear maps for Koopman approximation.\label{f:MLCF_4}}
\end{figure}
There exist many methods to approximate the Koopman tuples including DMD \cite{Rowley:17ARev,SchmidDMD:10}, EDMD \cite{Williams:15} and its kernel variant, Kernel DMD~ \cite{Williams:14arXiv} and generalized Laplace analysis (GLA) \cite{Mezic:05}.  DMD~\cite{SchmidDMD:10} employs observables which are linear functions of the state.  The multilayer architecture for DMD is shown in fig.\ref{f:MLCF_4} as a four-layer framework containing with both forward and backward maps, i.e. $X\rightarrow \bar{X}^1\Leftrightarrow\bar{Y}^1\leftarrow Y$ consisting of POD basis (via SVD~\cite{Trefethen:97} of the training data ($X$)) projections. Given data snapshots separated in time $X,Y$ as before, we use $X=C_{POD}{\bar X}^1$ and $Y=C_{POD}{\bar Y}^1$ to generate a Koopman approximation subject to $\Theta \bar X^1 = \bar Y^1$. 
The pairs ${\bar X}^1$ and ${\bar Y}^1$ have the structure shown in eqn.~\eqref{e:PODfeatures} where $a_i^j$ represent the $i^{th}$ POD coefficient of the $j^{th}$ snapshot. 
\begin{equation}\label{e:PODfeatures}
{\bar X}^1  = \begin{bmatrix}
$$a_{1}^{1}$$ & $$a_{1}^{2}$$ & $. . . . .$ & $$a_{1}^{M-1}$$ & $$a_{1}^{M}$$\\
$$a_{2}^{1}$$ & $$a_{2}^{2}$$ & $. . . . .$ & $$a_{2}^{M-1}$$ & $$a_{2}^{M}$$\\
$.$&$.$&$. . . . .$&$.$&$.$\\
$.$&$.$&$. . . . .$&$.$&$.$\\
$$a_{K}^{1}$$ & $$a_{K}^{2}$$ & $. . . . .$ & $$a_{K}^{M-1}$$ & $$a_{K}^{M}$$\\
\end{bmatrix}_{(K\textrm{x}M)}
\textrm{and}\ \ \ \
{\bar Y}^1 = \begin{bmatrix}
$$a_{1}^{2}$$ & $$a_{1}^{3}$$ & $. . . . .$ & $$a_{1}^{M}$$ & $$a_{1}^{M+1}$$\\
$$a_{2}^{2}$$ & $$a_{2}^{3}$$ & $. . . . .$ & $$a_{2}^{M}$$ & $$a_{2}^{M+1}$$\\
$.$&$.$&$. . . . .$&$.$&$.$\\
$.$&$.$&$. . . . .$&$.$&$.$\\
$$a_{K}^{2}$$ & $$a_{K}^{3}$$ & $. . . . .$ & $$a_{K}^{M}$$ & $$a_{K}^{M+1}$$\\
\end{bmatrix}_{(K\textrm{x}M)}
\end{equation}
Knowing $\Theta$ allows one to model the Markovian evolution of this dynamical system in the feature space. In this MSM framework, the data-driven learning is accomplished through a local optimization as shown in eqn.~\eqref{e:LDMD}, i.e. through minimizing the mapping error between the two immediate layers constituted by pairs of features $\pmb{a}^t,\pmb{a}^{t+1}$ which are the column vectors in ${\bar X}^1,{\bar Y}^1$. Estimating $\Theta$ that minimizes the Frobenius norm $\| \bar{Y}^1 - \Theta \bar{X}^1\|_F$ requires computing a least squares solution to eqn.~\eqref{e:LDMD} as $\Theta = \bar Y \ (\bar{X}+ \lambda I)^{+} $ where $()^{+}$ denotes the generalized Moore-Penrose pseudo-inverse \cite{Trefethen:97}. $\lambda$ is a $l_2$ regularization~\cite{Scholkopf:01} parameter to generate a unique solution for this overdetermined system (with $K < M$). 
\begin{equation}\label{e:LDMD}
{\bar Y}^1 =\begin{bmatrix}
$$\pmb{a}^{2}$$ & $$\pmb{a}^{3}$$ & $$\dots$$ &$$\pmb{a}^{i+1}$$ & $\dots$ & $$\pmb{a}^{M+1}$$
\end{bmatrix} = \Theta 
\begin{bmatrix}
$$\pmb{a}^{1}$$ & $$\pmb{a}^{2}$$ & $$\dots$$ &$$\pmb{a}^{i}$$ & $\dots$ & $$\pmb{a}^{M}$$
\end{bmatrix} =\Theta {\bar X}^1
\end{equation}
In this $4-$layer MSM architecture, the maps between any two layers are layer-wise optimal and the sequence of application, i.e. map direction becomes critical as $\Theta$ depends on $C_{POD}$, but not vice versa. 

\subsubsection{Extended DMD: A Six-level Multilayer Sequential Map (MSM) based Markov Model}\label{sss:EDMDLOC}

\begin{figure}
\begin{center}
\includegraphics[width=1.0\columnwidth]{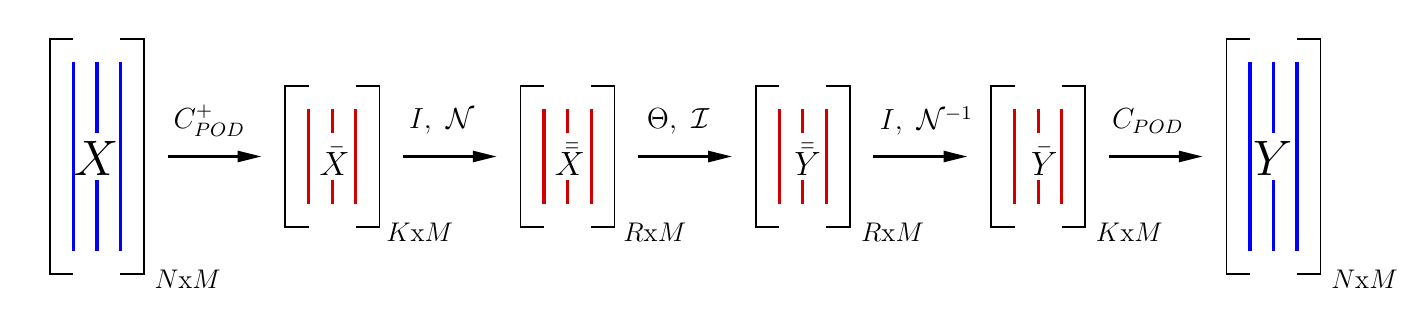}
\caption{Extended Dynamic Mode Decomposition (DMD): A six-level Koopman approximation MSM framework with nonlinear maps ($\mathcal{N}, \ \mathcal{N}^{-1}$). $I$ represents the identity linear operator and $\mathcal{I}$, the identity function. }
\label{f:MLCF_6}
\end{center}
\end{figure}
In the earlier DMD multilayer framework, the elemental maps $C$ were linear functions of the state (i.e. the POD features were computed form training data and not the the instantaneous flow state) which has difficulty modeling nonlinear dynamics\cite{Rowley:17ARev,Lu:18sparse}. Extensions to DMD such as EDMD \cite{Williams:15,Rowley:17ARev} help alleviate this problem to some extent by layering nonlinear maps ($\bar{X}^2=\mathcal{N}\left( I\bar{X}^1 \right)$ and $\bar{Y}^2=I\mathcal{N}\left(\bar{Y}^1 \right)$) over linear ones ($\bar{X}^1=\mathcal{I}\left(C_{POD}^+X\right)$ and $\bar{Y}^1=C_{POD}^+\mathcal{I}\left(Y\right)$)  as shown in fig.~\ref{f:MLCF_6}.
The architecture for the EDMD in fig.~\ref{f:MLCF_6} is a $6-$level framework ($4-$ layer without the POD-map for dimensionality reduction) with both forward and backward maps, i.e. $X\rightarrow \bar{X}^1\rightarrow \bar{X}^2\Leftrightarrow\bar{Y}^2\leftarrow \bar{Y}^1\leftarrow Y$ with the first and fifth layers representing a linear-map made up of POD-basis of the training data while the $2^{nd}$ and $4^{th}$ layers represent nonlinear functional maps operating on the corresponding features. In the schematic, we present a generalized representation where each map consists of linear operators, i.e. $C_{POD},I$ and functional maps, $\mathcal{N},\mathcal{I}$ with $I,\mathcal{I}$ representing the operator and functional forms of the identity map respectively. However, the architecture represented in fig.~\ref{f:MLCF_6} has a `forward' direction with the nonlinear mapping in the $4^{th}$ layer denoted by an inverse function $\mathcal{N}^{-1}$ that may not always be well behaved. In practice, the layers six to five to four flow backward, i.e. $Y\rightarrow \bar{Y}^1\rightarrow \bar{Y}^2$ ($\bar{Y^2}=I\mathcal{N}\left(\bar{Y}^1 \right)$ and $\bar{Y}^1=C_{POD}^+\mathcal{I}\left(Y\right)$) which helps bypass such issues.  The approximation to the Koopman operator, $\Theta$ is estimated as the optimal linear operator that relates the features $\bar{X}^2$ and $\bar{Y}^2$ in a least squares sense (i.e. find the $\Theta$ that minimizes the Frobenius norm $\|{\bar Y}^2- \Theta {\bar X}^2\|_F$) as was shown for the DMD framework. In this study, we present two variants of this method corresponding to different choices of $\mathcal{N}$, namely, EDMD-P \cite{Williams:15} which uses polynomial functions (eqn.\eqref{e:nonlinconv1}) of the features ($\bar{X^2}=\mathcal{N}\left(I \bar{X}^1 \right)$) and EDMD-TS which uses a tan-sigmoid nonlinearity (eqn.\eqref{e:nonlinconv2}) . 
\begin{eqnarray}
\label{e:nonlinconv1}
\bar{\pmb {a}}=\mathcal{N}(\pmb{a})&=&\begin{bmatrix}
\pmb{a} \\
\pmb{a}\otimes \pmb{a}
\end{bmatrix}\\
\bar{\pmb {a}}=\mathcal{N}(\pmb{a}) &=& \tanh{\left(\pmb a\right)}
\label{e:nonlinconv2}
\end{eqnarray}
Here $\bar{ \pmb{a}}$ represents the features in the $\mathcal{N}$ space, i.e. columns of $\bar{X}^1,\bar{Y}^1$. It is easily seen that EDMD-P with $2^{nd}$ order polynomials leads to a quadratic growth in the feature dimension and even worse when using higher order polynomials. On the other hand, EDMD-TS does not lead to increase in the number of features. Just as in DMD, the EDMD MSM architecture optimizes the maps only between the two immediate layers and direction of the map (sequence of application of operators) strongly influences the model, i.e. $C_{POD}$ and $\mathcal{N}$ determine $\Theta$ but not vice versa. As a consequence of this layer-wise treatment and symmetric formulation (i.e. $\pmb g = \pmb h$), the map is bi-directional which makes learning the linear Koopman operator efficient. In the following section, we will focus on end-to-end  learning of the map using neural networks.

\subsection{Feed Forward Neural Networks (FFNN): Multilayer End-to-End Map (MEM) based Markov Models}\label{ss:markovGOC}
In principle, multilayer map increases the number of design variables, i.e. the choice of nonlinear functions ($\mathcal{N}$), depth ($L$) and dimension ($K,R$) of the layers in the model. For the MSM framework described in section~\ref{ss:markovLOC}, we observe that the direction of the map, choice of the elemental operators and order of layering can generate different representations. This is also true in the case of a standard feed forward neural networks (FFNNs) as depicted in fig.~\ref{f:MGCF_6}. The figure shows a six-layer FFNN architecture so as to compare against the six-layer MSM framework (EDMD) in fig.~\ref{f:MLCF_6}. 
Here, each interior map between any two layers includes a linear map $\Theta_i,\ \left(i=1..5\right)$ and nonlinear transfer functions $\mathcal{N}_i,\ \left(i=1..5\right)$ with the latter predetermined for a given model. 

One can mimic the EDMD exactly using the FFNN framework by setting $\Theta_1=C_{POD}$, $\Theta_5=C_{POD}^+$ and $\Theta_2=\Theta_4=I$ where $I$ is the identity tensor, $\mathcal{N}_2=\mathcal{N}$, $\mathcal{N}_4=\mathcal{N}^{-1}$, $\mathcal{N}_1=\mathcal{N}_3=\mathcal{N}_5=\mathcal{I}$ where $\mathcal{I}$ is the identity map along with $\Theta_3=\Theta$, the Koopman operator. 
For this FFNN architecture that only supports forward maps, building a map with $\mathcal{N}^{-1}$ is not explored currently. This is because, for many common choices of $\mathcal{N}$, $\mathcal{N}^{-1}$ is not always bounded. It is for this reason, even in the MSM architectures, the backward operation is preferred. In this study, we use a tansigmoid function for $\mathcal{N}_{2,3,4}=\mathcal{N}$. Further, since we are dealing with high-dimensional flow datasets, we set $\mathcal{N}_{1,5}=\mathcal{I}$, $\Theta_1=C_{POD}$ and $\Theta_5=C_{POD}^+$ to reduce dimensionality of the interior layers. This leaves $\Theta_1,\Theta_2 \textrm{ and }\Theta_3$ to be determined from data. \cmnt{In addition to reducing the training cost, this also helps reduce overfitting.}  In addition, we did include a bias term to facilitate better comparison with conventional MSM architectures such as DMD and EDMD.

While similar in architecture, a key difference between the EDMD/DMD (MSM) and FFNN approaches is how they leverage the extended model hyperparameter space (e.g. elements of the Koopman operator) for learning from data. In the MSM framework, the linear parts of the map are either precomputed (i.e. $C_{POD}$) or assumed (i.e. $I$) for a given model design which allows estimation of the layerwise features before solving for the unknown Koopman operator $\Theta$ (size $R\times R$) using linear regression techniques. In the FFNN framework, the linear operators $\Theta_1,\Theta_2 \textrm{ and }\Theta_3$ are all unknowns while the nonlinear activation functions $\mathcal{N}_1,\mathcal{N}_2 \textrm{ and } \mathcal{N}_3$ are specified in the design. To learn the optimal solutions for $\Theta_{i},  i=1,2,3$, one needs constrain the resulting Markov model to the training data and solve a nonlinear regression problem~\cite{bengio2015deep}. In this way, the FFNN architecture takes advantage of the extended model hyperparameter space offered by the multilayer map by learning $K\times R+R\times R+R \times K$ parameters in $\Theta_{i},  i=1,2,3$ as against just $R \times R$ parameters in $\Theta$.  Such frameworks that incorporate `end-to-end' learning can be characterized as \emph{Multilayer End-to-end Map (MEM)} based Markov models. It is anticipated that MEM frameworks can offer improved representations of nonlinear dynamics as compared to MSM frameworks. It is well known that MSMs work well for predicting select dynamics but fail to model highly transient nonlinear systems. The downside of such FFNN/MEM framework includes: (i) increased training cost to estimate more unknowns than the MSM framework; (ii) propensity to generate non-unique solutions that require regularization and (iii) propensity to overfit data by learning more parameters, especially when using deeper networks which requires careful monitoring. 

\begin{figure}
	\begin{center}
		\includegraphics[width=1.0\columnwidth]{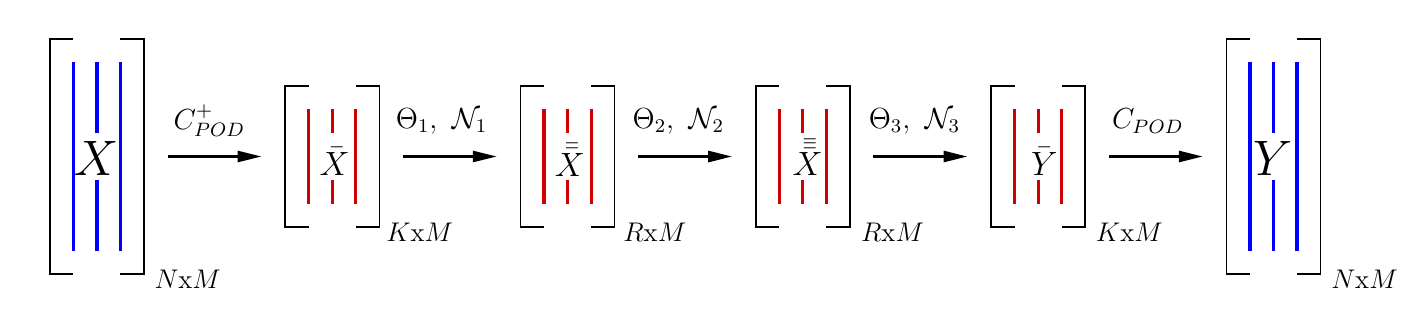}
		\caption{Feedforward Neural Network (FFNN) as a six-level Multilayer end-to-end map.  $\Theta_l, \ \mathcal{N}_l$ with the arrow represents the application of a linear operator followed by a nonlinear function.}
		\label{f:MGCF_6}
	\end{center}
\end{figure}

We briefly summarize the algorithm used for training the FFNN/MEM architecture. As before, $X,Y$ are the time dependent flow snapshot pairs and ${\bar X}^1,{\bar Y}^1$ represent the snapshots of time-dependent POD features as shown in eqn.~\eqref{e:LDMD} with columns $\pmb{a}^i$ and $\pmb{a}^{i+1}$ respectively. The effective nonlinear map is trained between $\bar{Y}^1$ and  $\bar{X}^1$  as shown below in eqns.~\eqref{e:FNN1}-\eqref{e:FNN3}:
\begin{equation}\label{e:FNN1}
{\bar{X}^2}=\mathcal{N}_2(\Theta_2 \bar{X}^1) 
\end{equation}
\begin{equation}\label{e:FNN2}
{\bar Y}^2=\mathcal{N}_3(\Theta_3 {\bar X}^2)
\end{equation}
\begin{equation}\label{e:FNN3}
{\bar Y}_p^1=\mathcal{N}_4(\Theta_4 {\bar Y}^2) 
\end{equation}
In general, a multilayer network is characterized by the recursive relationship $X_l=\mathcal{N}_l(\Theta_l X_{l-1})$ where $X_l,\Theta_l$ and $\mathcal{N}_l$ represent the mapped features, linear operator and nonlinear map relating the $l^{th}$ and $l+1^{th}$ layers. In this specific example, $\bar{Y}^1_{p}$ is the predicted features at the fifth layer to be compared with the ground truth, $\bar{Y}^1$, obtained from the training data as $\bar{Y}^1=C_{POD}Y$. The linear operator $\Theta_l,\ \textrm{with} \ l=2\dots (L-2)$, for a $L-$layer framework is estimated by minimizing the overall cost function as in eqn.~\eqref{e:regcost}:
\begin{equation}\label{e:regcost}
\begin{split}
\mathcal{J}(\Theta) = & \underbrace{\frac{1}{2M} \sum_{i = 1}^{M} \sum_{j = 1}^{K} (\bar{Y}_{p}(j,i) - 
	\bar{Y}(j,i) )^2}_\text{Feed forward Cost} \\ 
+ & \underbrace{\left(\frac{\lambda}{2M} \sum_{l = 
		2}^{L-2} \sum_{s = 1}^{S} \sum_{q = 1}^{Q} \left(\Theta_{l}(s,q)\right)^2 
	\right)}_\text{Regularization term}
\end{split}
\end{equation}
In the above, \cmnt{$\bar{Y}^1$ is the original data,} $S,Q$ represent the dimension of the features in layers $l \textrm{ and } l+1$ respectively. 
The optimal solution for $\Theta_l, \ l=2 \dots (L-2)$ is obtained using backpropagation with a gradient descent framework employing a Polack-Ribiere conjugate gradient algorithm~\cite{Golub:12matrix} that employs a Wolfe-Powell stopping criteria. This nonlinear inversion to estimate the $\Theta$'s is the most important distinction between MSM and MEM methods. The gradient descent framework requires $\mathcal{N}$ to be infinitely differentiable which is not always guaranteed when choosing $\mathcal{N}_l=\mathcal{N}^{-1}$.  
To minimize overfitting, we use $\mathcal{L}_2$ norm based regularization in the cost function in eqn.\eqref{e:regcost} with $\lambda$ as the tuning parameter. To characterize the dimension of intermediate layer features we use a factor $(N_f)$ that is multiplied with the input feature dimension, i.e., $S,Q=N_f\times \textrm{input feature dimension}$. For such FFNN architectures, designing a forward-backward map to learn the Koopman operator (as in MSM frameworks such as DMD \& EDMD) is hard to realize using regular backpropagation training. As shown in ~\cite{Shiva:18AIAA}, incorporating special feedback networks with some similarity to recurrent neural network offer a way forward. However, these aspects are beyond the scope of this article.

\begin{table}[]
	\centering
	\label{tab:compare}
	\begin{tabular}{|c|c|c|}
		\hline
		& \textbf{\begin{tabular}[c]{@{}c@{}}Multilayer Sequential Map\\ (MSM)\end{tabular}}                   & \textbf{\begin{tabular}[c]{@{}c@{}}Multilayer End-to-end Map\\ (MEM)\end{tabular}}                            \\ \hline
		\textbf{Interdependence of layers} & One-way dependence                                                                                   & Two-way dependence                                                                                            \\ \hline
		\textbf{Learning Paradigm}         & \begin{tabular}[c]{@{}c@{}}All the linear maps except \\ one are precomputed\end{tabular}            & \begin{tabular}[c]{@{}c@{}}All the linear maps across the layers \\ are computed  simultaensouly\end{tabular} \\ \hline
		\textbf{Real-time Learning Cost}   & \begin{tabular}[c]{@{}c@{}}Very efficient to train as most \\ layers are learnt offline\end{tabular} & \begin{tabular}[c]{@{}c@{}}Requires significantly more\\  time to train\end{tabular}                          \\ \hline
		\textbf{Map direction}             & Bi-directional (symmetric)                                                                           & Unidirectional (can be asymmetric)                                                                            \\ \hline
	\end{tabular}
\caption{Assessment of the sequential and end-to-end learning maps for generating Markov models }
\end{table}

\section{Numerical Experiments and Discussion}\label{s:results}

In this section we compare the predictive capabilities of MSM with MEM Markov models. While it is to be expected that learning an extended set of parameters by minimizing the training error cost function allows for improved predictions of time-series flow data, the dimension of this parameter set depends on nature of the model architecture. Consistent with the earlier sections, we adopt the nomenclature `L-Method-$\mathcal{N}-N_f$' to denote the different architectures and their respective parameters, where the `L' represents the total number of layers used to map from one flow state to another ($X\rightarrow Y$), $\mathcal{N}$ defines the choice of nonlinear function and $N_f$ represents the feature growth factor. For example, we can easily describe the EDMD framework in fig.~\ref{f:MLCF_6} as a 6-level multilayer-sequential map with a polynomial nonlinearity of order two as 6-MSM-P2-$N_f$ (EDMD-P2), while a 6-level EDMD with a tansigmoid nonlinearity is denoted by 6-MSM-TS-1 (EDMD-TS) where the number followed by TS represents the feature growth factor $(N_f)$ from the first layer to the next. A 4-level MSM representing the DMD architecture is denoted by 4-MSM-$\mathcal{I}$-1 (DMD), where $\mathcal{I}$ defines identity mapping and $M=1$ defines the feature growth factor. 
In this study, we have used FFNN as the MEM architecture with four different designs for comparative assessment. They are 6-MEM-TS-$N_f$ with $N_f\ = \ 1,\ 3,\ 9,\ 20$. The various model possibilities are delineated in section~\ref{ss:analframework}.  Section~\ref{ss:dataGen} details the generation of flow data from high fidelity computations for use in this study, namely the cylinder wake flow (sec.~\ref{sss:cylwake}) and the buoyancy-driven mixing flow (sec.~\ref{sss:buoymix}).

\subsection{Data generation}\label{ss:dataGen}

To assess the different modeling architectures and the learning algorithms, we build a database of snapshots of transient flow field data generated from high fidelity CFD simulations of a bluff body wake flow and a buoyancy-driven mixing layer. Both these flows are transient in their own way.  The cylinder wake flow evolves on a stable attractor and approaches limit-cycle behavior rather quickly while the buoyancy-driven flow is a transient mixing problem with dynamics that dies out in the long-time limit.  The former is an example of  `data-rich' situation where the training data requirement to predict the dynamics is limited. On the other hand, the latter represents a `data-sparse' situation where any amount of training data may not be sufficient to predict future evolution. We explore the performance of MSM and MEM architectures for both these situations. In the following section, we summarize the data generation process.

\subsubsection{Transient Wake Flow of a Cylinder}\label{sss:cylwake}

Studies of cylinder wakes~\cite{roshko54,williamson89,Noack:03hierarchy,Rowley:17ARev} have attracted considerable interest from the flow system learning community for its particularly rich physics that encompass many of the complexities of nonlinear dynamical systems and yet easy to compute. For this exploration into the performance of different data-driven modeling frameworks we leverage both the unstable transient and the stable limit-cycle dynamics of two-dimensional cylinder wake flow at a Reynolds numbers of hundred, i.e. $Re=100$. To generate two-dimensional cylinder flow data, we adopt a spectral Galerkin method~\cite{Cantwell:15nektar++} to solve incompressible Naiver-Stokes equations, as shown in Eq.~\eqref{eq:cylinderflow} below:
\begin{subequations} \label{eq:cylinderflow}
\begin{eqnarray}
\frac{\partial{u}}{\partial{x}}+\frac{\partial{u}}{\partial{y}}&=&0,\\
\frac{\partial{u}}{\partial{t}}+u\frac{\partial{u}}{\partial{x}}+v\frac{\partial{u}}{\partial{y}}&=&-\frac{\partial{P}}{\partial{x}}+\nu\nabla^2{u},\\
\frac{\partial{v}}{\partial{t}}+u\frac{\partial{v}}{\partial{x}}+v\frac{\partial{v}}{\partial{y}}&=&-\frac{\partial{P}}{\partial{y}}+\nu\nabla^2{v},
\end{eqnarray}
\end{subequations}

In the above system of equations, $u$ and $v$ are horizontal and vertical velocity components. $P$ is the pressure field, and $\nu$ is the fluid viscosity. The rectangular domain used for this flow problem is $-25D<x<45D$ and $-20D<y<20D$, where $D$ is the diameter of the cylinder. For the purposes of this study data is extracted from a reduced domain, i.e., $-2D<x<10D$ and $-3D<y<3D$, where the dynamics occur. The mesh with $\approx 24,000$ points was designed to sufficiently resolve the thin shear layers near the surface of the cylinder and transit wake physics downstream. 
The computational method employed fourth order spectral expansions within each element in each direction. The data snapshots were sampled at $\Delta t = 0.2$ non-dimensional time units, arranged as described in section \ref{ss:markovkoopman} and SVD of the flow state matrix was performed to obtain POD coefficients along the modes. The most dominant POD coefficients correspond to $St = 0.16 $ for $Re = 100$, from which we deduced that a single cycle corresponds to approximately $31$ data points in time. 
For this study we denote normalized time in as the number cycles to specify the width of the training regime.

Although, more than $15$ POD modes are required for capturing nearly $100 \%$ of the energy at $Re=100$, the large scale coherent structures which govern the flow dynamics are adequately represented within the first $3$ modes and account for approximately $95 \%$ energy as shown in fig.\ref{f:energyRE100}(a)\cmnt{ and \ref{f:energyRE1000}(a)}. The eigenfunctions corresponding to these three modes are presented in fig.~\ref{f:energyRE100}(b)\cmnt{and \ref{f:energyRE1000}(b) respectively and show qualitatively similar flow structures for both $Re=100$ and $Re=1000$}. In fig.~\ref{f:energyRE100}(c) \cmnt{and \ref{f:energyRE1000}(c)} we show the phase portrait for the flow dynamical system, wherein the flow transitions from a steady wake through an unstable growth phase and settles into a limit cycle regime. 
\begin{figure}
\begin{center}
\includegraphics[width=0.33\textwidth]{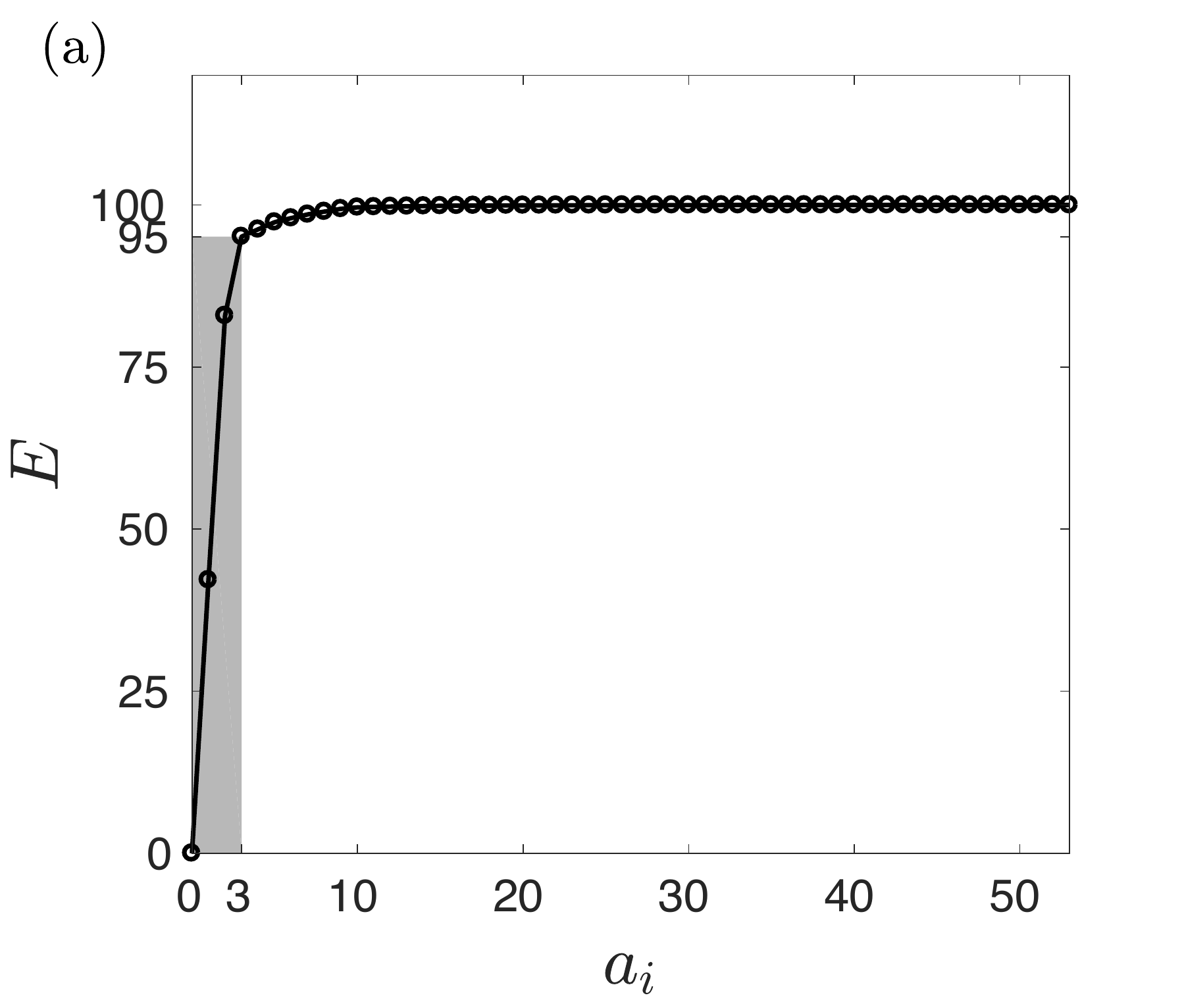}
\includegraphics[width=0.31\textwidth]{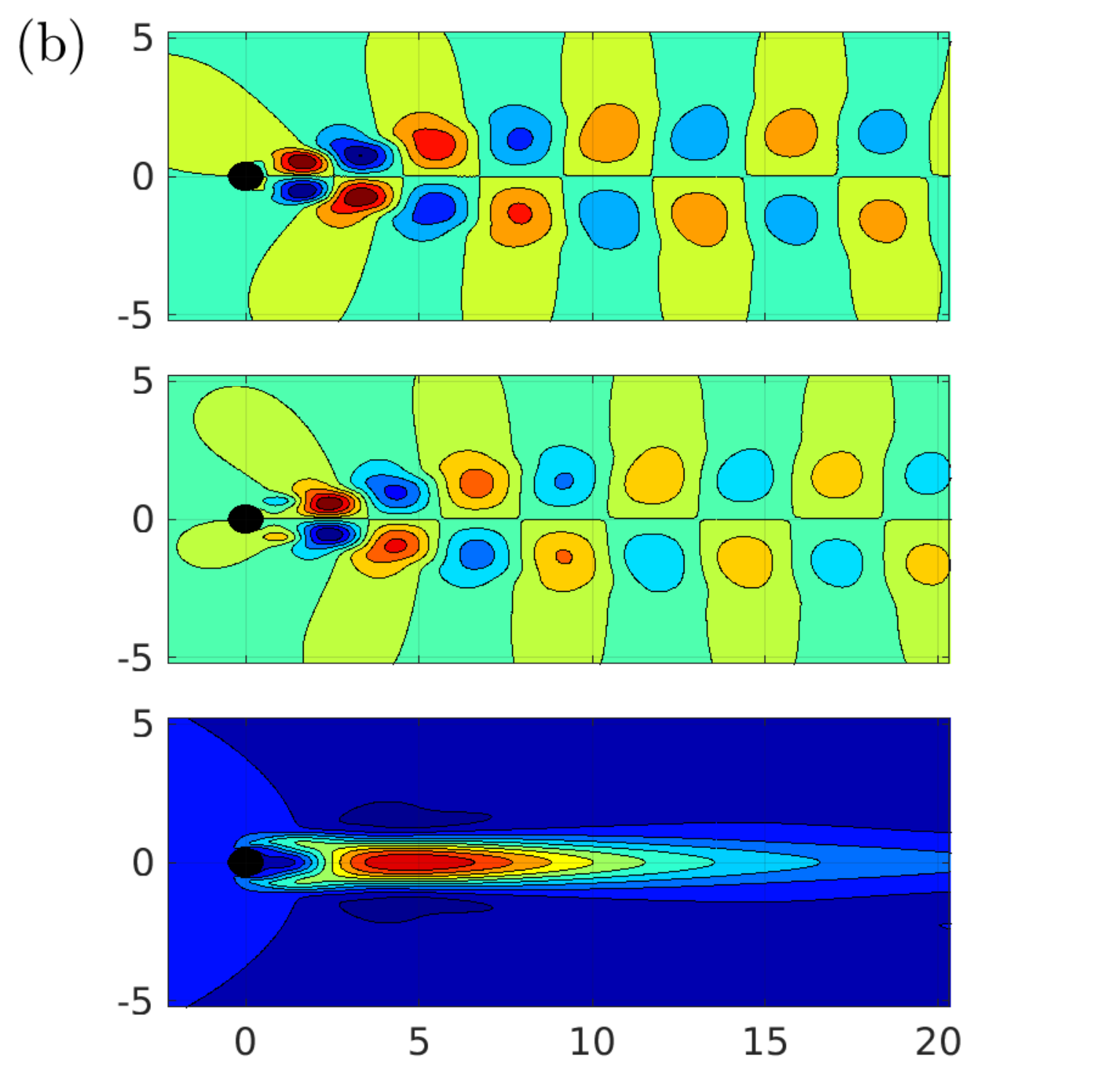}
\includegraphics[width=0.33\textwidth]{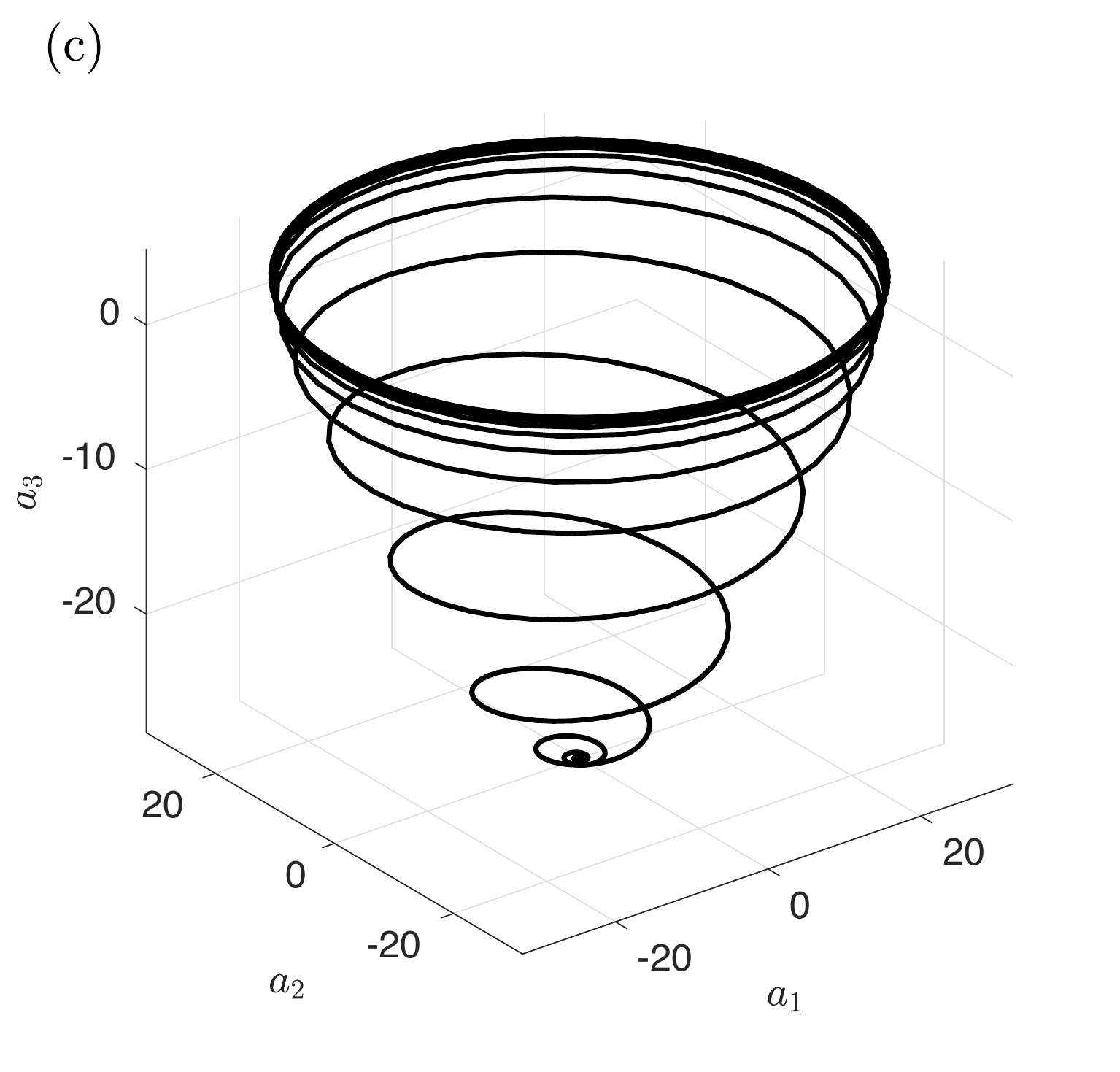}
\caption{Energy content in POD features selected (a) 3 coefficients (b) eigen modes/functions corresponding to 3 POD features (c) phase portrait of $Re=100$ flow.}
\label{f:energyRE100}
\end{center}
\end{figure}



\subsubsection{2D Buoyant Boussinesq Mixing Flow}\label{sss:buoymix}

The above discussion pertains to a nonlinear wake flow dynamical system that transitions from a steady wake into a stable limit-cycle attractor. Such systems have seen success in prediction from data-driven models with the availability of limited data as demonstrated in ~\cite{Lu:18sparse} . The instability-driven Bousinesq buoyant mixing flow~\cite{weinan98,liu03} exhibits strong shear and Kelvin-Helmholtz instabilities driven by thermal gradients.  The convective dynamics in such a system cannot be efficiently represented by data-driven POD modes. Further, the data-driven basis representing the low-dimensional manifold itself evolves temporally indicative of highly transient physics. Such systems are sensitive to noise in the initial state that produce very different trajectories and consequently, a very different dynamical system with its own basis space. This renders such dynamical systems hard to predict even if one were to leverage equation-driven models such as POD-Galerkin \cite{Noack:03hierarchy}. Earlier work from our team~\cite{Lu:18sparse} has shown that such problems are difficult to model accurately using MSM-based models. In this work, we compare these outcomes with those of the MEM-based Markov models.  
 
  The data is generated by modeling the dimensionless form of the two-dimensional incompressible flow transport equations\cite{liu03} augmented with buoyancy terms and thermal transport equations, as shown in Eq.~\ref{eq:bseq} on a rectangular domain that is $0<x<8$ and $0<y<1$. To achieve this,we use a $6^{th}$-order compact scheme~\cite{lele1992compact} in space and $4^{th}$-order Runge-Kutta method for the time-integration~\cite{gottlieb2001strong}.  
\begin{subequations} \label{eq:bseq}
\begin{eqnarray}
\frac{\partial{u}}{\partial{x}}+\frac{\partial{u}}{\partial{y}}&=&0,\\
\frac{\partial{u}}{\partial{t}}+u\frac{\partial{u}}{\partial{x}}+v\frac{\partial{u}}{\partial{y}}&=&-\frac{\partial{P}}{\partial{x}}+\frac{1}{Re}\nabla^2{u},\\
\frac{\partial{v}}{\partial{t}}+u\frac{\partial{v}}{\partial{x}}+v\frac{\partial{v}}{\partial{y}}&=&-\frac{\partial{P}}{\partial{y}}+\frac{1}{Re}\nabla^2{v}+Ri\theta,\\
\frac{\partial{\theta}}{\partial{t}}+u\frac{\partial{\theta}}{\partial{x}}+v\frac{\partial{\theta}}{\partial{y}}&=&\frac{1}{{Re}{Pr}}\nabla^2{\theta},\\
\end{eqnarray}
\end{subequations}
In the above system, $u$, $v$, and $\theta$ represent the horizontal, vertical velocity, and temperature field, respectively. The system is characterized by the following dimensionless parameters: Reynolds number, $Re$, Richardson number $Ri$, and Prandtl number, $Pr$ with values of $1000$, $4.0$, and $1.0$ respectively. The grid resolution employed is $256 \times 33$. The initial condition for the simulation is designed by vertically segregating the fluids at two different temperatures (uniformly distributed) at the middle of the domain. All the boundaries are adiabatic and friction generating walls. The thermal field evolution over the simulation duration of 32 non-dimensional time units as shown in fig. \ref{fig:buoyantmixing_time_evolution} illustrates the highly transient dynamics. To represent the system in a low-dimensional feature space, POD modes were computed from the entire $1600$ snapshots corresponding to $64$ time units. 
 The reduced feature set consisting of three POD features (capturing nearly 80\% of the total energy) representing a low resolution measurement is shown in fig. \ref{fig:buoyantmixing_PODWeights} is used to train the model and predict the trajectory.

\begin{figure}
	\centering
\begin{subfigure}{0.495\columnwidth}
\begin{center}
\includegraphics[width=\textwidth,clip]{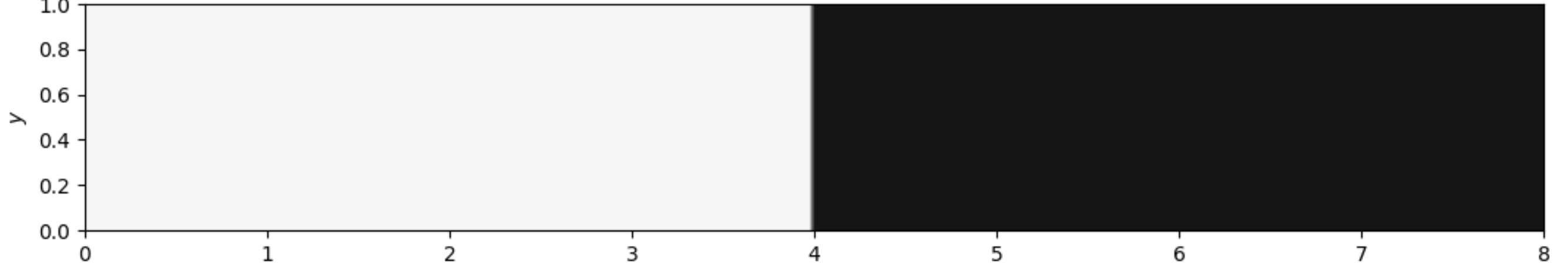}				
\caption{Time $=0$}
\end{center}
\end{subfigure}
\begin{subfigure}{0.495\columnwidth}
\begin{center}
\includegraphics[width=\textwidth,clip]{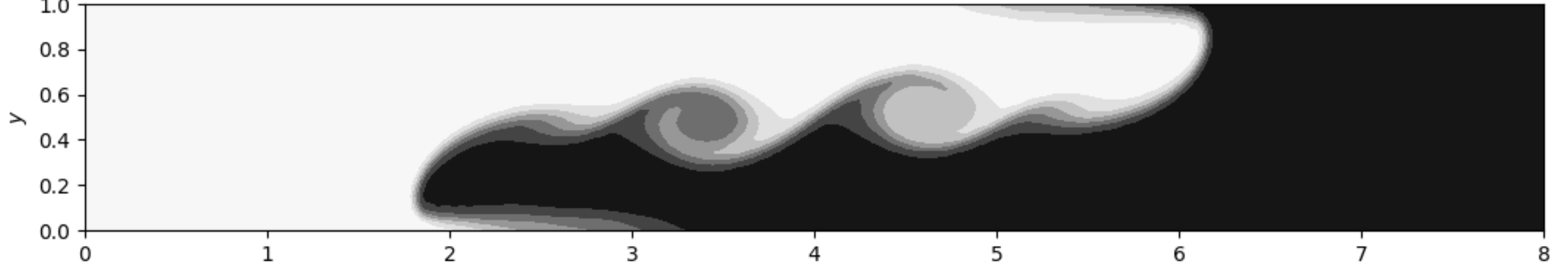}\\		
\caption{Time $=4$}
\end{center}
\end{subfigure}
\begin{subfigure}{0.495\columnwidth}
\begin{center}
	\includegraphics[width=\textwidth,clip]{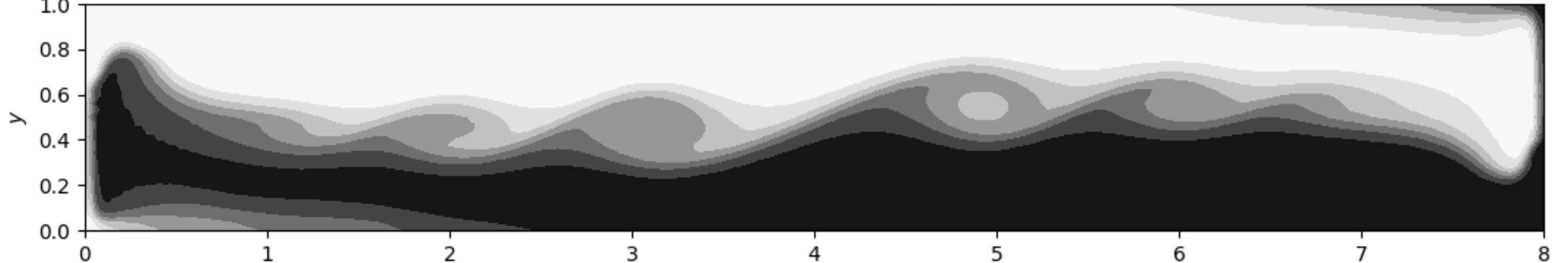}			
	\caption{Time $=8$}
\end{center}
\end{subfigure}
\begin{subfigure}{0.495\columnwidth}
\begin{center}
	\includegraphics[width=\textwidth,clip]{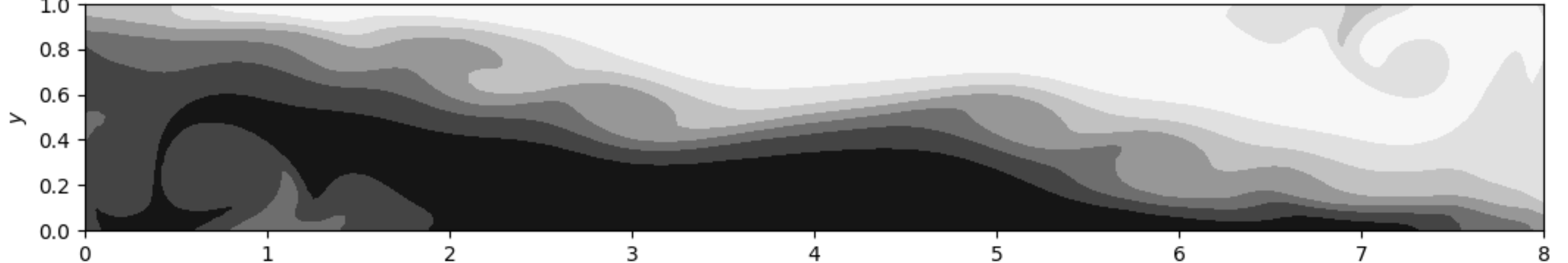}	\\		
\caption{Time $=12$}
\end{center}
\end{subfigure}
\begin{subfigure}{0.495\columnwidth}
\begin{center}
	\includegraphics[width=\textwidth,clip]{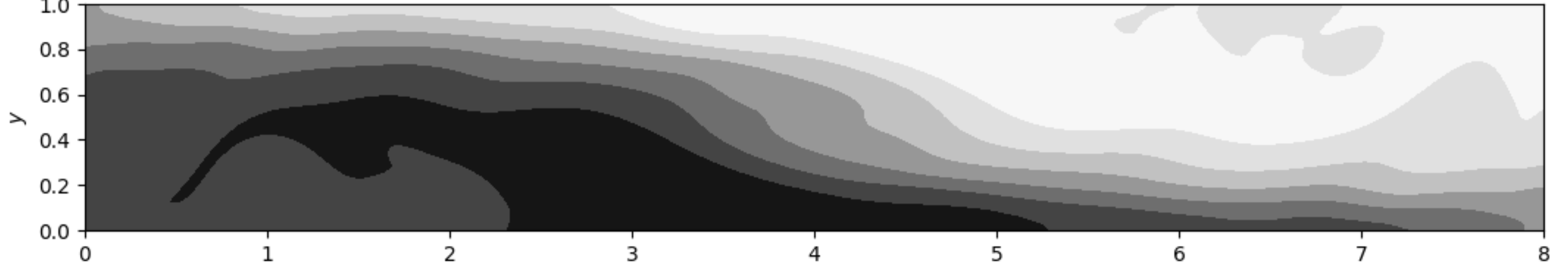}			
\caption{Time $=16$}
\end{center}
\end{subfigure}
\begin{subfigure}{0.495\columnwidth}
\begin{center}
	\includegraphics[width=\textwidth,clip]{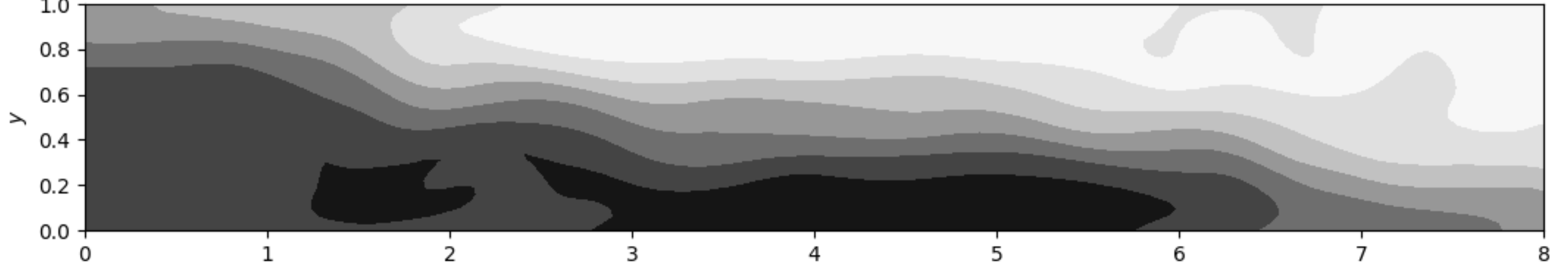}\\
\caption{Time $=20$}
\end{center}
\end{subfigure}	
\begin{subfigure}{0.495\columnwidth}
\begin{center}
	\includegraphics[width=\textwidth,clip]{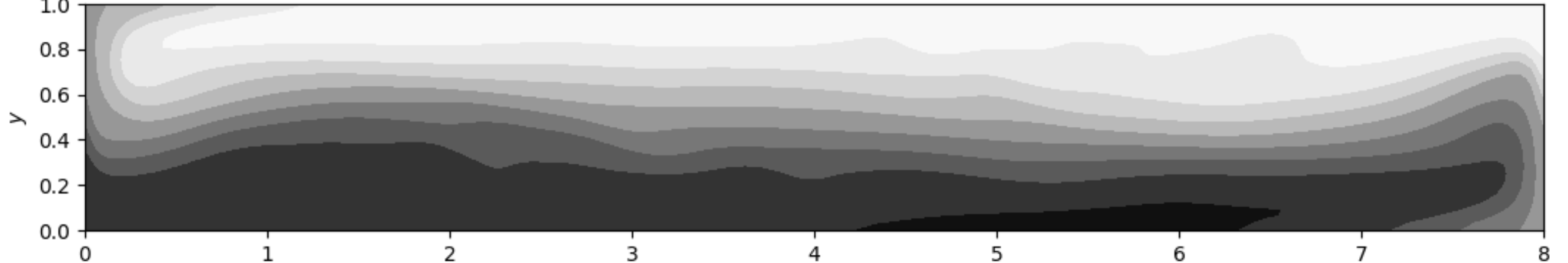}			
\caption{Time $=24$}
\end{center}
\end{subfigure}	
\begin{subfigure}{0.495\columnwidth}
\begin{center}
	\includegraphics[width=\textwidth,clip]{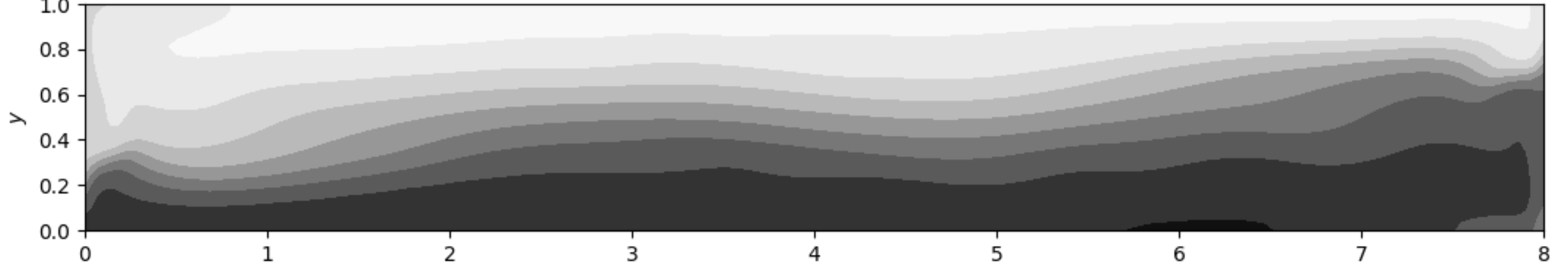}\\		
\caption{Time $=28$}
\end{center}
\end{subfigure}
\begin{subfigure}{0.495\columnwidth}
\begin{center}
	\includegraphics[width=\textwidth,clip]{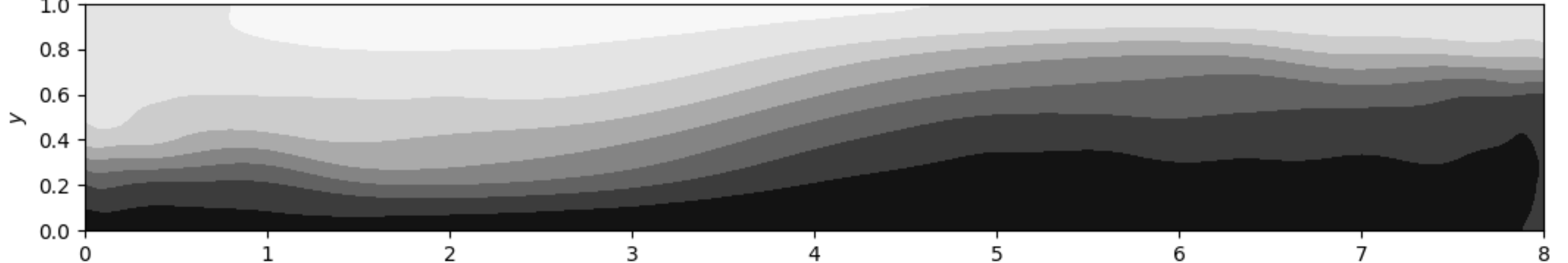}			
\caption{Time $=32$}
\end{center}
\end{subfigure}
	\caption{Time evolution of the isocontours of the temperature field in the 2D buoyant Boussinesq mixing layer is shown over a period of $32$ non-dimensional time units. }
	\label{fig:buoyantmixing_time_evolution}
\end{figure}

\begin{figure}[]
	\centering
	\includegraphics[width=1.0\columnwidth, scale=1.0]{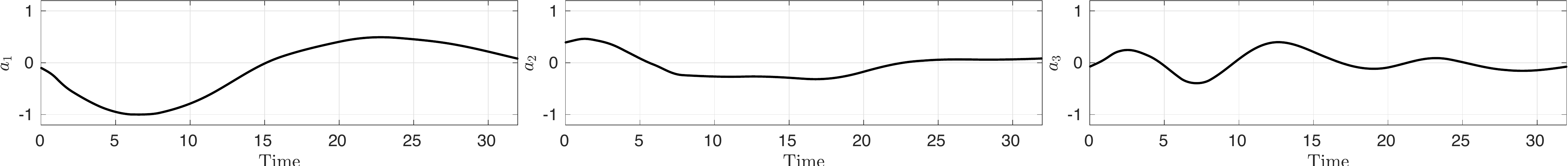}
	\caption{Time evolution of the POD weight features, $a_i^t$ for the buoyant mixing flow.}
	\label{fig:buoyantmixing_PODWeights}
\end{figure}

\subsection{Analysis Framework}\label{ss:analframework}

In this section we summarize the different candidate model architectures and learning algorithms. Table~\ref{t:methods}, lists the different MSM architectures and the comparable MEM architectures along with the total number of learning parameters ($\mathcal{LP}$) to be estimated. The first column under each class of sequential and end-to-end map in table~\ref{t:methods} verbalizes the multilayer structure and the second column represents the feature dimension of the different layers. For readability and conciseness, we have excluded the first and last layers corresponding to the input and output state vectors whose dimension is reduced by projecting onto a  POD basis. Here we remind the reader of the nomenclature used to denote the different architectures as `L-Method-$\mathcal{N}-N_f$'. 
As noted earlier, $N_f$ represents the feature dimension growth factor from the $2^{nd}$ to the $3^{rd}$ layers in the FFNN architectures. For the MSM maps, we invariably denote $N_f=1$ as the 
dimension of the inner layer features are determined by the choice of nonlinear map $\mathcal{N}$.
The rightmost column represents the total number of learning parameters ($\mathcal{LP}$) to be estimated from training. For example, when using 6-MSM-P2-1 or EDMD-P2 in table~\ref{t:methods} we learn an operator ($\Theta$) with $81\ (9\textrm{x}9)$ parameters and similarly, when using 6-MEM-TS-3 we learn operators($[\Theta_1, \Theta_3, \Theta_3]$ as in section~\ref{ss:markovGOC}) totaling $135\ (27+81+27)$ parameters. The six-layer EDMD-P (i.e. 6-MSM-P2-1 with quadratic polynomial features) method generates $9$ features in the intermediate layer which is then used to learn a linear map between the $9$ features at the next intermediate layer followed by reverse map to the penultimate layer with $3$ POD features. A similar construct is designed for the FFNN (MEM) framework using 6-MEM-TS-3 (i.e. $N_f = 3$). While the EDMD-P requires learning $81$ parameters, the FFNN with 6-MEM-TS-3 requires estimating $135$ parameters. In the following analysis of the predictive performance, we find that using just $3$ POD features with a 2nd order polynomial expansion in EDMD-P does not produce accurate results. So, in addition to P2, we also explore higher order polynomial basis explore if better predictions can be realized. 

\begin{table}
	\begin{center}
		\ra{1.6}
		\begin{tabular}{@{}lcccccccccc@{}}
			\toprule
			\phantom{a}&\phantom{a}&\multicolumn{2}{c}{\bf \it Sequential Maps}& \phantom{a}& $\mathcal{LP}$ & \phantom{a}& \multicolumn{2}{c}{\bf \it End-to-End Maps}& \phantom{a}& $\mathcal{LP} $\\
			\cmidrule{3-4} \cmidrule{6-6} \cmidrule{8-9} \cmidrule{11-11}
			\addlinespace
			$1$&& \multicolumn{2}{c}{\bf DMD}&&    && \multicolumn{2}{c}{\bf FFNN-Linear} &&   \\
			&& 4-MSM-$\mathcal{I}-1$       & 3-3 && 9 && 6-MEM-$\mathcal{I}-1$ & 3-3-3-3 && 27 \\
			\midrule
			\addlinespace
			$2$&& \multicolumn{2}{c}{\bf EDMD-TS} &&  && \multicolumn{2}{c}{\bf FFNN ($N_f=1$)}&& \\
			&& 6-MSM-TS-1 & 3-3-3-3 && 9 && 6-MEM-TS-1 & 3-3-3-3 &&27\\
			\addlinespace
			$3$&& \multicolumn{2}{c}{\bf EDMD-P} &&  &&\multicolumn{2}{c}{\bf FFNN ($N_f=3,9,20$)}&& \\
			&& 6-MSM-P2-1	 & 3-9-9-3 &&  81   && 6-MEM-TS-3 & 3-9-9-3&&  135 \\
			&& \multirow{2}{*}{6-MSM-P7-1} & \multirow{2}{*}{3-125-125-3}&& \multirow{2}{*}{15,625} && 6-MEM-TS-9 & 3-27-27-3&& 891  \\
			&&           				    &&  &            && 6-MEM-TS-20 & 3-60-60-3&& 3960 \\   
			\bottomrule
		\end{tabular}
		\caption{Overview of the different model architectures used as part of this analysis. The dimensions of different layers correspond to that used for cylinder wake flow. }
		\label{t:methods}
	\end{center}
\end{table}


\begin{figure}[H]
	\begin{center}
		\begin{subfigure}{\columnwidth}
			\begin{center}
				\includegraphics[width=0.6\columnwidth]{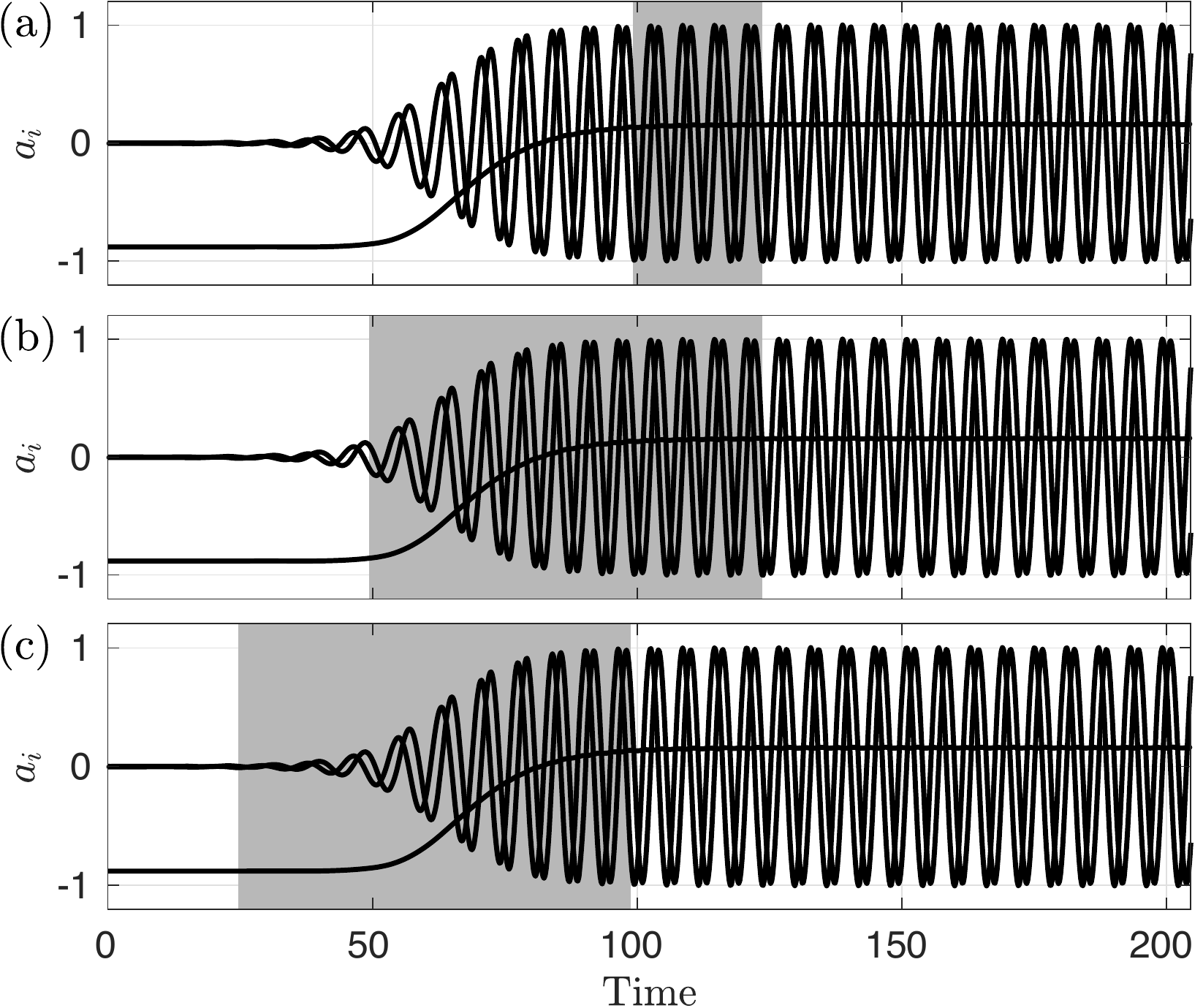}
				\caption{Times series plot of the weights corresponding to the three most energetic POD modes with different training regions (a) Limit cycle ($16-20$): 124 data points, (b) transient region-I ($8-20$): 372 data points and (c) Transient region-II ($4-16$): 372 data points, where each cycle consists of $31$ data points.}
				\label{f:datawindRE100}
			\end{center}
		\end{subfigure}\\
		\caption{Schematic showing the different training regions chosen for prediction using the different models. }
	\end{center}
\end{figure}

\subsection{Training, Validation and Error Quantification in \emph{Posteriori} Predictions}\label{ss:trainValidate}
A key aspect of data-driven modeling is to minimize overfitting so that realistic learning can be realized. To achieve this the data generated from computer simulations described above are separated into training and testing regimes. The training data set is used for learning the optimal $\Theta$'s using which \emph{a posteriori} predictions are computed with the earlier prediction(s) alone as the input to mimic a practical usage of the model. For this study , we assess model performance based on both qualitative representation of the dynamics and \emph{posteriori} prediction errors unlike the \emph{a priori} error estimates used in machine learning community. We quantify model errors using the $\mathcal{L}_{2}$ norm of the \emph{posteriori} prediction error from the data-driven model relative to the truth which requires accurate specification of only the initial condition $\pmb{a}_{0}$. To bypass the complexities of computing the $2-$norm, we instead compute the Forbenius norm of the error as in eqn.~\eqref{e:l2error}. 
\begin{equation}
\mathcal{E}_{t,p} = \frac{1}{2M_{i}} \| \bar{Y}^1_{p} - \bar{Y}^1 \|_{2}^{2}.
\label{e:l2error}
\end{equation}
In the above equation $\bar{Y}^1_{p}$ represents the posteriori prediction of the data-driven model and $\bar{Y}^1$ the true data. We make separate quantifications of the posteriori error in the training region where the data-driven model is operating in reconstruction mode and in the testing region where the model operates in a prediction or extrapolation role.  The posteriori error in the training region is denoted by $\mathcal{E}_t$  and combined error in both the training and testing regions is denoted by $\mathcal{E}_p$. 

To assess and characterize the robustness of the different architectures (table \ref{t:methods}) we train the models across various data regimes corresponding to different  dynamics of the flow, i.e. transient unstable wake or stable limit cycle regime with periodic vortex shedding. To this end, we identified three different training regions (see fig.~\ref{f:datawindRE100}) highlighted by windows shaded in grey. A stiff test for any data-driven model is to learn the underlying dynamics using information from the steady wake regime as shown in fig.~\ref{f:datawindRE100} and predict the growth of instability which ultimately stabilizes into limit cycle. From our experience models that use only training information from the steady wake region to predict the vortex shedding dynamics are highly unstable. Consequently, we designed two different training regions (TR I and TR II) where the flow transitions across flow regimes, but with different proportions of limit-cycle (vortex shedding) and steady wake content.
The figures in row (a) represent a training region in the limit-cycle regime and rows (b) and (c) correspond to regions in the transition part of the dynamics and denoted by region I (or TR-I) and region II (or TR-II). 
  In the following sections we will highlight and discuss the key results from our data-driven aposteriori predictions. 
These training regions are further divided into training data ($\approx 70\%$) and validation data ($\approx 30\%$) uniformly as shown in fig.\ref{f:TrainTestdata} to assess and validate (check for overfitting) learning performance. In fig.\ref{f:Cost8_20}, we show the learning cost evolution for TR-I obtained for the FFNN with a design specified by 6-MEM-TS and $N_f = 1,3,9,20$. We see that the learning cost for training dataset and validation dataset are same, which signifies model generalization. A similar trend was observed for all the MEM models used in this study. We also note that for all these posteriori predictions, a regularization parameter in range $(1e^{-12} - 1e^{-8})$ was used. The FFNN models learned in this study show difference between \emph{a priori} and \emph{a posteriori} predictions as shown in figs.~\ref{f:predictions} and \ref{f:Timeseriespredictions}.  For the \emph{a priori} predictions, we see that the dynamics are predicted accurately while for the \emph{a posteriori} case there exists deviations from the true data. It is worth reminding that while the \emph{a priori} analysis (Fig.\ref{f:apriori}) is directly correlated to the learning cost, the posteriori analysis(fig.\ref{f:apriori}) shows how the accumulated error interacts with the learned model. From the timeseries of priori and posteriori predictions in figs.\ref{f:aprioriTimeseries} and \ref{f:aposterioriTimeseries}, we see that the posteriori error growth impacts the shift POD mode the most. The shift mode~\cite{Noack:03hierarchy} represents the shift in trajectory of the system from an unstable regime to a neutrally stable regime. We have seen that including a bias term in the FFNN (6-MEM-TS-3) models decrease this prediction error (see Appendix.\ref{s:Appendix1}).  In the following sections we highlight the key results from our data-driven \emph{a posteriori} predictions. 


\begin{figure}[ht!]
\begin{center}
\begin{subfigure}{0.485\columnwidth}
\begin{center}
\includegraphics[width=0.9\columnwidth]{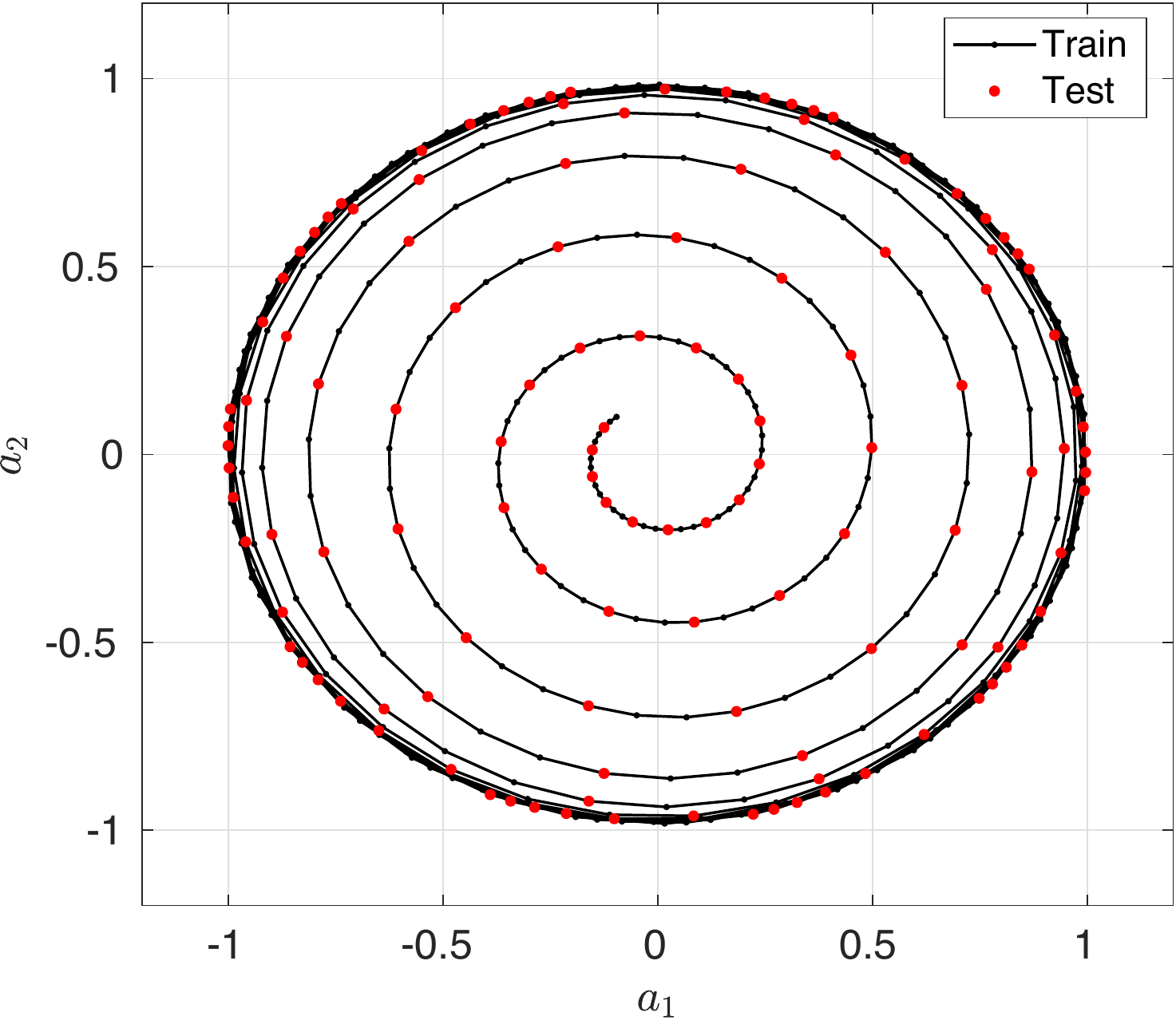}
\caption{TR-I region uniformly divided into training and \\ validation data. The green dots denote the validation \\ dataset and black line with dots denote training dataset.}
\label{f:TrainTestdata}
\end{center}
\end{subfigure}%
\begin{subfigure}{0.485\columnwidth}
\begin{center}
\includegraphics[width=0.9\columnwidth]{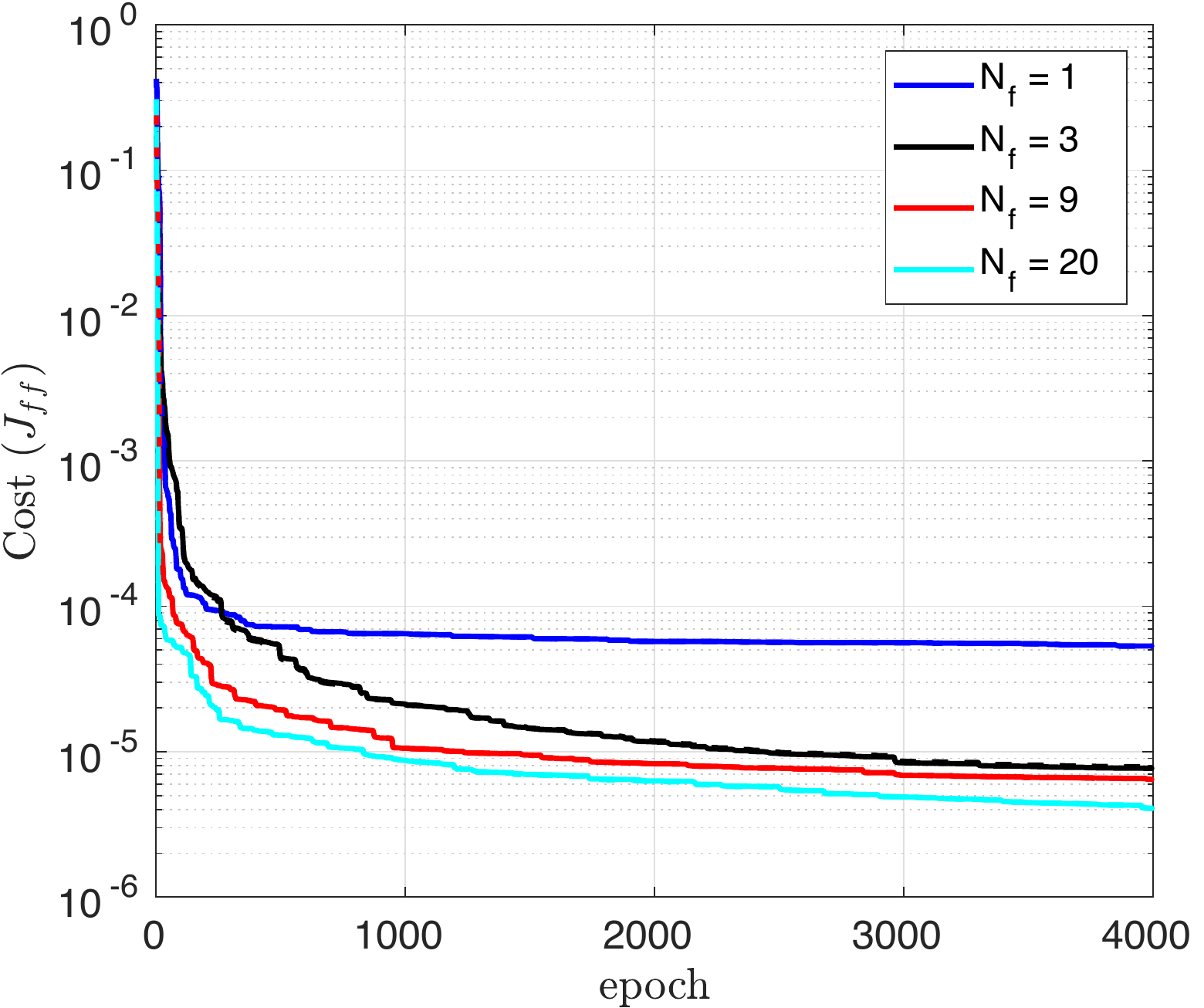}
\caption{Learning cost evolution with respect to epochs on TR-I using FFNN (6-MEM-TS-$N_f$ with $N_f = 1,3,9,20$). The symbols represent the error cost of the test dataset.}
\label{f:Cost8_20}
\end{center}
\end{subfigure}
\caption{Schematic showing training performance of FFNNs. (a) Separation of input data into training and testing sets on a phase plot; (b) Comparison of learning performance for the transient regime training region I. }
\end{center}
\end{figure}

\begin{figure}[H]
\begin{center}
\begin{subfigure}{0.485\columnwidth}
\begin{center}
\includegraphics[width=0.9\columnwidth]{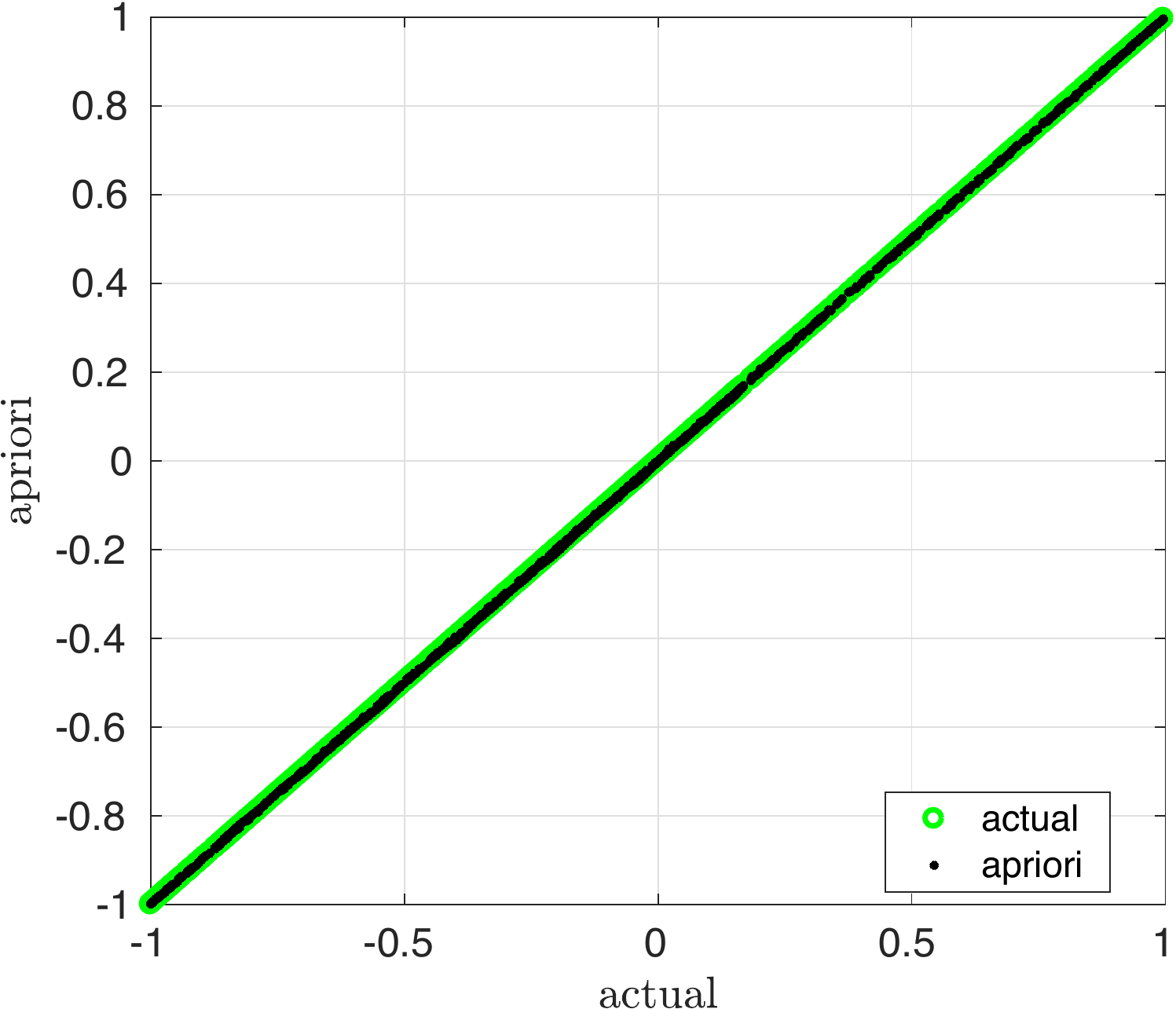}
\caption{\emph{a priori} prediction of TR-I using 6-MEM-TS3.}
\label{f:apriori}
\end{center}
\end{subfigure}%
\begin{subfigure}{0.485\columnwidth}
\begin{center}
\includegraphics[width=0.9\columnwidth]{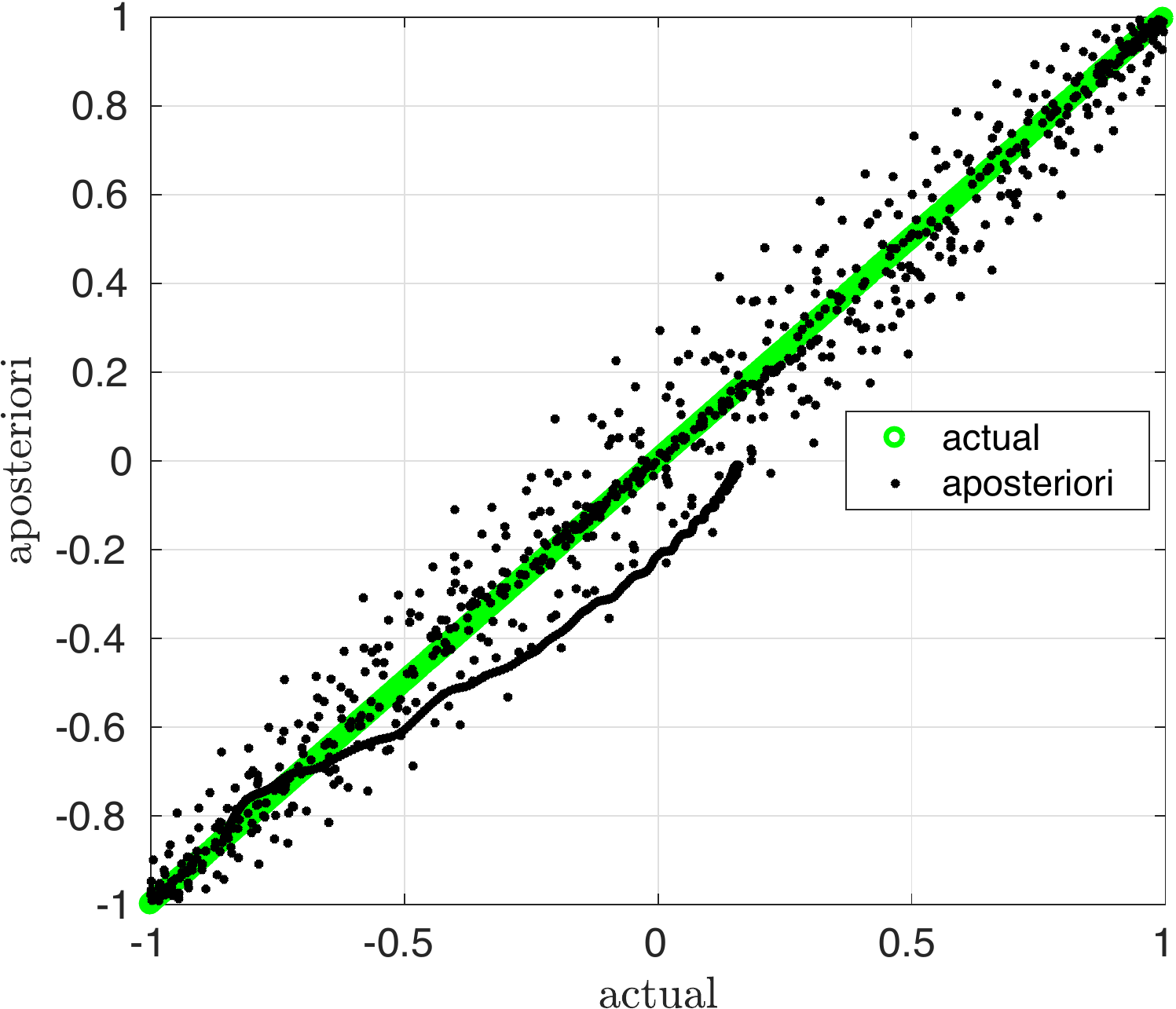}
\caption{\emph{a posteriori} prediction of TR-I using 6-MEM-TS3. The black dots denote the entire set of features $[a_1,a_2,a_3]$}
\label{f:aposteriori}
\end{center}
\end{subfigure}
\caption{\emph{a Priori} vs \emph{a Posteriori} predictions using FFNNs. }
\label{f:predictions}
\end{center}
\end{figure}

\begin{figure}[H]
\begin{center}
\begin{subfigure}{\columnwidth}
\begin{center}
\includegraphics[width=1.05\columnwidth]{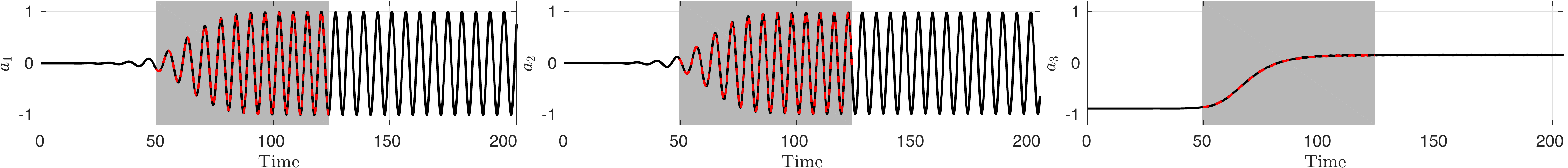}
\caption{\emph{a priori} prediction of TR-I using 6-MEM-TS3.}
\label{f:aprioriTimeseries}
\end{center}
\end{subfigure}\\
\begin{subfigure}{\columnwidth}
\begin{center}
\includegraphics[width=1.05\columnwidth]{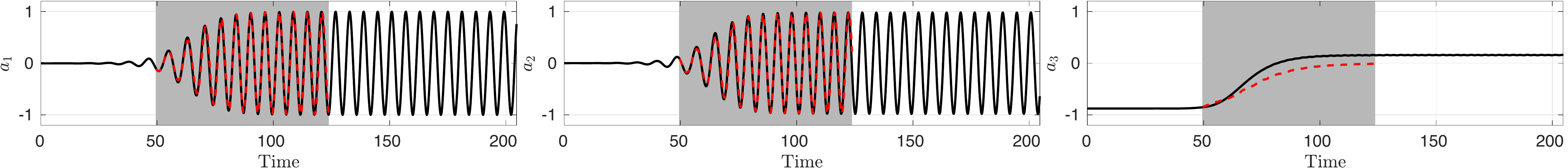}
\caption{\emph{a posteriori} prediction of TR-I using 6-MEM-TS3.}
\label{f:aposterioriTimeseries}
\end{center}
\end{subfigure}
\caption{Timeseries of features from \emph{a priori} and \emph{a posteriori} prediction compared with the true data. }
\label{f:Timeseriespredictions}
\end{center}
\end{figure}

\subsection{\emph{a Priori} Learning and \emph{a Posteriori} Prediction of Limit-cycle Cylinder Wake Dynamics}\label{ss:limitcycle}

The focus of this section is to learn from limit-cycle training data and predict the corresponding limit-cycle physics over long durations. Successful prediction of this case is considered a benchmark for data-driven models. The underlying theme in this article is to explore whether iterative end-to-end learning of the model parameters ($\mathcal{LP}$) can outperform one-time sequential learning of the  model parameters for predictions of fluid flows. To verify this we compare the following four models namely: DMD (4-MSM-$\mathcal{I}$-1, FFNN-linear (6-MEM-$\mathcal{I}$-1), EDMD-TS (6-MSM-TS-1) and FFNN-TS (6-MEM-TS-1).  The FFNN-linear architecture can be viewed as a multilayer neural network analogue of linear map-based methods such as DMD. Similarly, EDMD-TS can be viewed as a MSM analogue of the standard FFNN architecture. Therefore, these set of two pairs of architectures can provide useful insight into the role of learning methodology (sequential vs end-to-end) and nonlinear functions in multilayer maps. It is well known that DMD performs well in the limit cycle region as shown in \cite{Lu:18sparse,Rowley:17ARev} and under performs in the strongly nonlinear transient regimes on account of being a linear model of the state. In fig.~\ref{f:ltcyc}, the time series posteriori predictions of the first three POD features are shown with rows $1-4$ (top-to-bottom) representing outcomes from the learned parameters ($\Theta$s) obtained using the DMD, FFNN-linear (6-MEM-$\mathcal{I}$-1), EDMD-TS (6-MSM-TS-1)  and FFNN (6-MEM-TS-1) architectures respectively. 
Specifically, we assess the role of sequential versus end-to-end optimization of the  parameters as well as the impact of nonlinear mapping on model prediction.

The first major observation is that both the sequential and end-to-end models with linear mapping predict the overall dynamics relatively accurately while the sequential model with nonlinear sigmoid mapping damps the POD features over time.
 The second observation is that all the models show gradual error growth with time except the standard FFNN (6-MEM-TS-1) architecture. The plots in fig.~\ref{f:ltcyc} convey that a nonlinear mapping is not essential to capturing the limit-cycle dynamics, but if used, should be carefully designed. For example, it was shown in \cite{Lu:18sparse} that EDMD-P2 can predict such dynamics very well while the current results (fig.\ref{f:ltcyc}(c)) show that the same architecture with a tansigmoid function (EDMD-TS) produces errors. The TS function is primarily used in machine learning for classification and has a \emph{squashing} nature to it, i.e. it has the effect of compressing the features which explains its inability to predict the dynamics. 
 A plausible reason could be that the TS nonlinearity does not extend the space of learning parameters in contrast to polynomial basis. Nevertheless, when the TS nonlinearity (using the $\mathcal{N}$) is combined with an end-to-end framework such as the well known FFNN, the prediction drastically improves as learning the parameters in $\Theta_1,\Theta_2,\Theta_3$ simultaneously while applying the TS nonlinearity produces a compensatory and powerful outcome. Further, this FFNN model can predict over long times without growth in error as seen from the evolution of third POD feature (shift mode) in fig.\ref{f:ltcyc}(d).

\begin{figure}
\includegraphics[width=0.33\columnwidth]{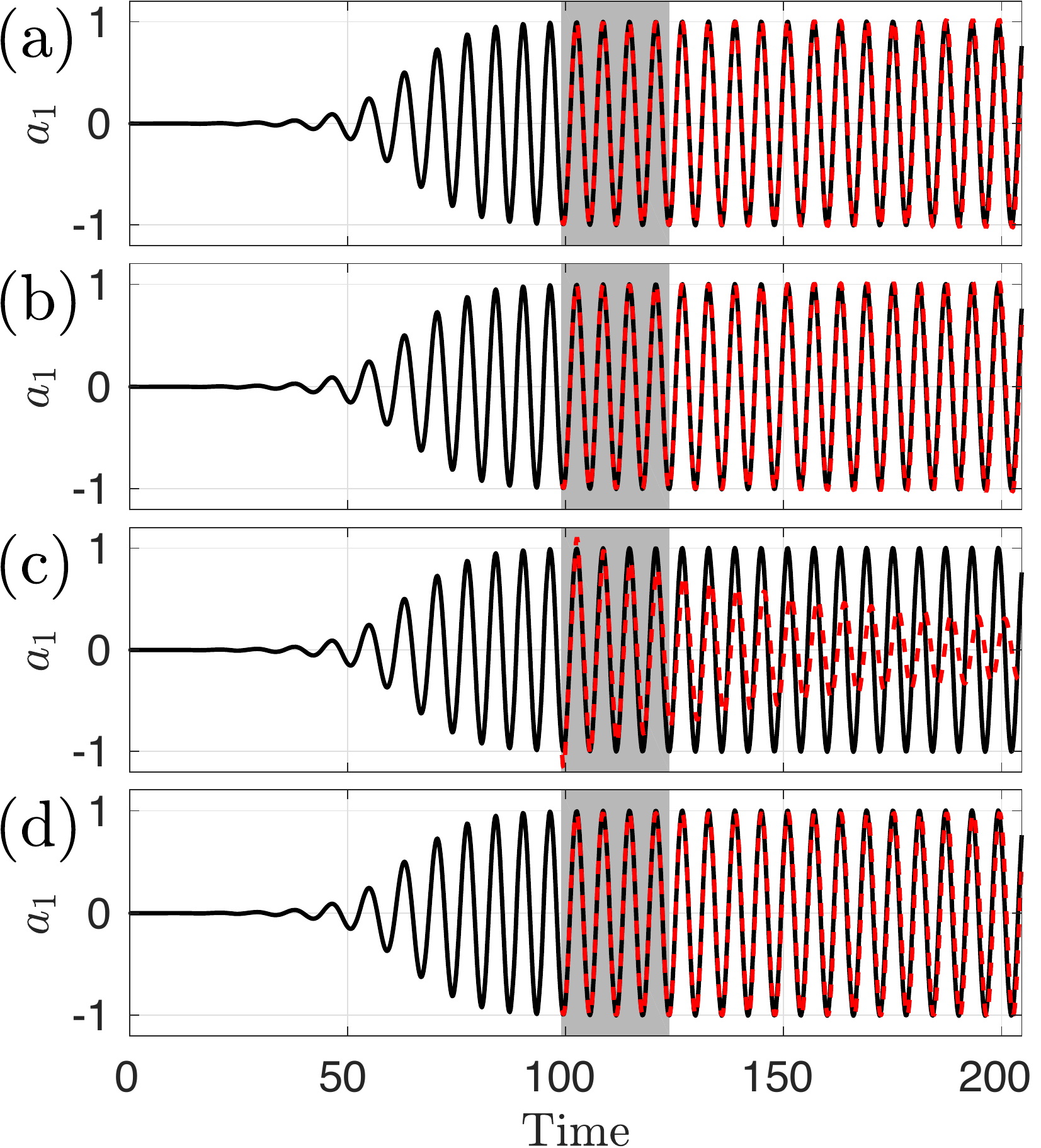}\hfill
\includegraphics[width=0.32\columnwidth]{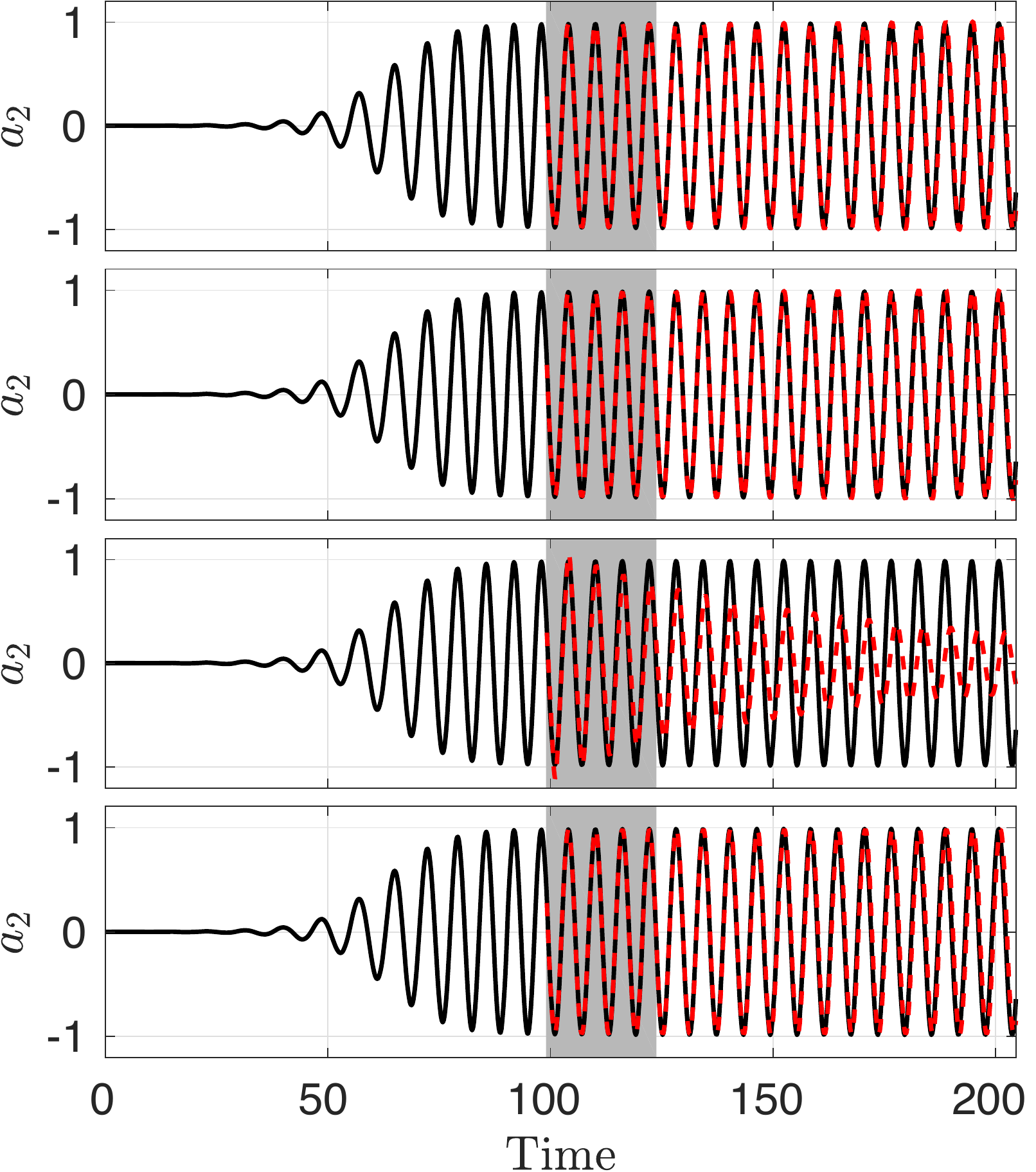}\hfill
\includegraphics[width=0.32\columnwidth]{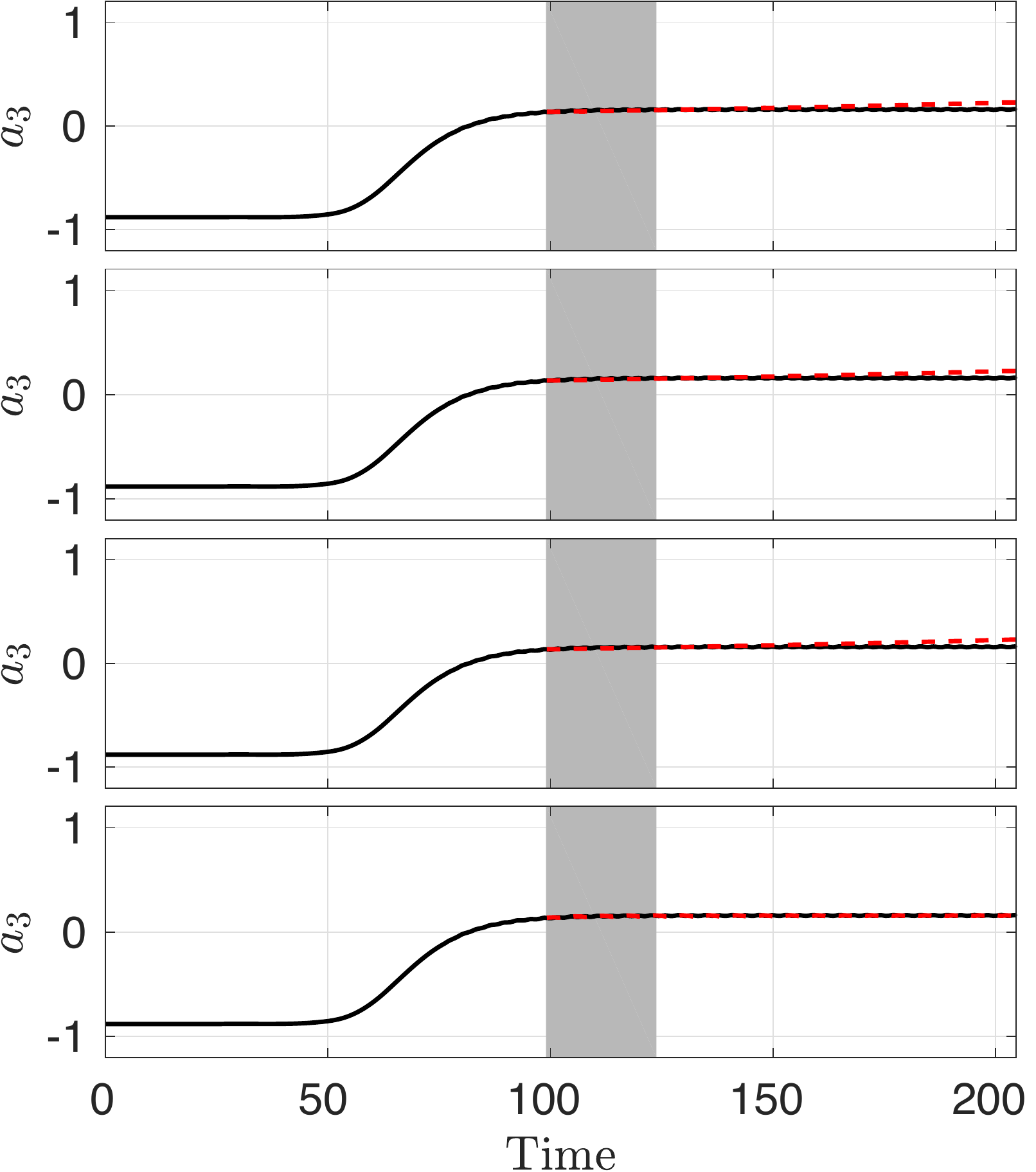}
\caption{Times series of \emph{posteriori} predicted POD features (\xdash[0.5em] \xdash[0.5em] \xdash[0.5em]) obtained from (a) DMD (4-MSM-$\mathcal{I}$-1), (b) FFNN-linear (6-MEM-$\mathcal{I}$-1), (c)  EDMD-P (6-MSM-TS1-1) and (d)  FFNN (6-MEM-TS-1) are plotted with their respective original data (\xdash[1.5em]) in the limit cycle regime.}
\label{f:ltcyc}
\end{figure}

We had mentioned earlier that the success of the FFNN/MEM frameworks possibly comes from learning an extended parameter ($\mathcal{LP}$) space, but the following discussion shows that this is true only in the presence of a nonlinear function as part of the mapping. In the DMD framework, there are $9$ learning parameters in $\Theta$ to predict the limit cycle dynamics as compared to $27$ parameters for FFNN-linear \cmnt{6-MEM-$\mathcal{I}$-1} architecture. However, in the absence of an nonlinear function in the map, the linear operator computed from the two methods turned out to be the same, i.e. the product of the different $\Theta_l\ \textrm{for } l=1-3$ from FFNN-linear is same as the $\Theta$ learned from DMD.  
 In fig.\ref{f:ltcyc}, we use 4-cycles of (124 points) data in the limit cycle region for training and predict upto 17 cycles (527 data points). We see that the predictions obtained using DMD and FFNN-linear \cmnt{(6-MEM-$\mathcal{I}$)} in fig.\ref{f:ltcyc}(a) and (b) are similar as the same linear transition operator is estimated. However, with limited training data, the predictions start to diverge from the truth over large times as is clearly seen from the evolution of the third POD feature, $a_3$.

While the addition of nonlinear functions in the map aids the prediction of nonlinear dynamics, employing this formulation with a local optimization of the $\mathcal{LP}$ does not always guarantee good results. We see an illustration of this in the performance of the EDMD-TS \cmnt{(6-MSM-TS-1}} architecture as seen from fig.\ref{f:ltcyc}(c), where all the three input features are incorrectly predicted in contrast to predictions by the FFNN \cmnt{(6-MEM-TS1)} in fig.\ref{f:ltcyc}(d). The \emph{a posteriori} prediction error quantifications for the limit-cycle regime in the training and testing regions are shown in the first two rows of the table \ref{t:predstatRe100}. These show that the DMD and FFNN \cmnt{(6-MEM-TS-1}} produce error magnitudes of $7.4\times 10^{-3}$ and $1.6\times 10^{-2}$ respectively outside the training region. These errors are higher than the $O(1e^{-4})$ values in the training region as one would expect. In spite of generating more prediction errors, the MEM models cap their growth which is a desirable feature.  As additional benchmarks we also include testing region errors for other architectures including EDMD-P2 \cmnt{(6-MSM-P2-1)}, FFNN \cmnt{(6-MEM-TS-1)} and FFNN with $N_f=3$ \cmnt{(6-MEM-TS-3)} which generate comparable prediction accuracy with EDMD-P2 \cmnt{(6-MSM-P2-1)} being the smallest. In summary, except for the EDMD-TS \cmnt{(6-EDMD-TS-1)}  all the other models display reasonable accuracy for this limit-cycle dynamics in both the training and prediction regimes. However, we observe gradual error growth \cmnt{of the third feature} in all the models except for the FFNN \cmnt{(6-MEM-TS-1)} which has implications for long-time predictions.  

\begin{table*}
\begin{center}
{
\ra{1.6}
\begin{tabular}{@{}lccccccccccccc@{}}
\toprule
Train & \phantom{a}& \multicolumn{1}{c}{DMD}& \phantom{a}&\multicolumn{1}{c}{EDMD-TS} & \phantom{a}& \multicolumn{2}{c}{EDMD-P} & \phantom{a} & \multicolumn{4}{c}{FFNN}\\
arch & \phantom{a}& \multicolumn{1}{c}{(4-MSM-$\mathcal{I}$-$N_f$)}& \phantom{a}&\multicolumn{1}{c}{(6-MSM-TS-$N_f$)} & \phantom{a}& \multicolumn{2}{c}{(6-MSM-Pp-$N_f$)} & \phantom{a} & \multicolumn{4}{c}{(6-MEM-TS-$N_f$)}\\
\cmidrule{3-3} \cmidrule{5-5} \cmidrule{7-8}  \cmidrule{10-13}
 cycles   			    & $\mathcal{E}$  & $N_f = 1$  && $N_f = 1$  &&  p = $2$   & p =$7$     &&$N_f =1$    & $N_f =3$   &  $N_f =9$  & $N_f =20$ \\ \midrule\\
\multirow{2}{*}{$16-20$}&$\mathcal{E}_t$ &$1.6e^{-4}$ &&$3.3e^{-2}$ &&$6.7e^{-5}$ &$--$        &&$2.7e^{-4}$ &$2.1e^{-4}$ &$--$        &$--$\\
{(LC)}&$\mathcal{E}_p$ &$7.4e^{-3}$ &&$0.269$     &&$3.5e^{-4}$ &$--$        &&$1.6e^{-2}$ &$8.1e^{-3}$ &$--$        &$--$\\ 

\multirow{2}{*}{$08-20$}&$\mathcal{E}_t$ &$0.417$     &&$0.467$     &&$0.475$     &$6.4e^{-6}$ &&$0.320$     &$3.7e^{-2}$ &$1.9e^{-2}$ &$2.1e^{-2}$\\
{(TR-I)}&$\mathcal{E}_p$ &$0.513$     &&$0.483$     &&$0.776$     &$3.9e^{-4}$ &&$0.686$     &$0.146$     &$0.153$     &$0.148$\\ 
					   
\multirow{2}{*}{$04-16$}&$\mathcal{E}_t$ &$0.246$     &&$0.238$     &&$0.223$     &$0.191$     &&$--$        &$0.106$     &$0.182$     &$0.385$\\
{(TR-II)}&$\mathcal{E}_p$ &$0.551$     &&$0.530$     &&$0.683$     &$0.977$     &&$--$        &$0.883$     &$0.948$     &$0.720$\\
\\ 
\bottomrule
\end{tabular}}
\caption{\emph{a Posteriori} Prediction error estimates for the different MSM and MEM architectures for $Re = 100$ data across training regimes. \label{t:predstatRe100}}
\end{center}
\end{table*}

\subsection{\emph{a Priori} Learning and \emph{a Posteriori} Prediction of Transient Cylinder Wake Dynamics }\label{ss:transientdata}

In the earlier section, we highlighted the importance of nonlinearity in the map and its combination with a MEM framework for stable long-time predictions.  In this section, we focus on learning from transient wake flow data and predict the resulting limit-cycle system. It is well known that DMD performs better on limit cycle problems and underperforms in the transient regime due to its inability to handle the enhanced nonlinear instability growth that characterizes the underlying dynamical system. In particular, if the limit-cycle dynamics represents a nonlinearity of order $k$ then the transient wake regime corresponds to a nonlinearity of order $\geq k+1$~\cite{Noack:03hierarchy}. 
Consequently, models that incorporate nonlinearity in the map such as the EDMD-P with polynomial basis~\cite{Williams:15} or the corresponding kernel representation~\cite{Williams:14arXiv} perform better for such problems, but only when using significant number of input features. In this section, we show that end-to-end learning of a nonlinear multilayer map provides much better  prediction capabilities from as little input data as three features which is the minimum needed to capture the wake instability behind a cylinder~\cite{Noack:03hierarchy}.  


\begin{figure}
\begin{center}
\includegraphics[width=0.33\columnwidth]{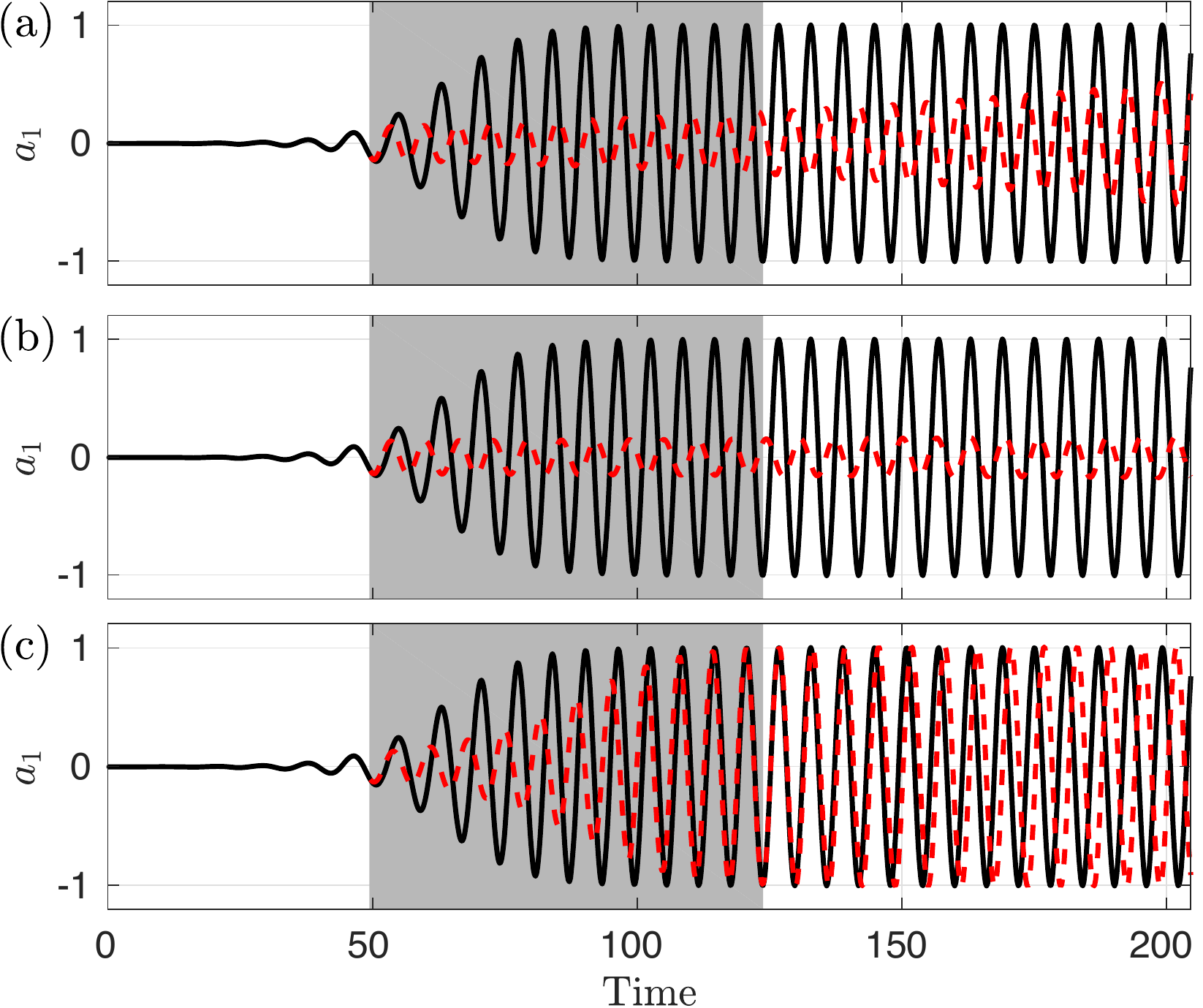}
\includegraphics[width=0.32\columnwidth]{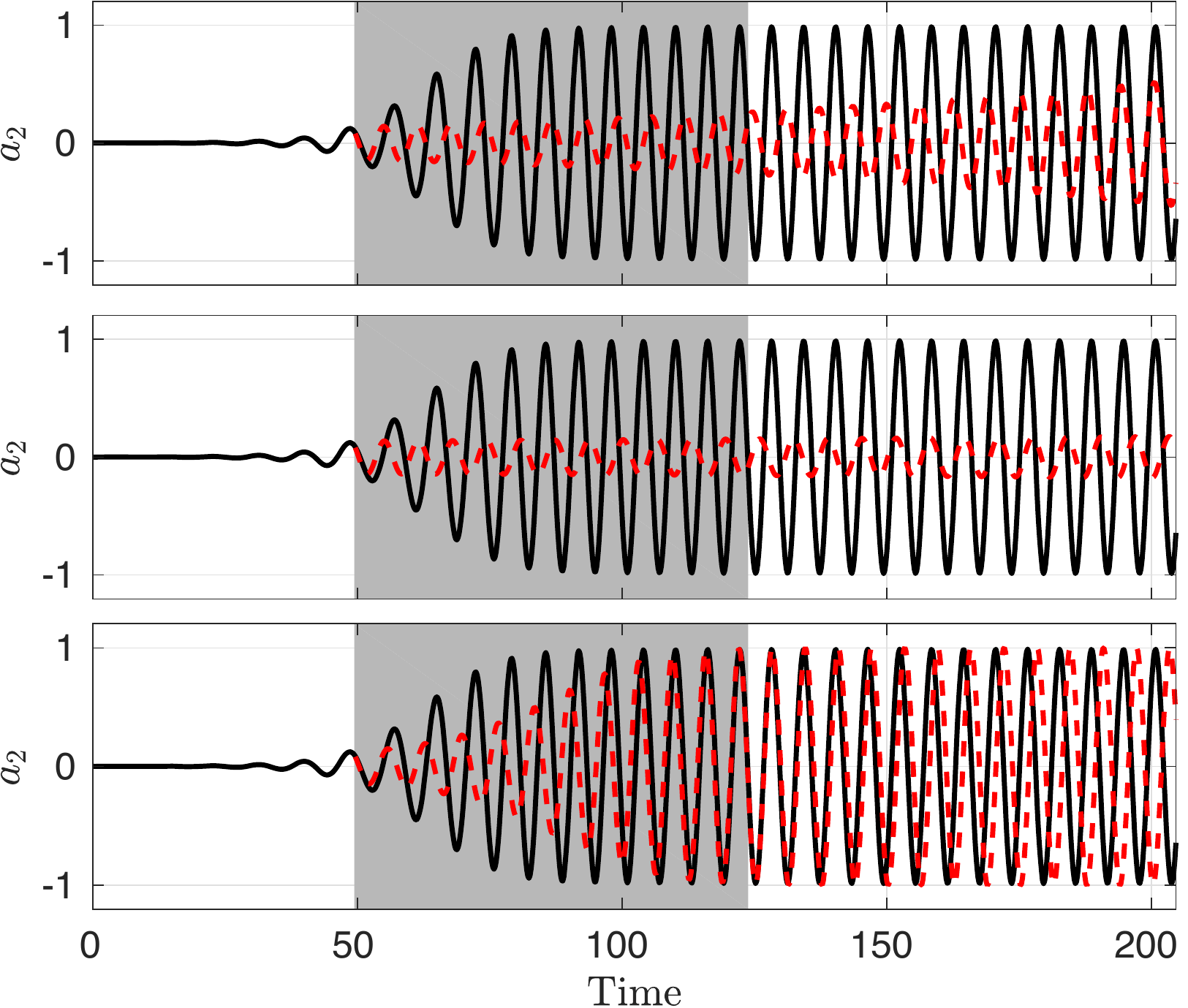}
\includegraphics[width=0.32\columnwidth]{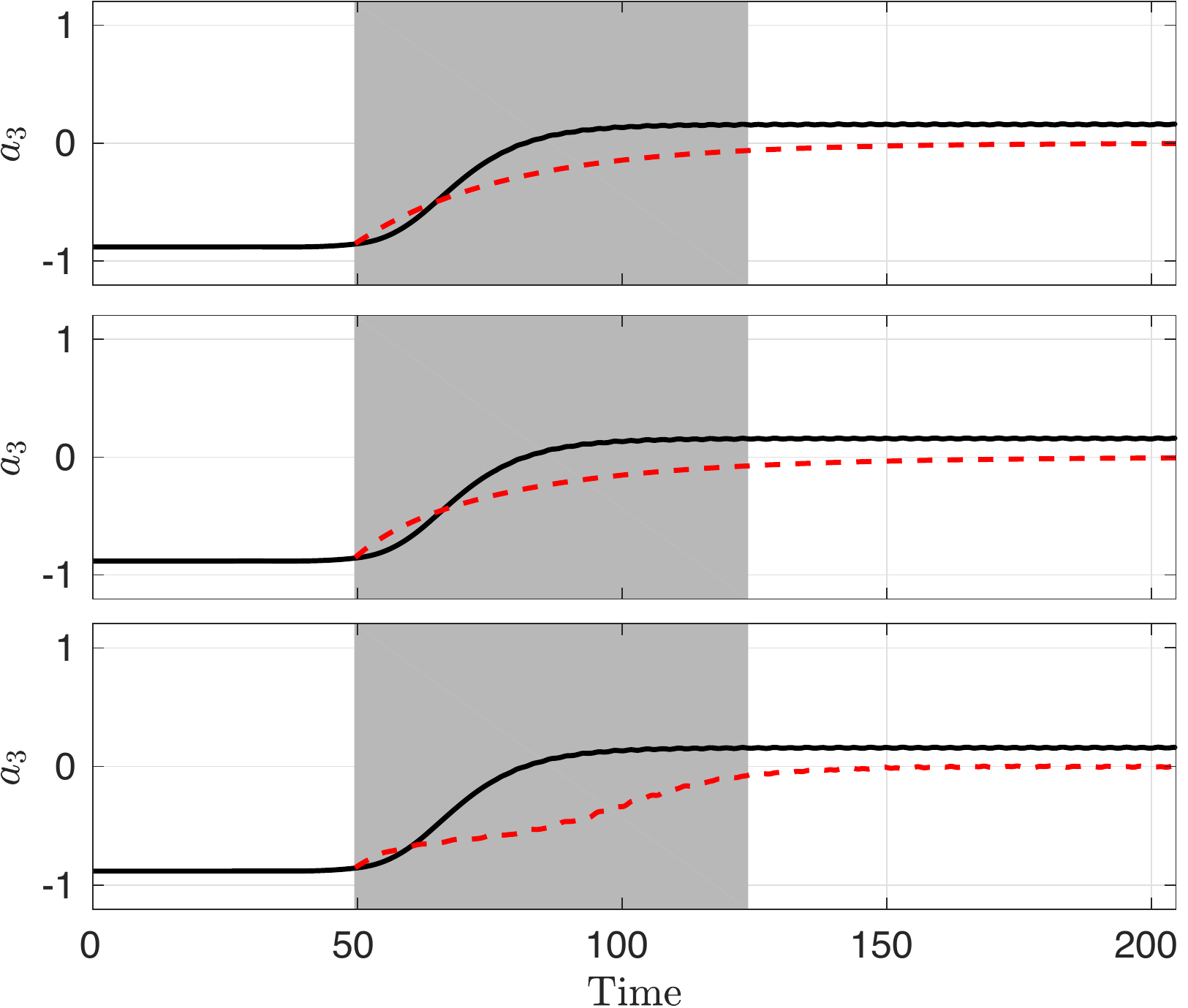}
\caption{Times series of predicted POD features obtained from (a) DMD (4-MSM-$\mathcal{I}$-1), (b) EDMD-TS (6-MSM-TS1) and (c) FFNN (6-MEM-TS-1) for TR-I as training region.}
\label{f:MT_F1}
\end{center}
\end{figure}    

\subsubsection{Choice of Training Data:} For this analysis, we used two training regions in the unstable transition regime, namely transient region-I (TR-I) and transient region-II (TR-II) as shown in figure \ref{f:datawindRE100} corresponding to $8-20$ and $4-16$ cycles respectively with both regions consisting of 372 data points. TR-I is relatively less challenging as almost all of the training data incorporates vortex shedding, but with an amplitude that is growing. In TR-II the first $30\%$ of the training data includes a stable wake with onset of instability that grows in amplitude all through the regime. This has implications for predictions using machine learning models where the training data almost always determines what kind of dynamics the model can predict. If one were to rank the level of difficulty in predicting the resulting limit-cycle dynamics from different sets of training data then the most difficult would be TR-II followed by TR-I and lastly, the limit-cycle training data\cmnt{ used in the previous section}. 

\subsubsection{Posteriori Predictions with Training Region I (TR I)}
\paragraph{\underline{Posteriori Predictions with Insufficient Nonlinearity and $\mathcal{LP}$ Dimension} :}  Figure~\ref{f:MT_F1} shows the predictions obtained from the different multilayer sequentially maps such as DMD and EDMD-TS and multilayer end-to-end FFNN  for the TR-I training region. We see that all these methods fail to learn the nonlinear dynamics and predict the resulting limit-cycle system to varying levels of inaccuracy with MEM being the least. This can be attributed to the lack of sufficient nonlinearity in the models and insufficient learning parameters to capture the dynamics. Highly transient systems with instability do not adhere to a point spectrum and require many eigenmodes to represent the unstable growth phase of the dynamics. However, once it settles into a limit cycle, a discrete spectrum is sufficient to represent the system. This correlates with a need for nonlinearity and increase in learning parameters in the data-driven architecture for modeling such systems.  It is worth pointing out that the EDMD-TS does not extend the $\mathcal{LP}$ space as against its polynomial variant, EDMD-P2. Also, the choice of P2 basis is physics-driven to account for the quadratic nonlinearity of the POD features as embedded within the Navier-Stokes equations that describe the flow. On the other hand, for the MEM architectures, a logical way to extend the $\mathcal{LP}$ space is to increase the number of features in the intermediate layers by increasing $N_f$. Consequently, we use EDMD-P2 as the baseline case and design a MEM architecture with similar sized $\mathcal{LP}$ space with feature factor, $N_f=3$. This approach of choosing $N_f$ based on the dimension of the quadratic polynomial features is a logical way to design MEM architectures as against more \emph{ad hoc} choices. 
  For EDMD-P2 (6-MSM-P2-1), the three input POD features are mapped onto a polynomial basis space with nine features. In the FFNN (6-MEM-TS-3), the three input features are mapped onto an unknown basis space, but guaranteed to be optimal for the chosen architecture and given training data. In this spirit of exploration, we also try a $7^{th}$-order polynomial feature map, i.e. a EDMD-P7 (6-MSM-P7-1) and corresponding MEM architectures with an increased $\mathcal{LP}$ dimension ($N_f=9$ and $20$) to assess the effect of $\mathcal{LP}$ dimensionality on the predictions. A downside to increasing the  $\mathcal{LP}$ dimension is a tendency to overfit the data which we will address.   
  
%
\par
\begin{figure}
\begin{center}
\includegraphics[width=0.33\columnwidth]{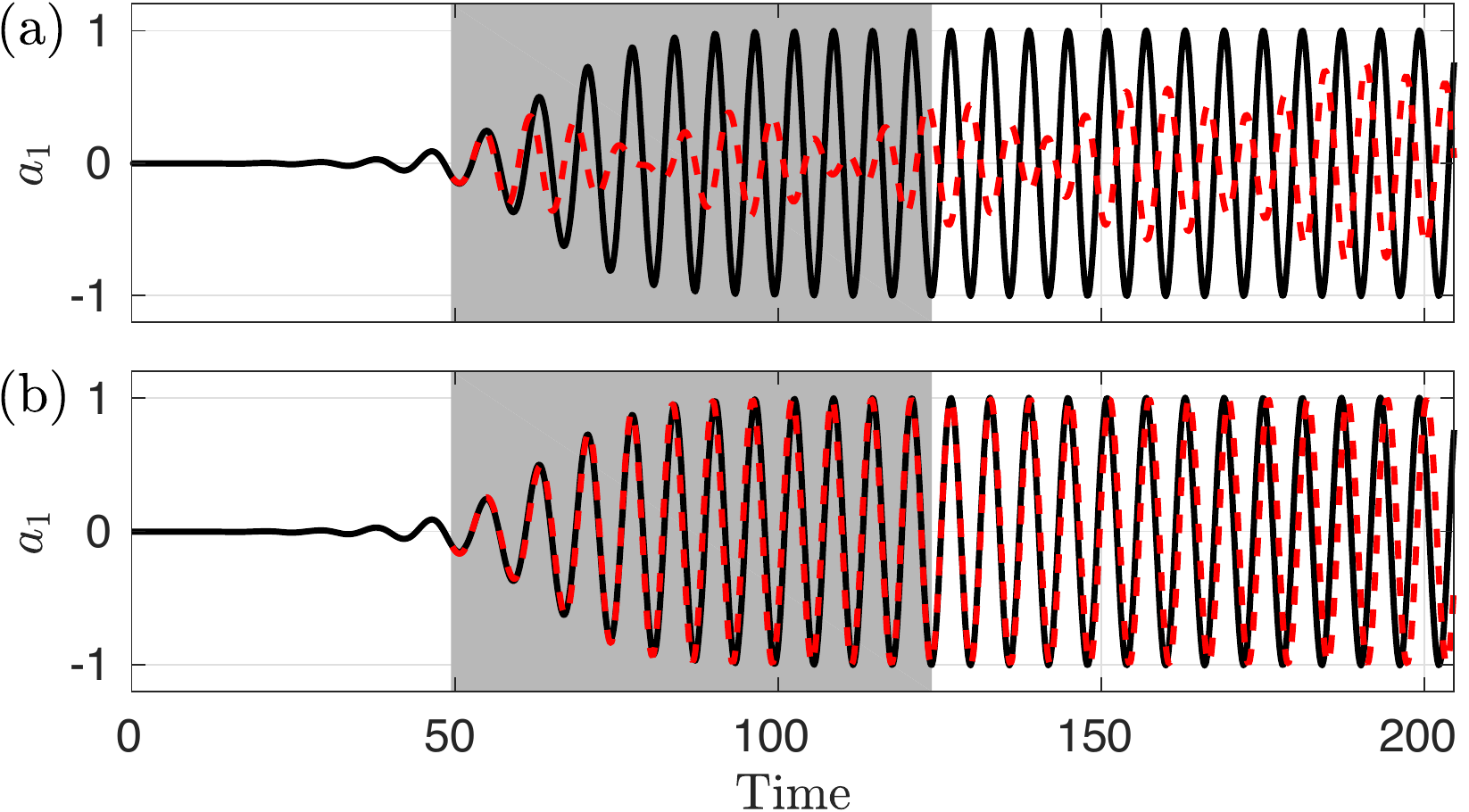}
\includegraphics[width=0.32\columnwidth]{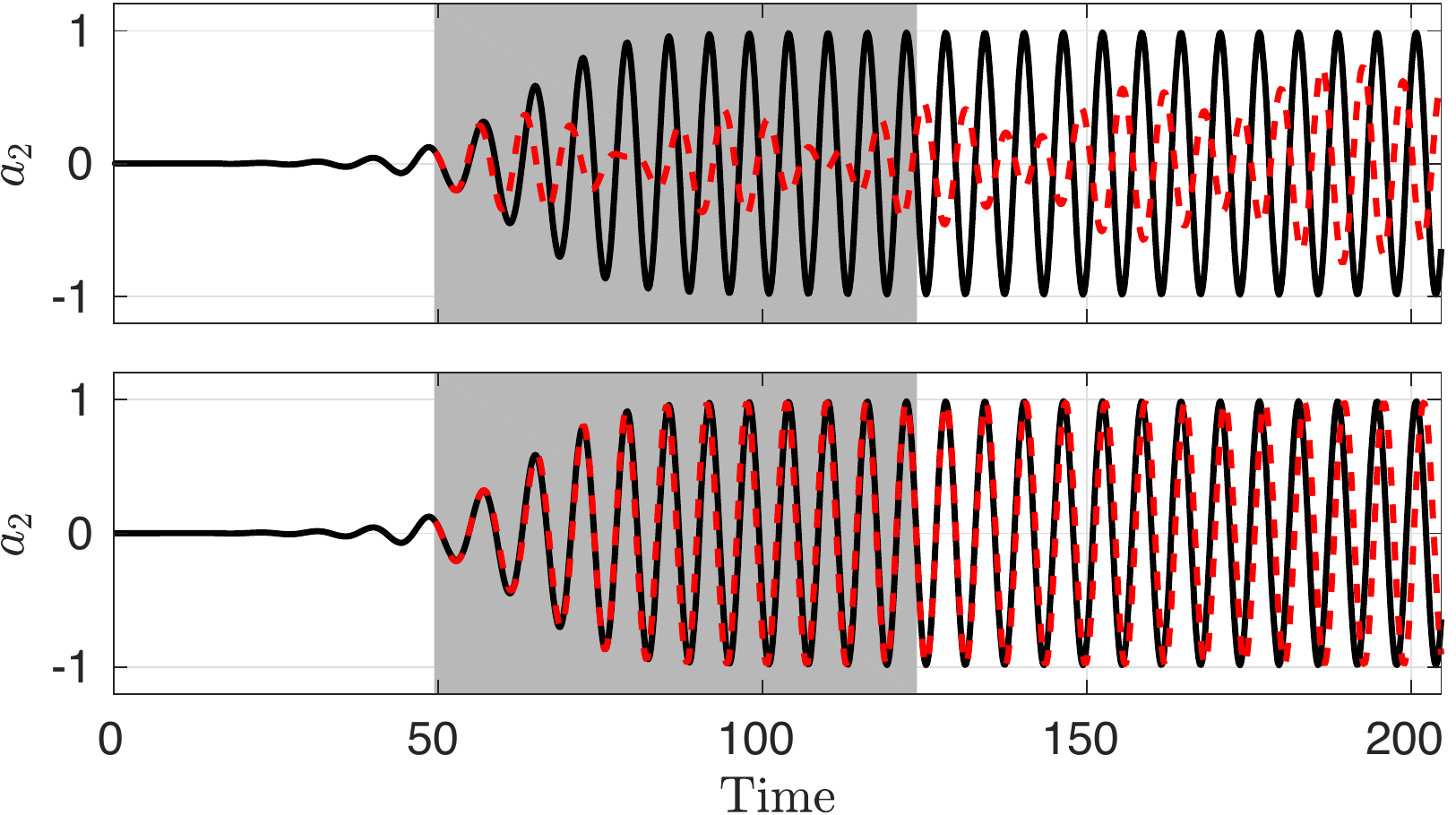}
\includegraphics[width=0.32\columnwidth]{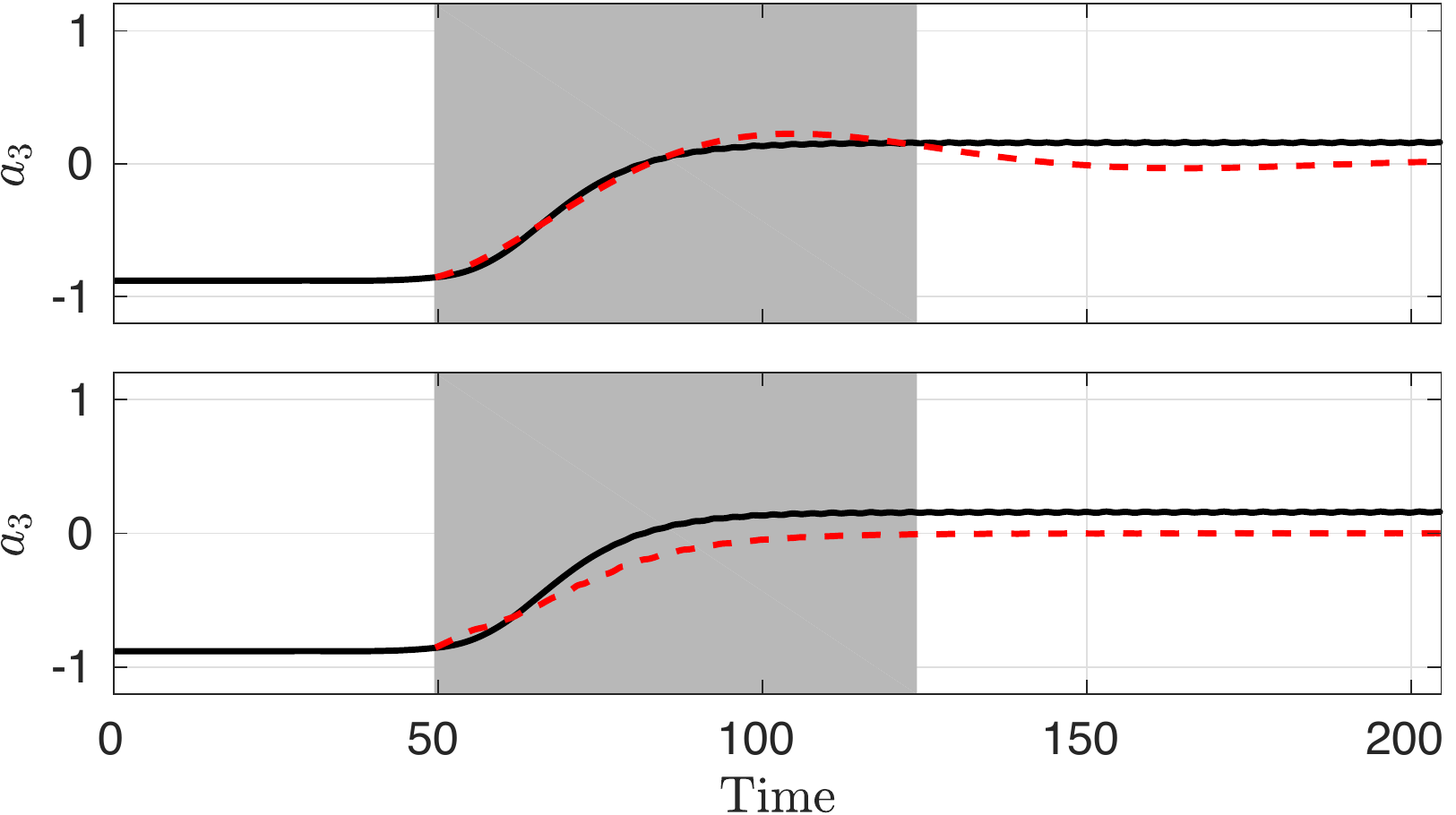}
\caption{Times series of predicted POD features obtained from (a) 6-EDMD-P2, (b) 6-MEM-TS3 for TR-I as training region. }
\label{f:MT_F2}
\end{center}
\end{figure}

 Figure \ref{f:MT_F2} shows the predictions from EDMD-P2\cmnt{(6-MSM-P2-1)} and FFNN ($N_f=3$)\cmnt{(6-MEM-TS-3)} using TR-I data. In spite of the embedded quadratic nonlinearity, the EDMD-P2 fails to the predict the correct limit-cycle dynamics using just three input features. On the other hand, FFNN ($N_f=3$)\cmnt{(6-MEM-TS-3)}   with a similar architecture but with end-to-end learning predicts the dynamical evolution of the  more accurately. These prediction error trends are quantified in table \ref{t:predstatRe100}. This is consistent with our expectation that a increasing $\mathcal{LP}$ dimension improves predictions as FFNN ($N_f=3$)\cmnt{(6-MEM-TS-3)}  learns $135$ parameters compared to $27$ for the FFNN ($N_f=1$)\cmnt{(6-MEM-TS-1)}  case. On the other hand, the EDMD-P2\cmnt{(6-MSM-P2-1)} case with $\mathcal{LP}=81$ fails to even predict qualitatively accurate results in spite of the added nonlinearity through the quadratic features.  In a related work by Jayaraman et al.~\cite{Lu:18sparse}, we have observed that EDMD-P2 with nearly fifty input features (with $1325$  quadratic nonlinear features and $\mathcal{LP}=1.7\times10^6$) can predict this transient instability driven growth of the wake.  It is also worth noting that $N_f=3$)\cmnt{(6-MEM-TS-3)} predicts the first two POD features accurately (see fig.~\ref{f:MT_F2}), but the third coefficient is biased towards a zero magnitude. We have found that this can mitigated through the use of a bias term which when incorporated into the MEM architectures corrects for this systematic deviation as discussed and shown in fig.~\ref{f:WBias_820Re100} included in Appendix \ref{s:Appendix1}. 

\begin{figure}
\begin{center}
\includegraphics[width=0.33\columnwidth]{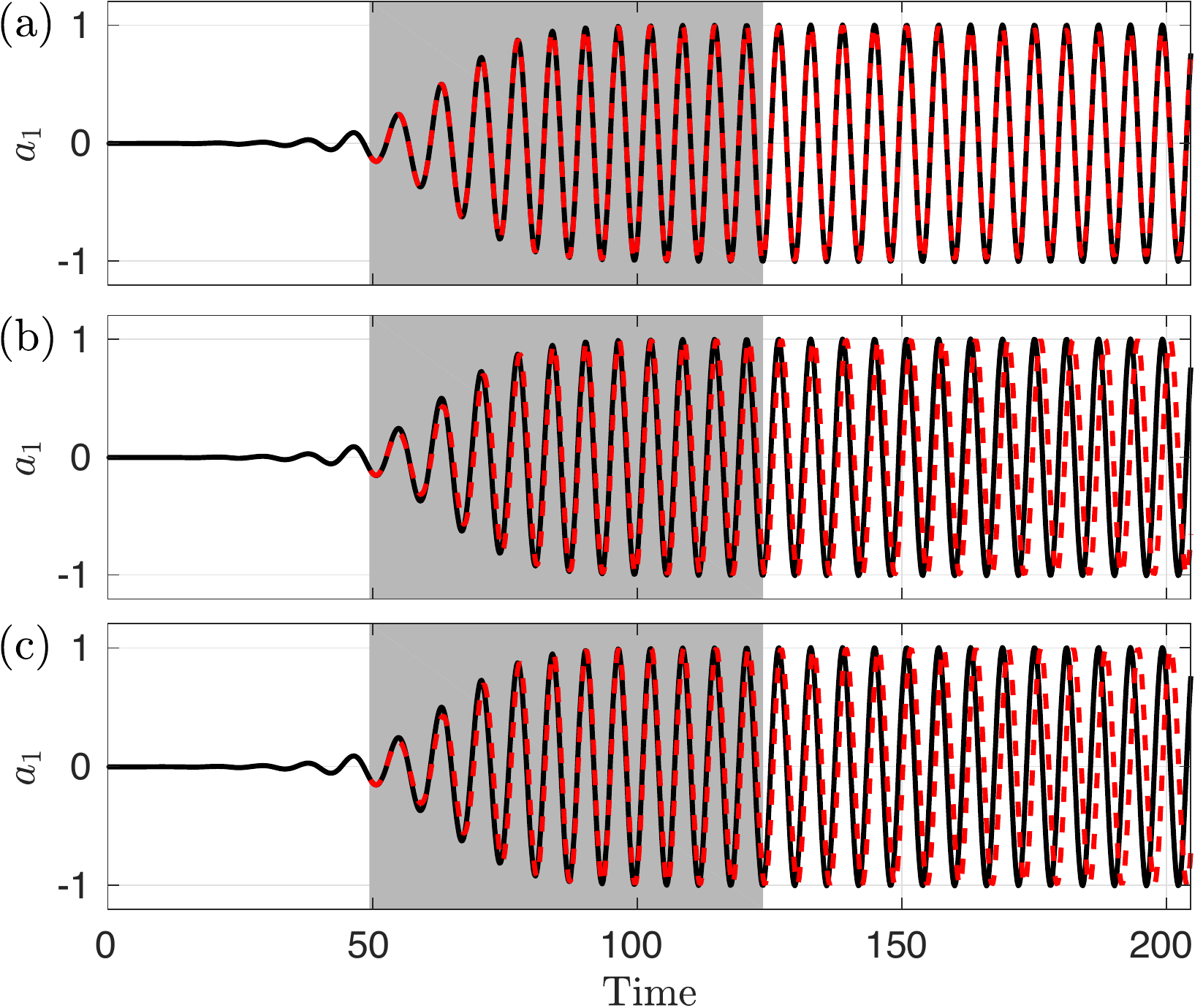}
\includegraphics[width=0.32\columnwidth]{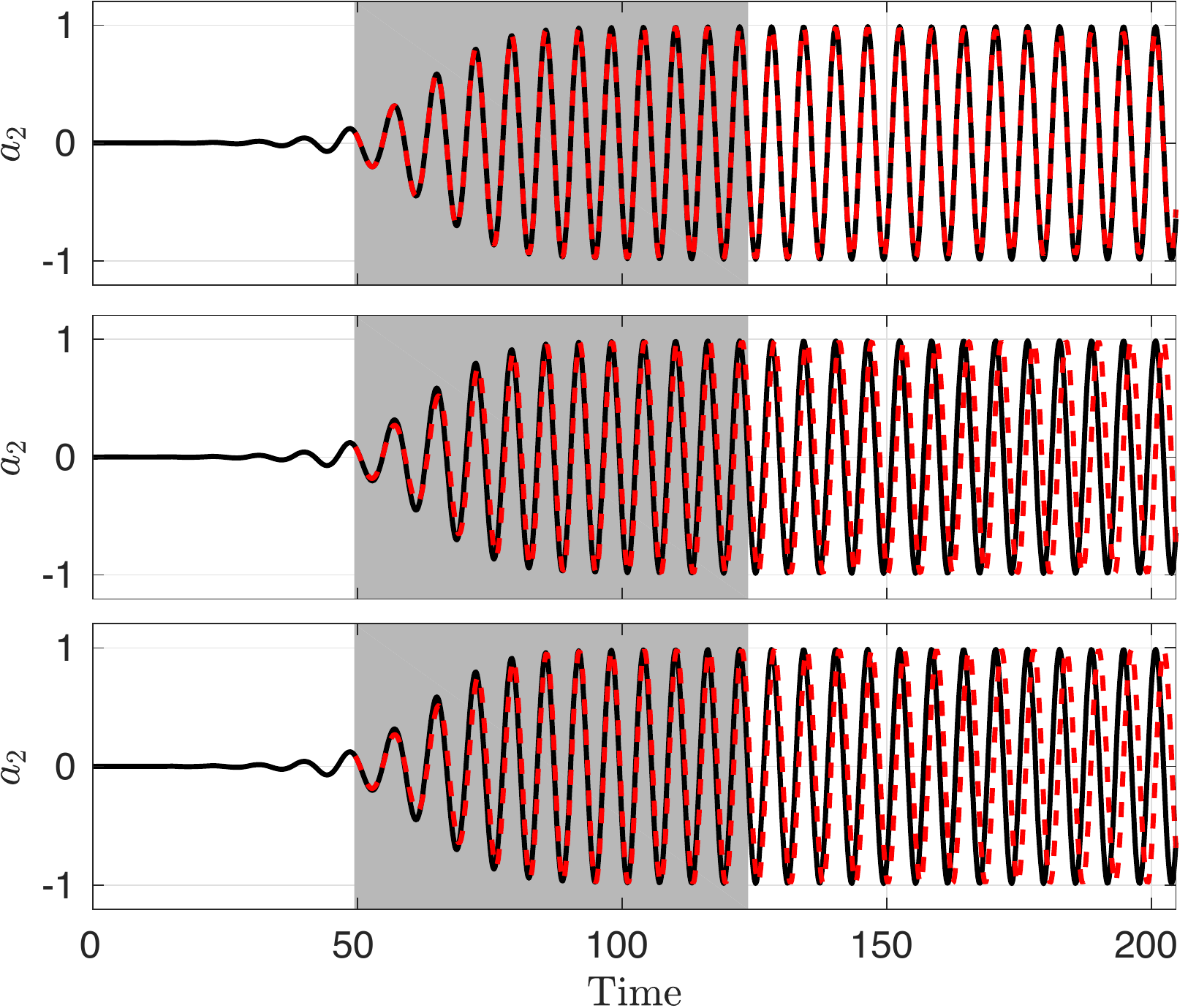}
\includegraphics[width=0.32\columnwidth]{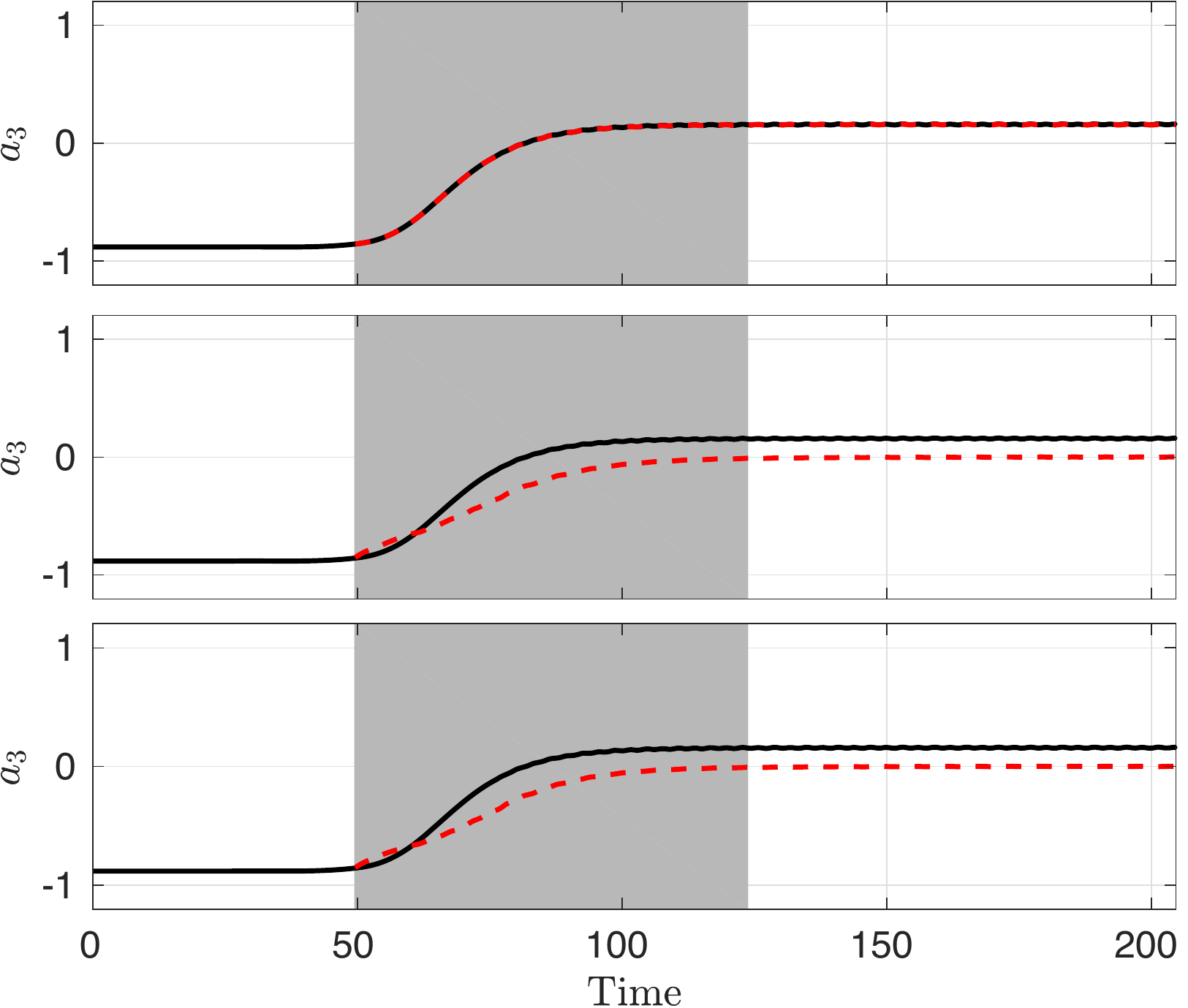}
\caption{Times series of predicted POD features obtained from (a) 6-EDMD-P7, (b) 6-MEM-TS9 and (c) 6-MEM-TS20 for TR-I as training region.}
\label{f:MT_F3}
\end{center}
\end{figure}

\paragraph{\underline{Effect of Increased $\mathcal{LP}$ Dimension on Posteriori Predictions:}}
Here, we explore the effect of expanding the $\mathcal{LP}$ dimension with just $3$ input features on the model performance. We accomplish this by increasing the order of polynomial to $7^{th}$-degree for the MSM i.e. we consider a EDMD-P7 with the architecture denoted by 6-MSM-P7-1. For this method, using just $3$ POD features results in an $\mathcal{LP}$ dimension of $15,625$- a nearly $\approx 200$ time increase as compared EDMD-P2.  With just $372$ snapshots being used, learning a linear operator $\Theta$ of size $15,625\times 15,625$ leads to overfitting and is expected to generate good results. In fact, increasing the $\mathcal{LP}$ dimension by a couple of orders of magnitude produces accurate predictions of the nonlinear dynamics as shown in fig.~\ref{f:MT_F3}a. We note that choices of polynomial smaller than degree seven did not produce accurate predictions although there may exist an isolated regularized solution that is reasonably accurate.  

We also explore the effect of increasing the $\mathcal{LP}$ dimension for the MEM architectures by changing $N_f$ as shown in table~\ref{t:methods}. The predictions obtained using FFNN ($N_f=9$)\cmnt{ (6-MEM-TS-9)}  and  FFNN ($N_f=20$)\cmnt{ (6-MEM-TS-20)}  (see figs.\ref{f:MT_F3}(b) and (c)) with $\mathcal{LP}$ dimension of $891$ and $3960$ respectively (factors of $\approx 10 \ \& \ 40$) also showed improved performance and compare favorably to the outcomes from the  EDMD-P7 architecture. To address concerns of overfitting associated with these large $\mathcal{LP}$ dimension, we performed validation of the learning process by splitting the data into training and testing sets as discussed in section~\ref{ss:trainValidate}. The outcomes shown in fig.~\ref{f:Cost8_20} clearly indicate that the error cost between training and testing remain consistent indicative of little overfitting for the MEM models with $N_f=1,3,9,20$. Another indication of how MEM models reduce overfitting as compared to the MSM (i.e. EDMD-P7) is how the prediction saturates as one increases the $\mathcal{LP}$ dimension.
In summary, both the sequential and end-to-end architectures work better by increasing the $\mathcal{LP}$ dimension and introducing nonlinearity. However, MEM requires relatively modest increases in $\mathcal{LP}$ dimension for substantial increases in performance.  Contrastingly, the MSM frameworks require large growth in features and $\mathcal{LP}$ dimension for performance improvement and is prone to overfitting the data. In a way, this result reinforces the underlying principles behind the success of deep learning architectures~\cite{bengio2015deep}. The MSM can be viewed as a two-layer shallow learning architecture requiring larger intermediate layer dimensions while the MEM is its deep learning counterpart requiring smaller number of intermediate layer features, but across multiple layers which in turn reduces overfitting.

\subsubsection{Posteriori Predictions with Training Region II (TR II)}
\begin{figure}
\begin{center}
\includegraphics[width=0.33\columnwidth]{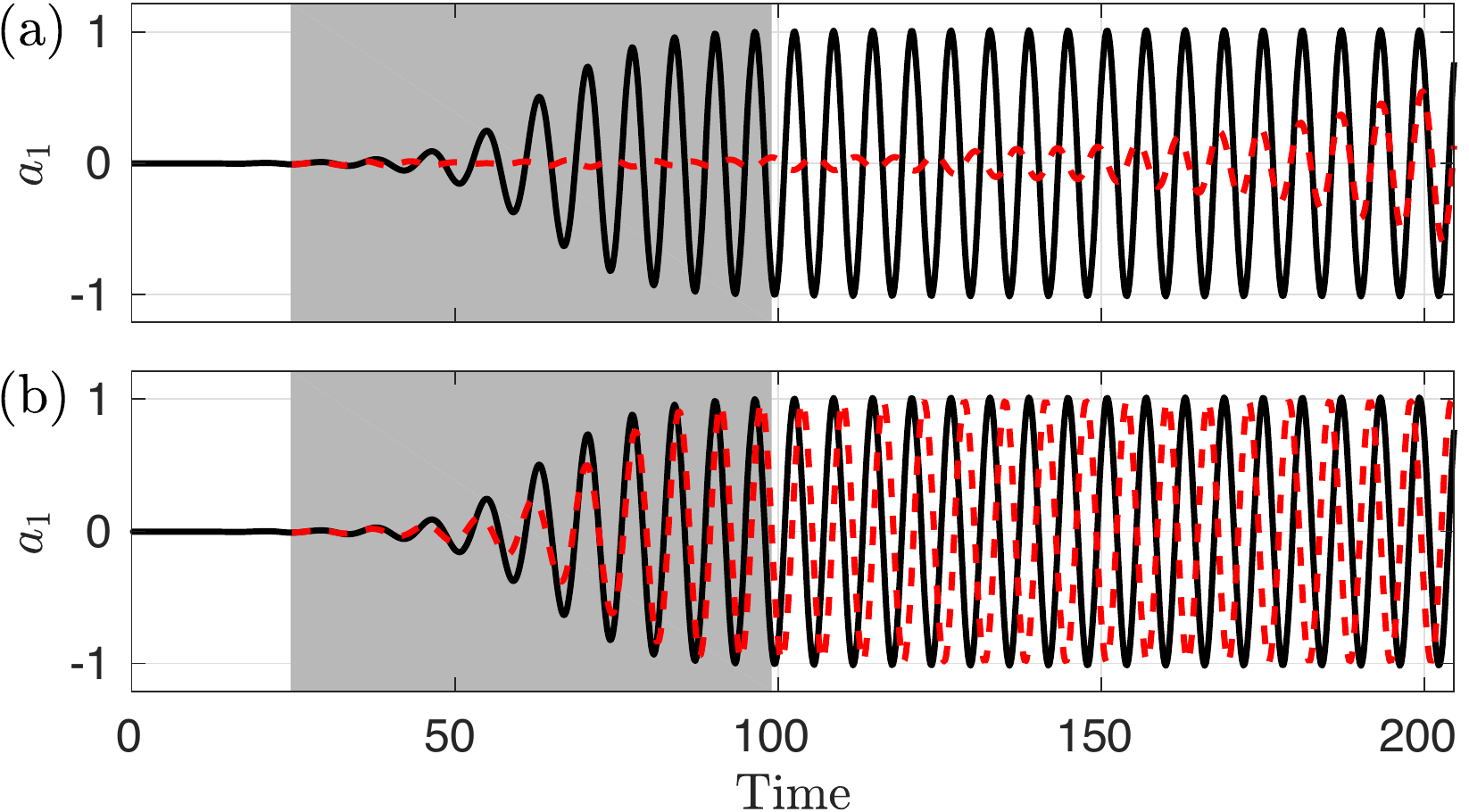}
\includegraphics[width=0.32\columnwidth]{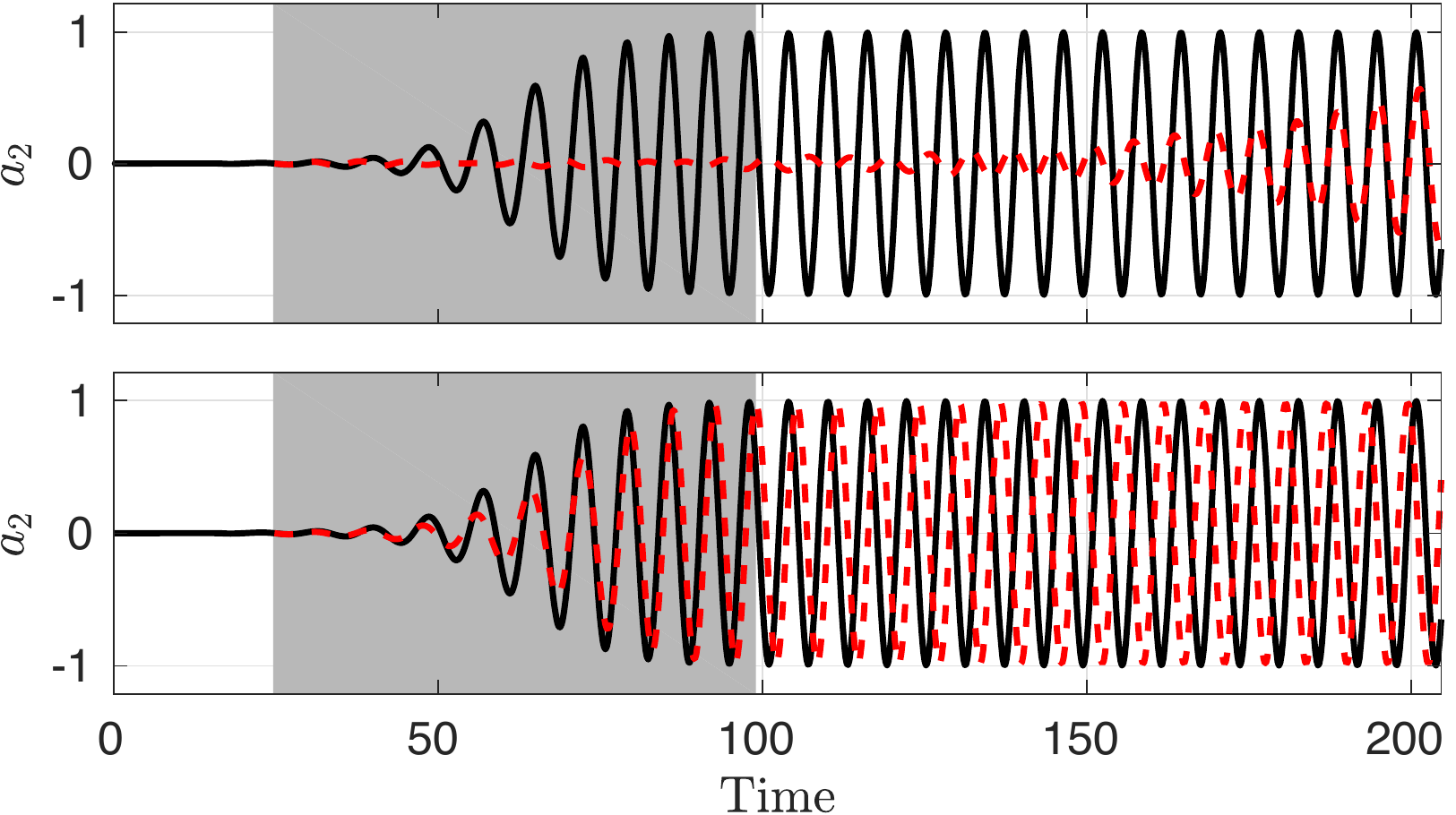}
\includegraphics[width=0.32\columnwidth]{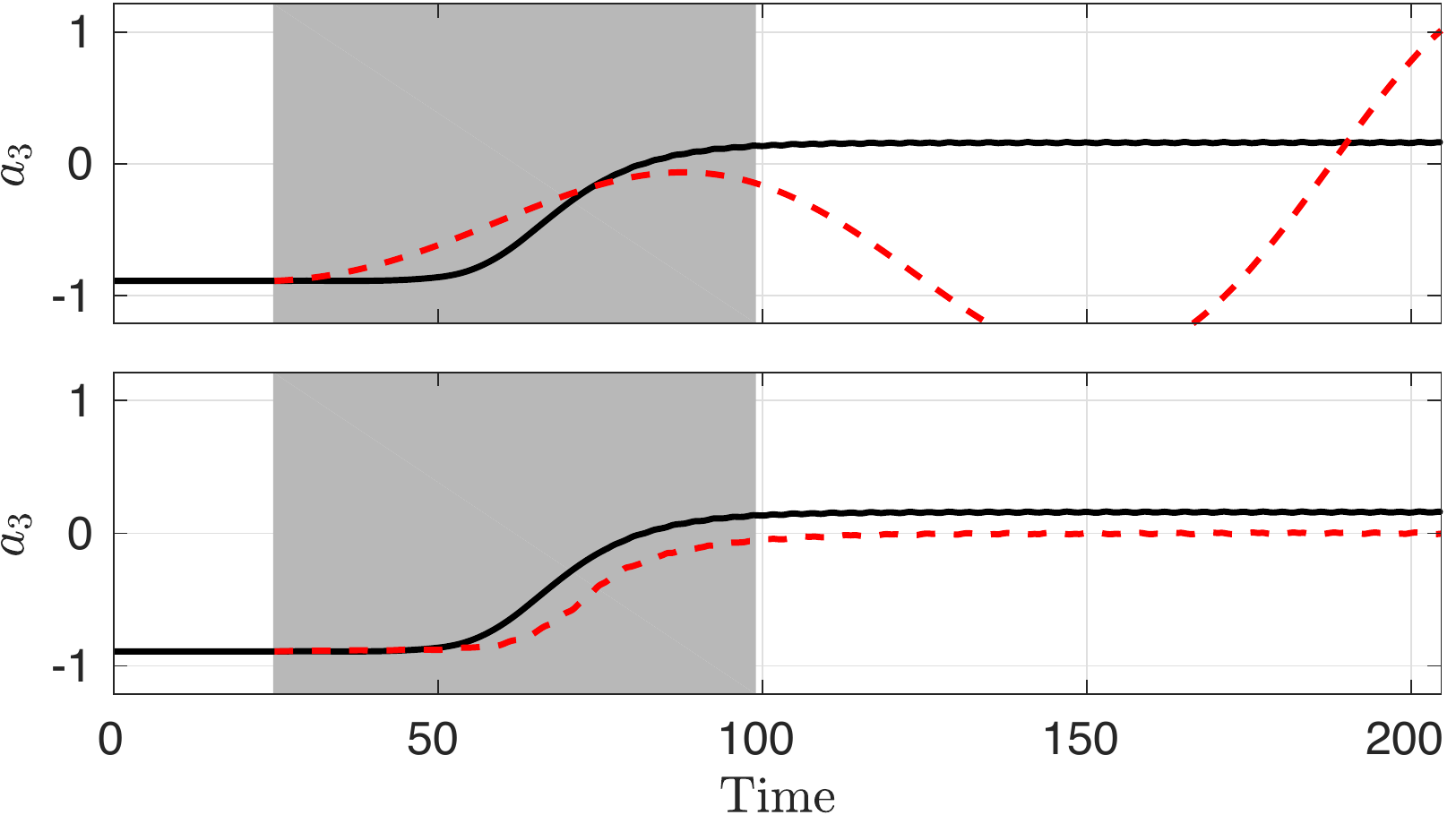}
\caption{Times series of predicted POD features obtained from (a) 6-EDMD-P2, (b) 6-MEM-TS3 for TR-II data. }
\label{f:FT_F2}
\end{center}
\end{figure}

\begin{figure}
\begin{center}
\includegraphics[width=0.33\columnwidth]{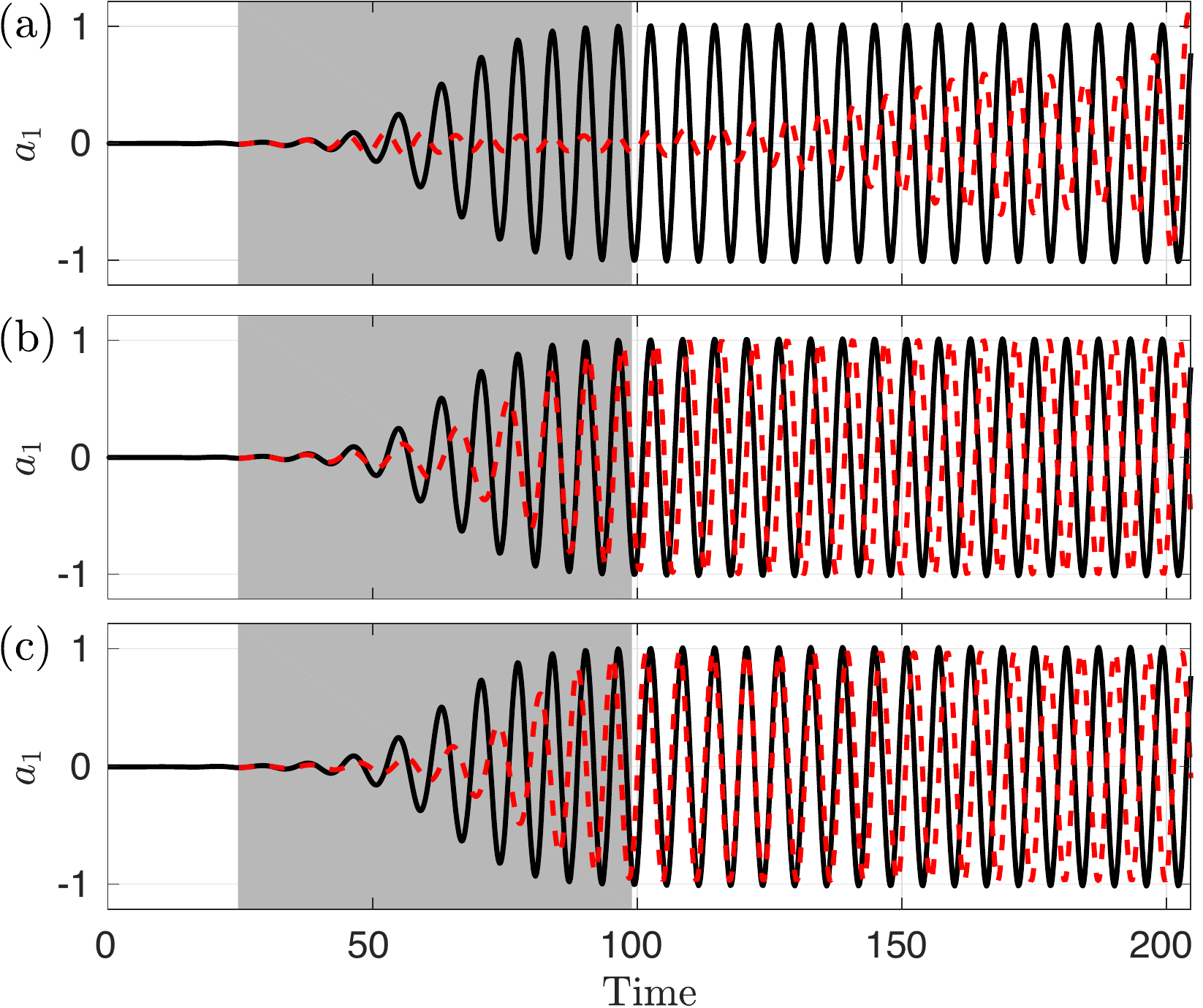}
\includegraphics[width=0.32\columnwidth]{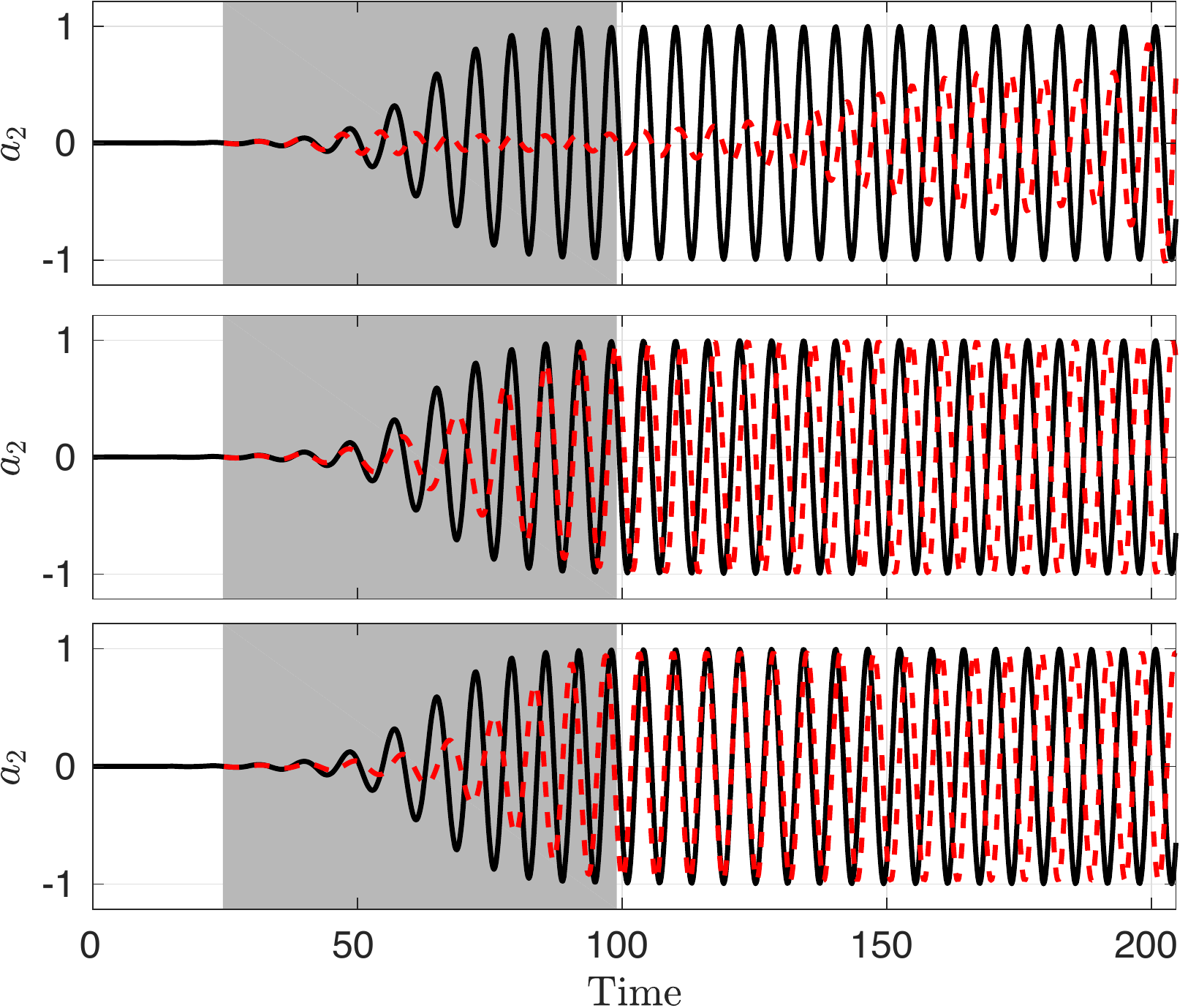}
\includegraphics[width=0.32\columnwidth]{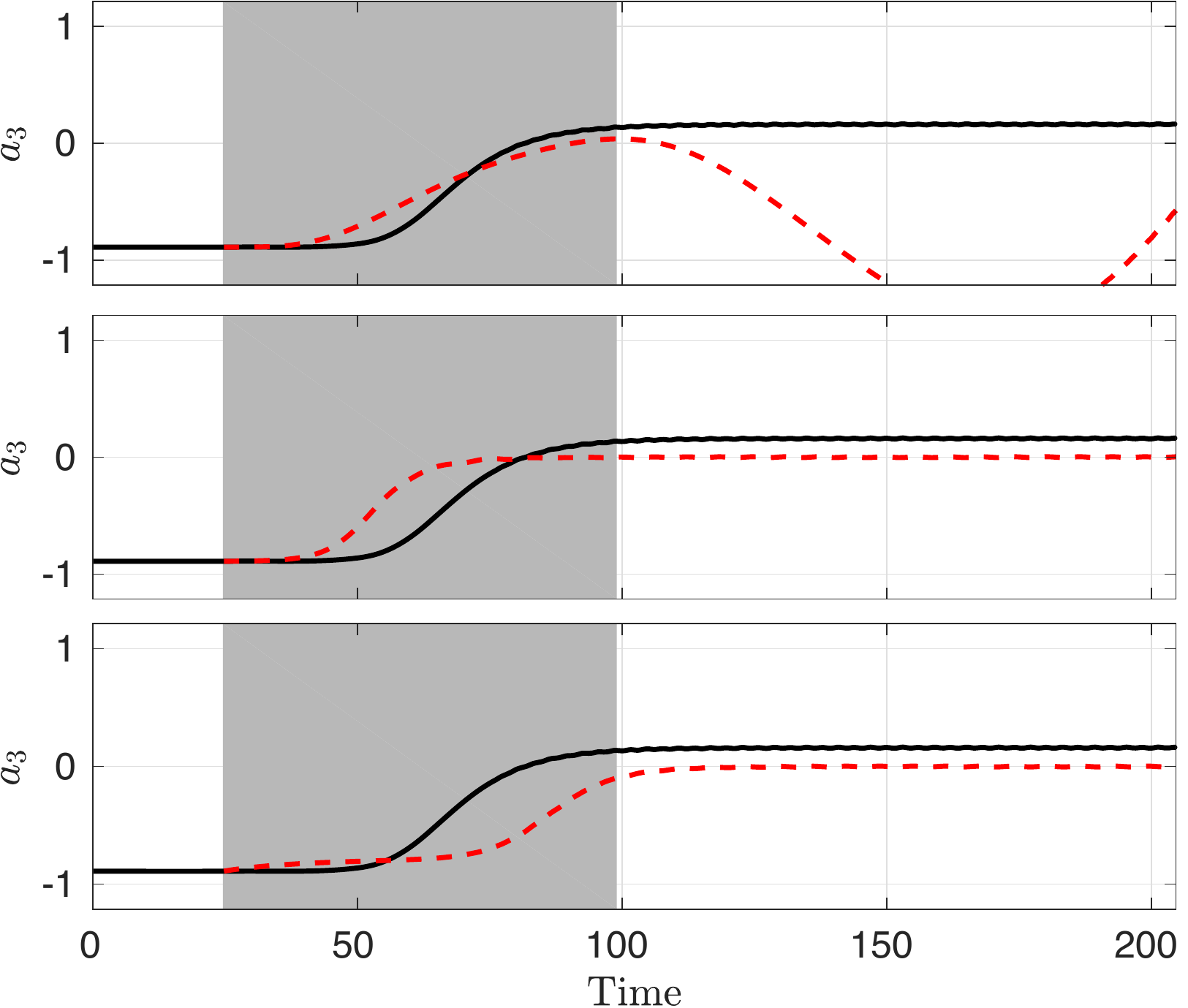}
\caption{Times series of predicted POD features obtained from an extended $\mathcal{LP}$ space (a) 6-EDMD-P7, (b) 6-MEM-TS9 and (c) 6-MEM-TS20 for TR-II data.}
\label{f:FT_F3}
\end{center}
\end{figure}

We use the same modeling architecture's as before for this challenging TR-II dataset and the resulting predictions of the POD features are shown in figures \ref{f:FT_F2} and \ref{f:FT_F3}. In this case both the MSM architectures, i.e. EDMD-P2 and EDMD-P7 perform inadequately in spite of the increased $\mathcal{LP}$ dimension. On the other hand, predictions obtained using FFNN (MEM) offer better qualitative results and predict the limit cycle dynamics, but display perceptible quantitative inaccuracy without a bias term and is insensitive to extension of learning parameter space (see table~\ref{t:predstatRe100}). However, as before, we observe that this quantitative inaccuracy, especially in the third POD feature is mitigated through the inclusion of a bias term as the plots clearly show in fig.~\ref{f:WBias_416Re100} in Appendix~\ref{s:Appendix1}.

\subsubsection{Analysis of Prediction Errors}
\begin{figure}
\begin{center}
\includegraphics[width=0.95\columnwidth]{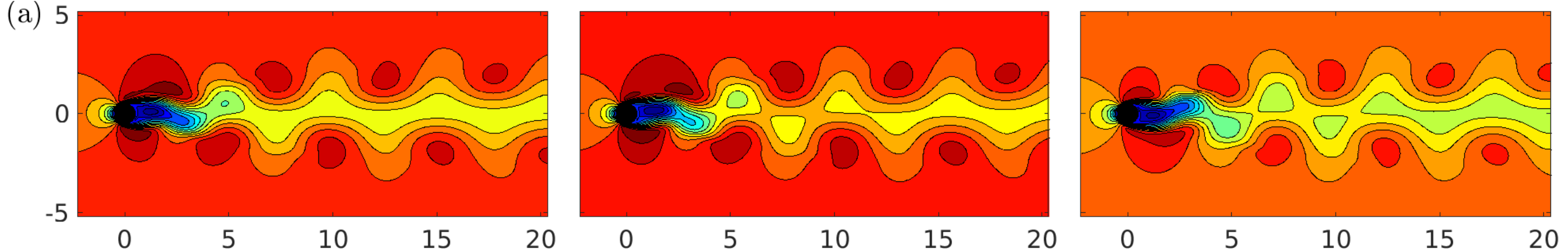}
\includegraphics[width=0.95\columnwidth]{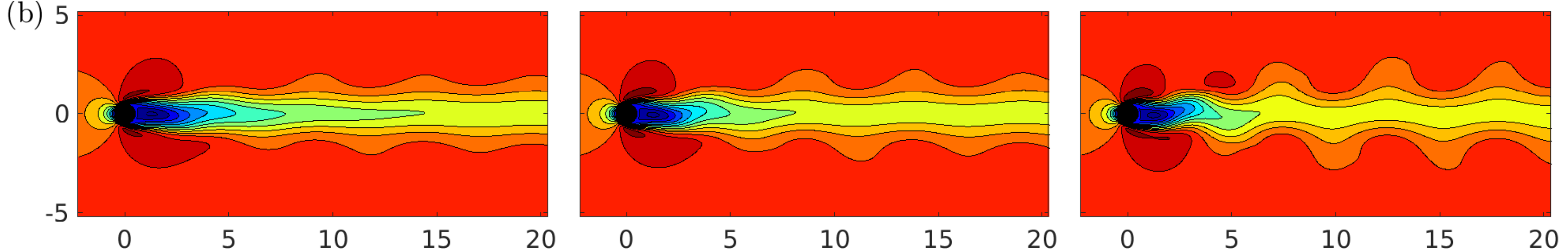}
\includegraphics[width=0.95\columnwidth]{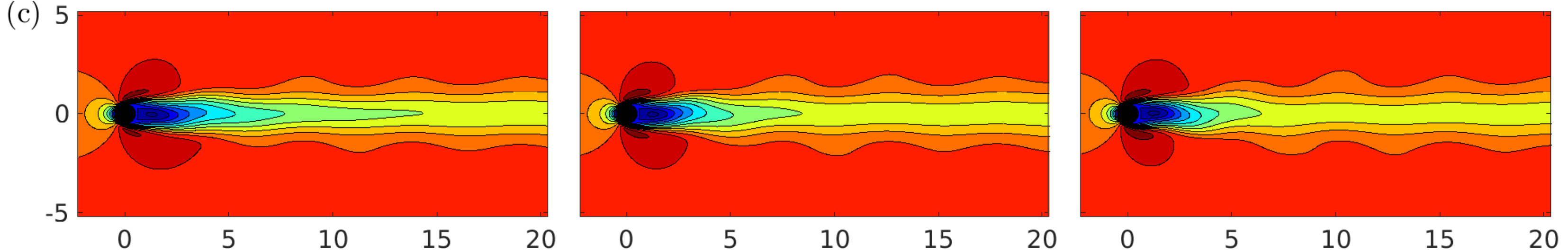}
\includegraphics[width=0.95\columnwidth]{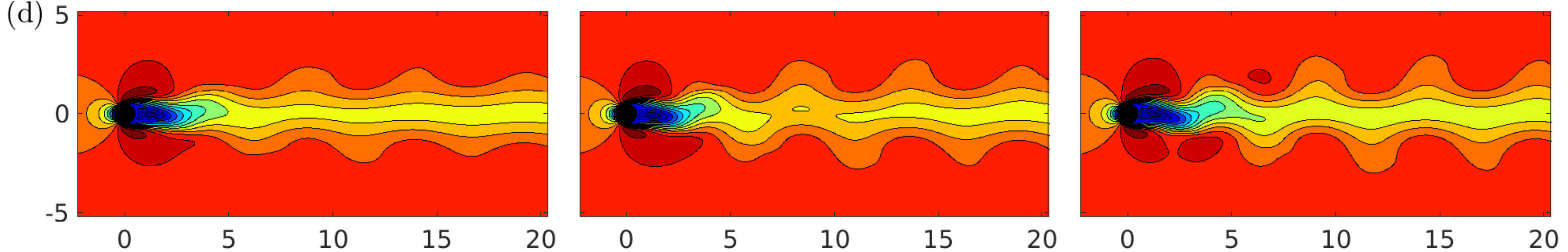}
\includegraphics[width=0.95\columnwidth]{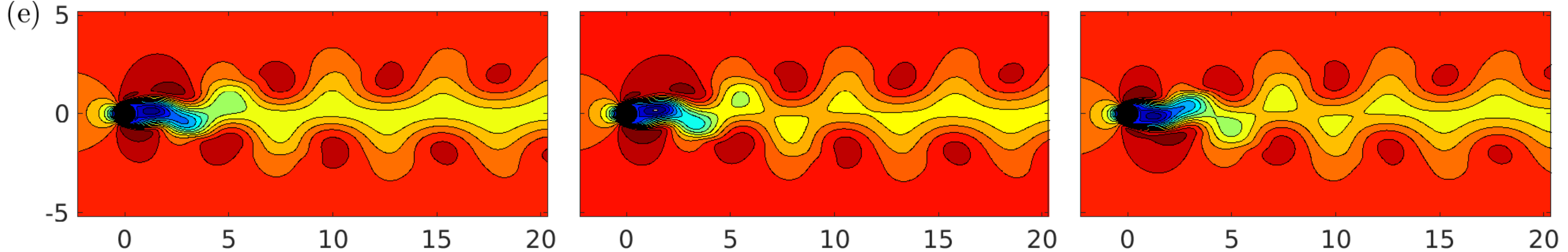}
\includegraphics[width=0.95\columnwidth]{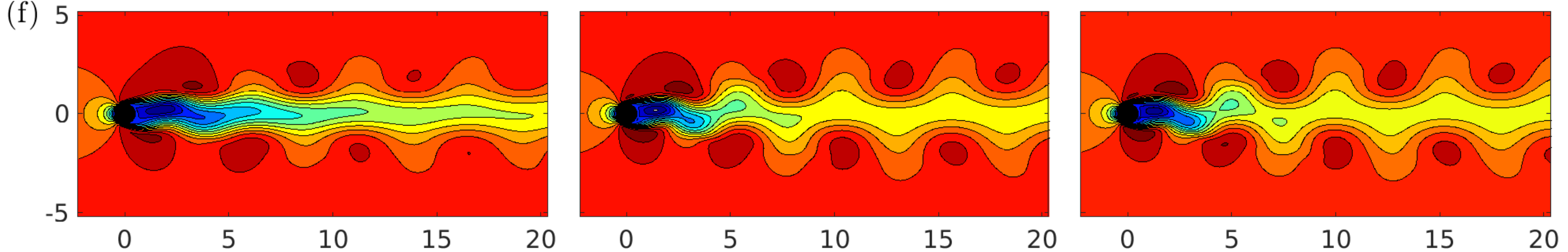}
\includegraphics[width=0.95\columnwidth]{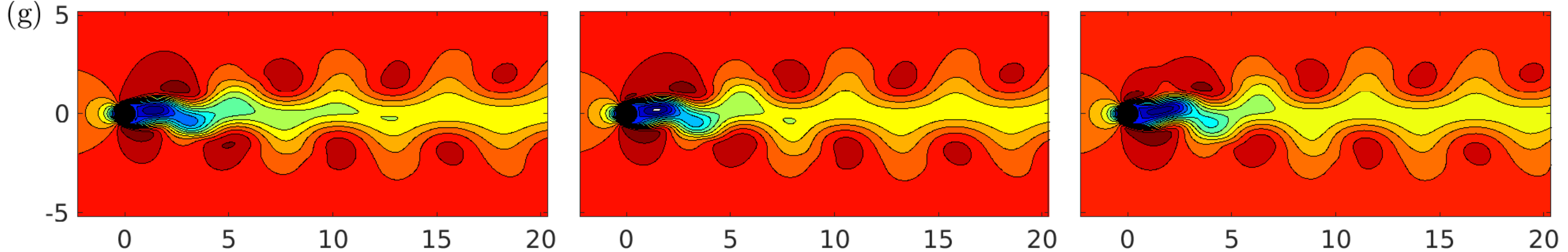}
\includegraphics[width=0.95\columnwidth]{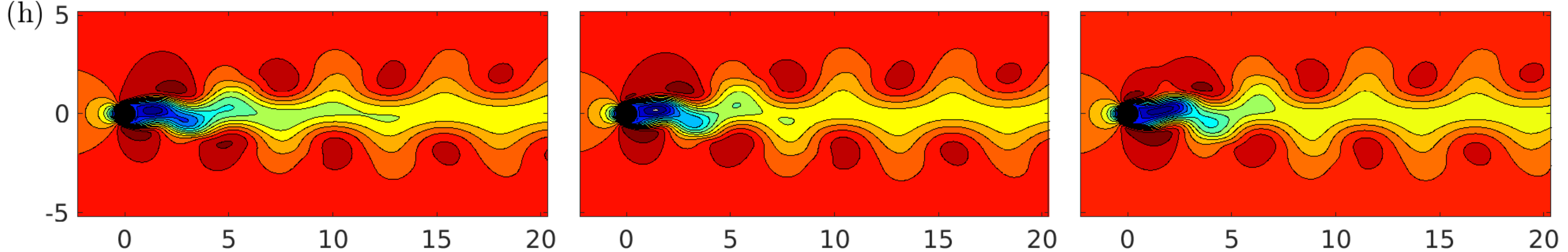}
\caption{Reconstruction of $Re100$ flow field based on predicted POD features obtained from (a) Actual data, (b) DMD (4-MSM-$\mathcal{I}$-1) (c) EDMD-TS (6-MSM-TS-1) (d) EDMD-P2 (6-MSM-P2-1)  (e) EDMD-P7 (6-MSM-P7-1)  (f) FFNN with $N_f=1$ (6-MEM-TS-1)  (g) FFNN with $N_f=3$ (6-MEM-TS-3) (h) FFNN with $N_f=9$ (6-MEM-TS-9). The contours levels are given by fifteen equally spaces values ranging between $(-0.2645,1.2963)$ .} 
\label{f:Reconst}
\end{center}
\end{figure}

We note that computing the error metrics using a simple $L_2$ norm does not adequately represent the qualitative nature of the predictions accurately for such repetitive limit-cycle dynamics. For example, the predictions which qualitatively mimic the dynamics, but with incorrect phase tends to show larger errors than some of the non-qualitative predictions.  The other aspect worth mentioning is that learning is based on \emph{a priori} prediction cost minimzation and not the \emph{a posteriori} predictions (as shown in section~\ref{ss:trainValidate} and figs.~\ref{f:apriori} \& \ref{f:apriori}) that takes into account error propagation. We can understand this clearly by studying the compilation of the error metrics in table \ref{t:predstatRe100}.  While the learning cost ($\mathcal{J}$) for the different FFNN architectures is $O(1e^{-6})$ (see fig.\ref{f:Cost8_20}), the associated posteriori prediction errors are of $O(1e^{-1})$. It is well known that classical machine learning is based on \emph{a priori} prediction cost and is not designed for time-series estimation where error propagation is significant. Recent approaches~\cite{pan2018long} propose improved regularizations that account for error growth through the use of a Jacobian of the cost function. To relate the observed deviations in the POD features to the predicted flow field of interest, we show in  fig.\ref{f:Reconst} the reconstructed solution (i.e. the actual predicted state vector) for $Re=100$ obtained using the different methods considered in this paper. These plots are generated based on learning and prediction using TR-I $(cycles: \ 8-20)$ data, and shown at $\approx T=86.2$ (first column) which is the midpoint of the training region. Columns 2 and 3 in fig.\ref{f:Reconst} represent predictions at $T=124$, the last point in TR-I and $T=205$, the last point in the prediction regime. These results clearly show that the MSM frameworks with low  $\mathcal{LP}$ dimension such as DMD, EDMD-TS and EDMD-P2 show delayed onset of wake instability and incorrect vortex shedding while the FFNN for all the different architectures predict the instability growth accurately.

%

\subsection{\emph{A priori} Learning and \emph{A posteriori} Prediction of a Transient 2D Buoyant Boussinesq Mixing Flow }
\label{ss:evo_transientdata}

Unlike the low-dimensional limit-cycle attractor modeled in the earlier sections, here we explore a non-stationary and higher-dimensional buoyant Boussinesq mixing flow discussed in sec.~\ref{sss:buoymix}. In fact, we observed previously that prediction of the transient instability growth and subsequent stabilization of the cylinder wake dynamics is highly sensitive to the choice of training data. In addition, learning and predictability of these dynamics are also dependent on the training data including sufficient information for accurate prediction. For this study, we chose to retain just $80\%$ of the total energy of the system captured in the CFD generated data snapshots (similar to a low resolution measurement) resulting in just $3$ POD features in the $2^{nd}$ layer of the MSM and MEM architectures. Sensitivity to these aspects is stronger when trying to predict non-stationary phenomena that may settle into an unknown attractor over long times. The training data is almost always insufficient to represent all the possible dynamics for such systems and may not overtly show any evidence of the existence of such an attractor.  Such instability-driven non-stationary problems are challenging for data-driven techniques that do not leverage knowledge of the underlying governing system and employ black box machine learning.  Even if one were to diversify the training data-set with multiple realizations of the system, performance improvements are not guaranteed as the underlying dynamics will depend a lot on the initial state. For this study, we choose a single realization of such a data-sparse and low-dimensional representation of a system for assessment of the different MSM and MEM architectures . 

\begin{figure}
\begin{center}
\includegraphics[width=0.81\columnwidth]{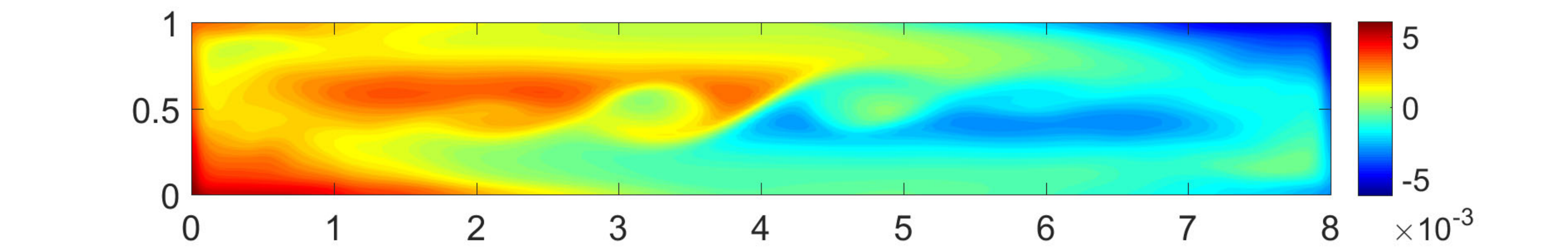}
\includegraphics[width=0.81\columnwidth]{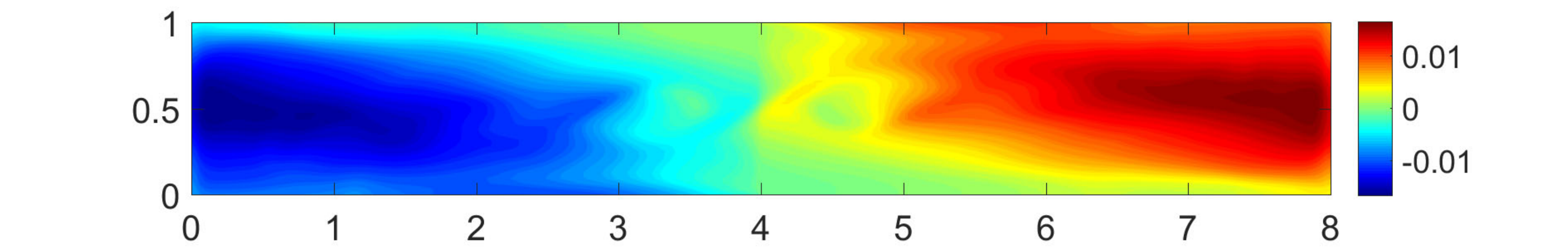}
\includegraphics[width=0.81\columnwidth]{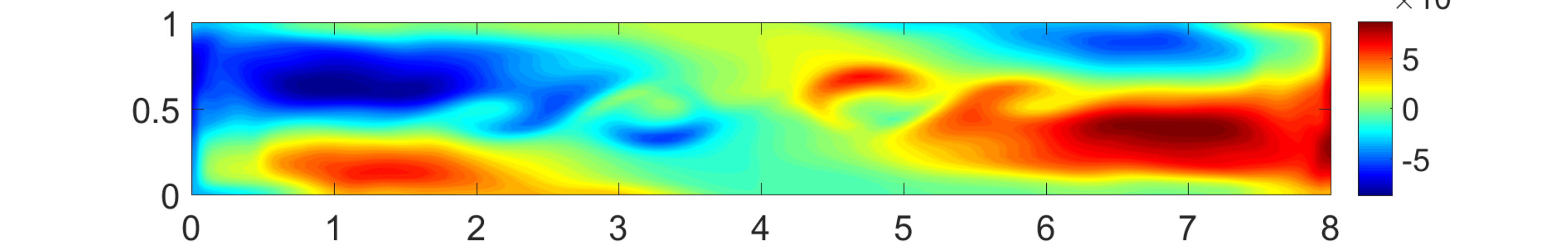}
\caption{Visualization of the first three POD basis (in decreasing order of energy content) used to model the dynamics with the data-driven models.}
\label{f:lockexmodes}
\end{center}
\end{figure}

\begin{figure}
\begin{center}
\includegraphics[width=0.33\columnwidth]{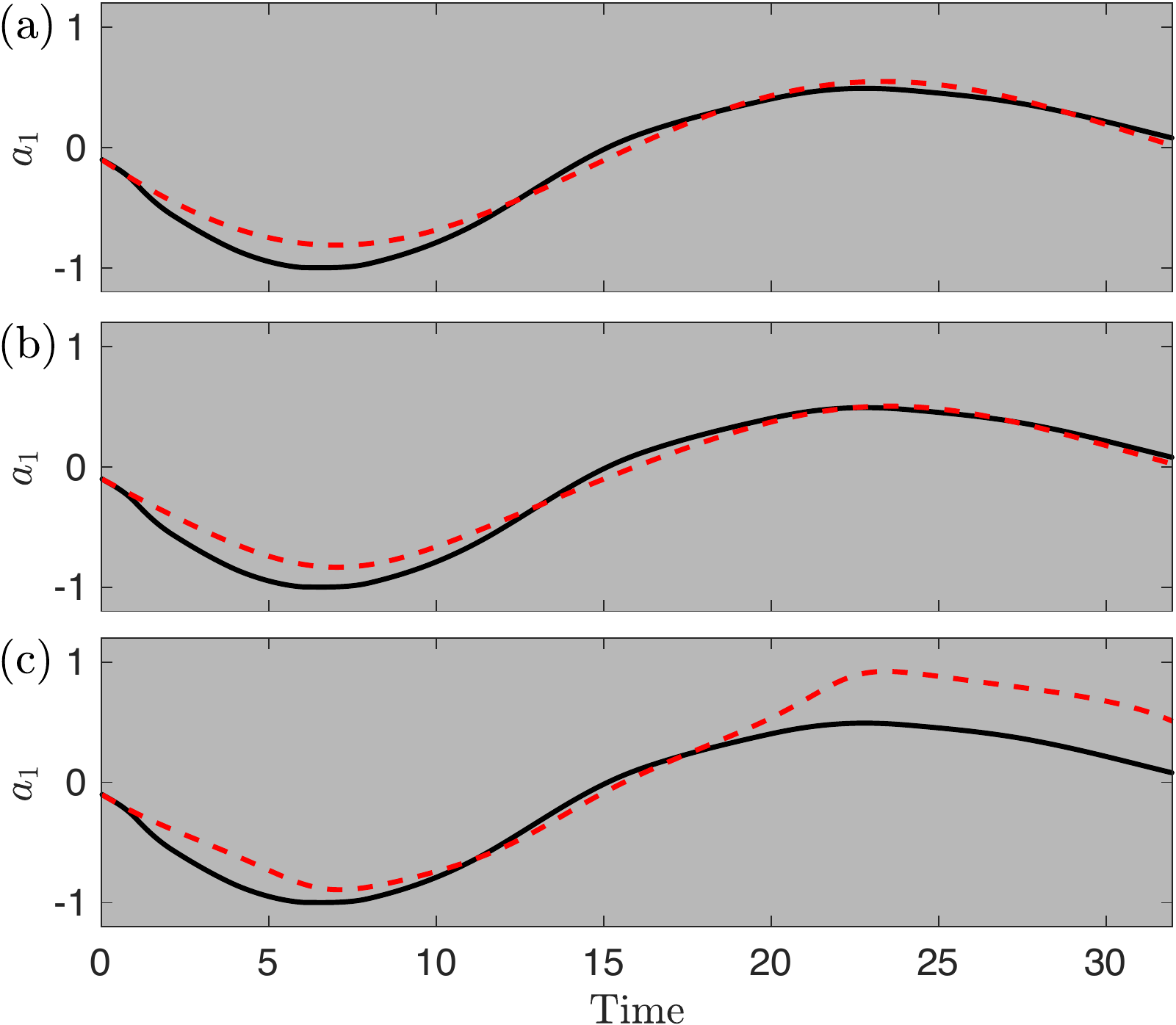}
\includegraphics[width=0.32\columnwidth]{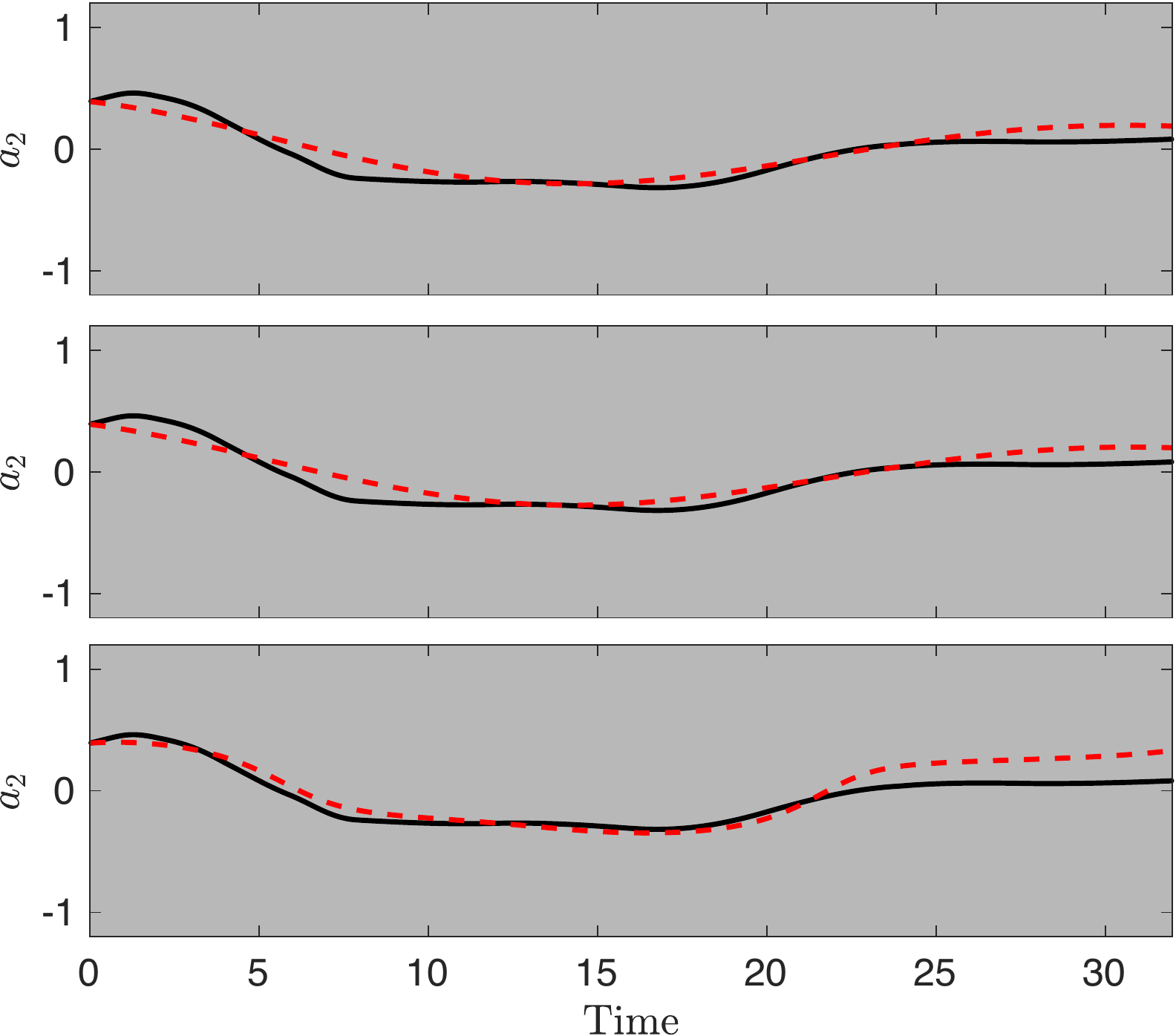}
\includegraphics[width=0.32\columnwidth]{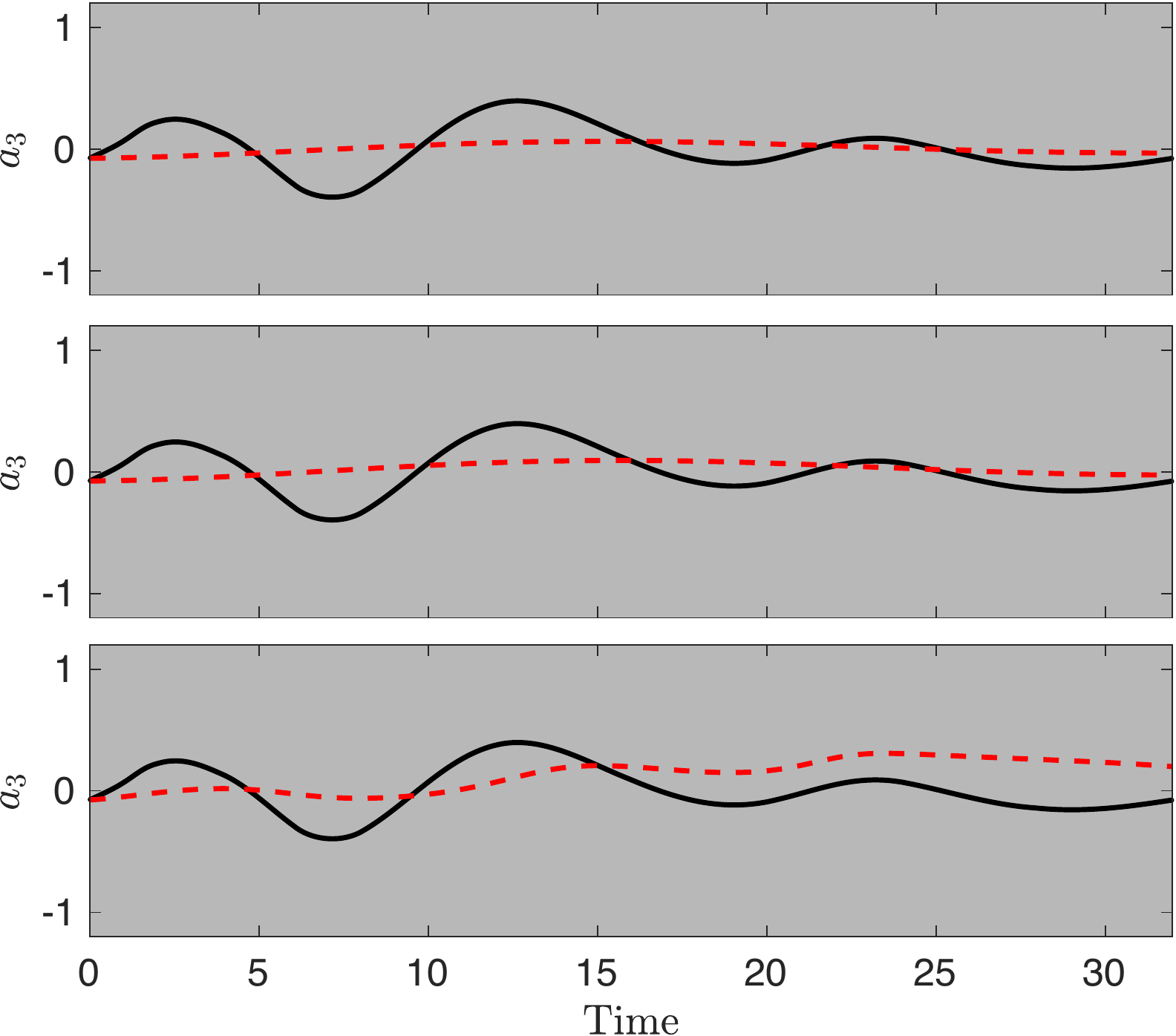}
\caption{Comparison of the time evolution of the posteriori prediction of the $3$ POD features generated from the different modes with the true data using all the $1600$ snapshots for training. The different plots correspond to: (a)DMD, (b)  EDMD-TS1  and (c) FFNN with $N_f=1$.}
\label{f:F1_1600}
\end{center}
\end{figure}

\begin{figure}
\begin{center}
\includegraphics[width=0.33\columnwidth]{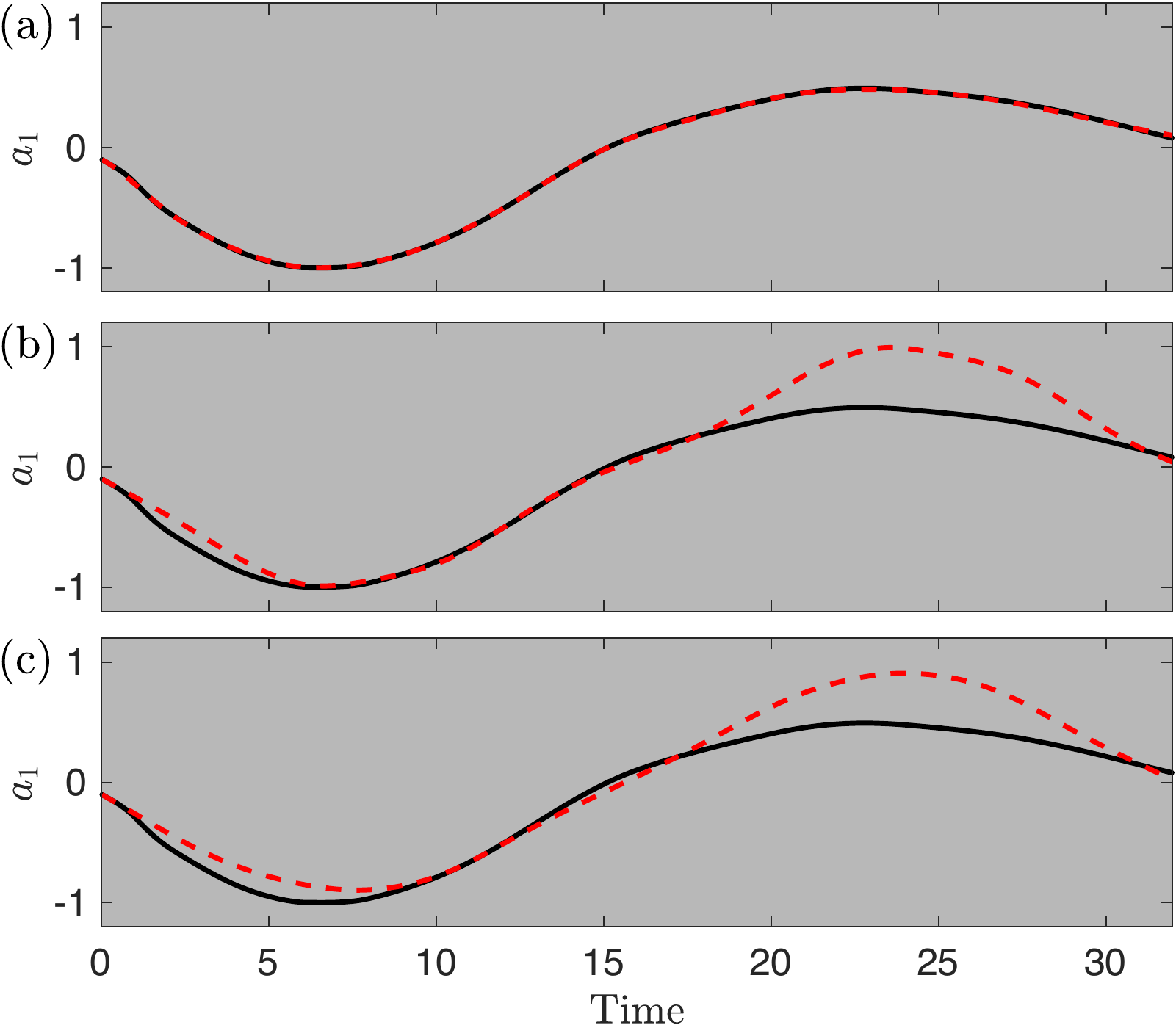}
\includegraphics[width=0.32\columnwidth]{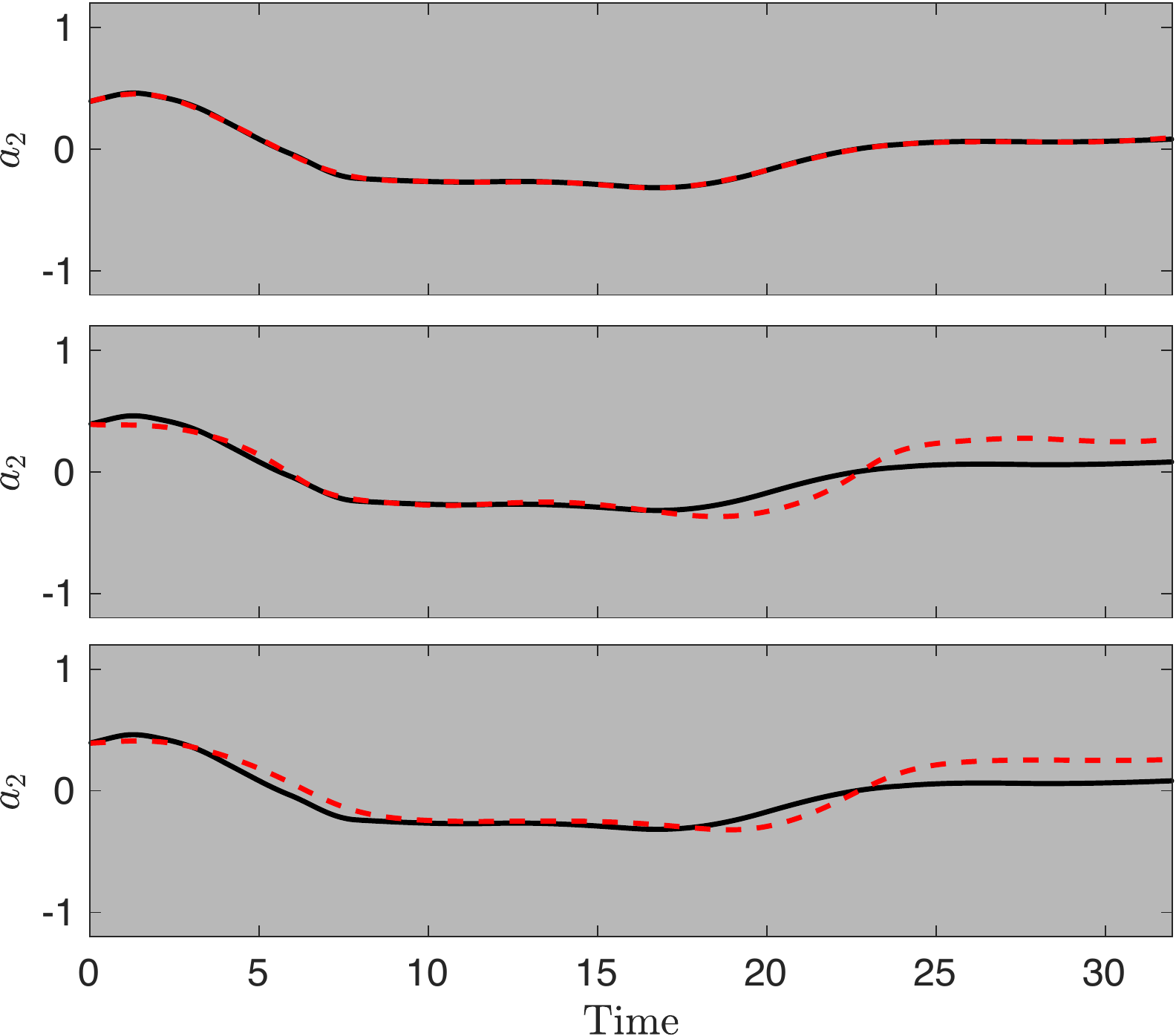}
\includegraphics[width=0.32\columnwidth]{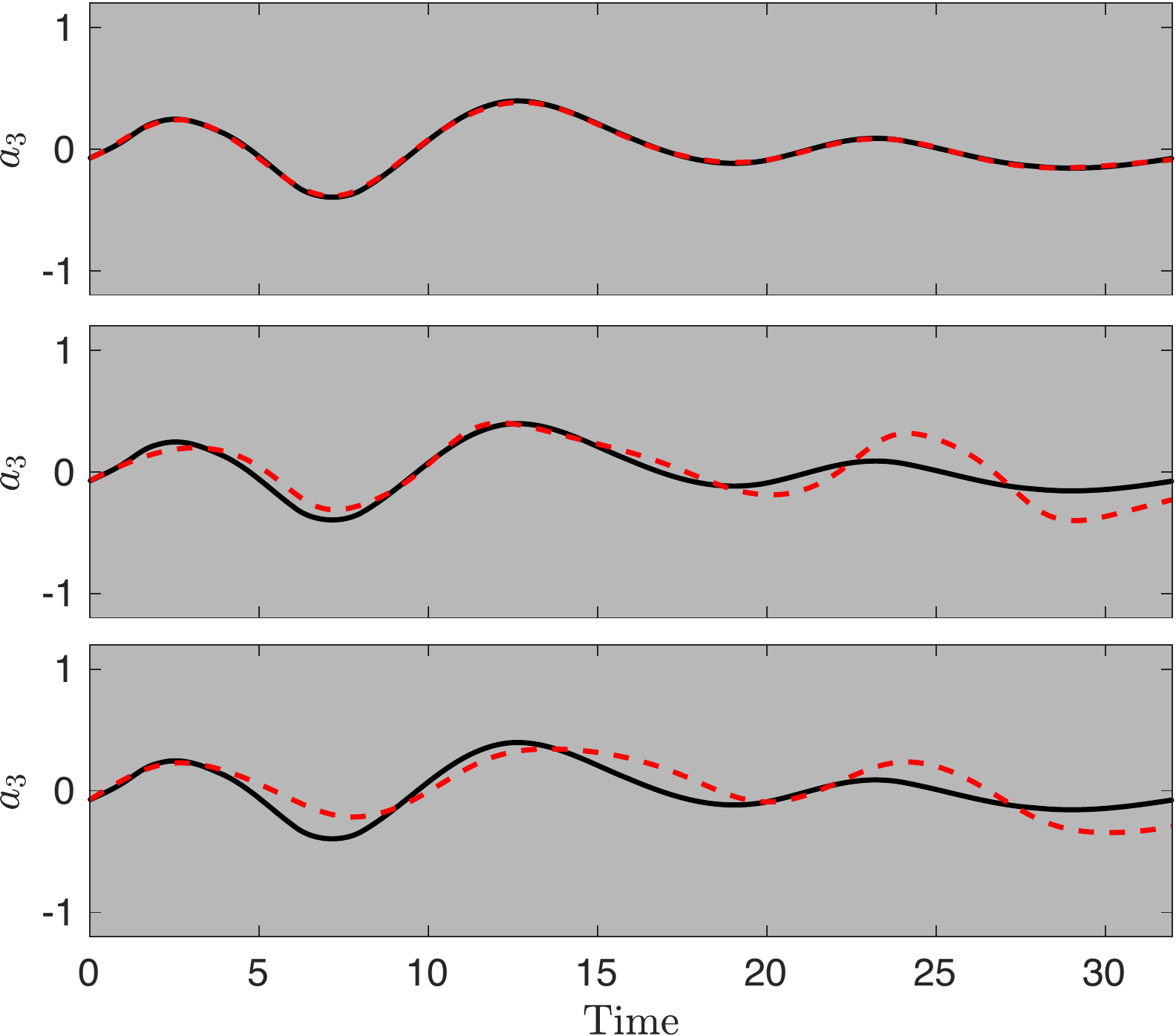}
\caption{Comparison of the time evolution of the posteriori prediction of the $3$ POD features generated from the different modes with the true data using all the $1600$ snapshots for training. The different plots correspond to:  (a) EDMD-P3 , (b)  FFNN with $N_f=3$  and (c) FFNN with $N_f=5$.}
\label{f:F3_1600}
\end{center}
\end{figure}

In particular, we consider DMD,  EDMD-TS and  EDMD-P for the MSM class of methods and contrast these with different FFNN (MEM with $N_f = 1,3,5$ ) models that incorporate a growing $\mathcal{LP}$ dimension. As a preliminary step, we use the entire available data for training and assess the reconstruction performance of these models. Figure \ref{f:F1_1600} compares the results for the DMD with the nonlinear EDMD-TS1 and  FFNN ($N_f=1$) models with small number of learning parameters ($3,9$ and $27$ respectively). Contrary to findings from  the earlier sections for the transient cylinder wake, all the MSM frameworks including the linear DMD and nonlinear EDMD-TS1 compare favorably to the FFNN/MEM with $N_f=1$. All three models fail to predict the dynamics of the third POD feature which represents the secondary eddies from the Kelvin-Helmholtz instability generated by the mixing layer dynamics (see bottom plot in fig.~\ref{f:lockexmodes}). The MSM models generate slightly better outcomes as compared to MEM for the first two POD features that represent transverse and vertical mixing (top two plots in fig.~\ref{f:lockexmodes}). To improve the predictions of the third POD feature, we enhance the learning parameter ($\mathcal{LP}$) dimension by employing EDMD-P3 (6-MSM-P3-1), FFNN with $N_f=3$ (6-MEM-TS-3) and FFNN with $N_f=5$ (6-MEM-TS-5) as shown in fig.~\ref{f:F3_1600}. Consistent with earlier observations, this increase in $\mathcal{LP}$ improves the prediction of the third feature for both the multilayer sequential and end-to-end learning methods with the former performing better. Similar performance was also realized with the EDMD-P2 architecture and is not reported here for brevity.  This shows that for reconstructing the dynamics, MSM architectures are more accurate as compared to the MEM frameworks that leverage  nonlinear regression techniques. An aspect that is relatively under-explored in the study of FFNNs is role of nonlinear mapping, $\mathcal{N}$ on learning performance. For example, this current study shows that for the MSM class of methods, EDMD-TS performs inadequately relative to the different variants of  EDMD-P thus hinting that a polynomial basis being better suited to approximate this data. Give this, it is only natural to speculate whether MEM architectures would perform better with other choices of nonlinear maps although such exploration is beyond the scope of this study. 
 
%
%

\begin{figure}
\begin{center}
\includegraphics[width=0.33\columnwidth]{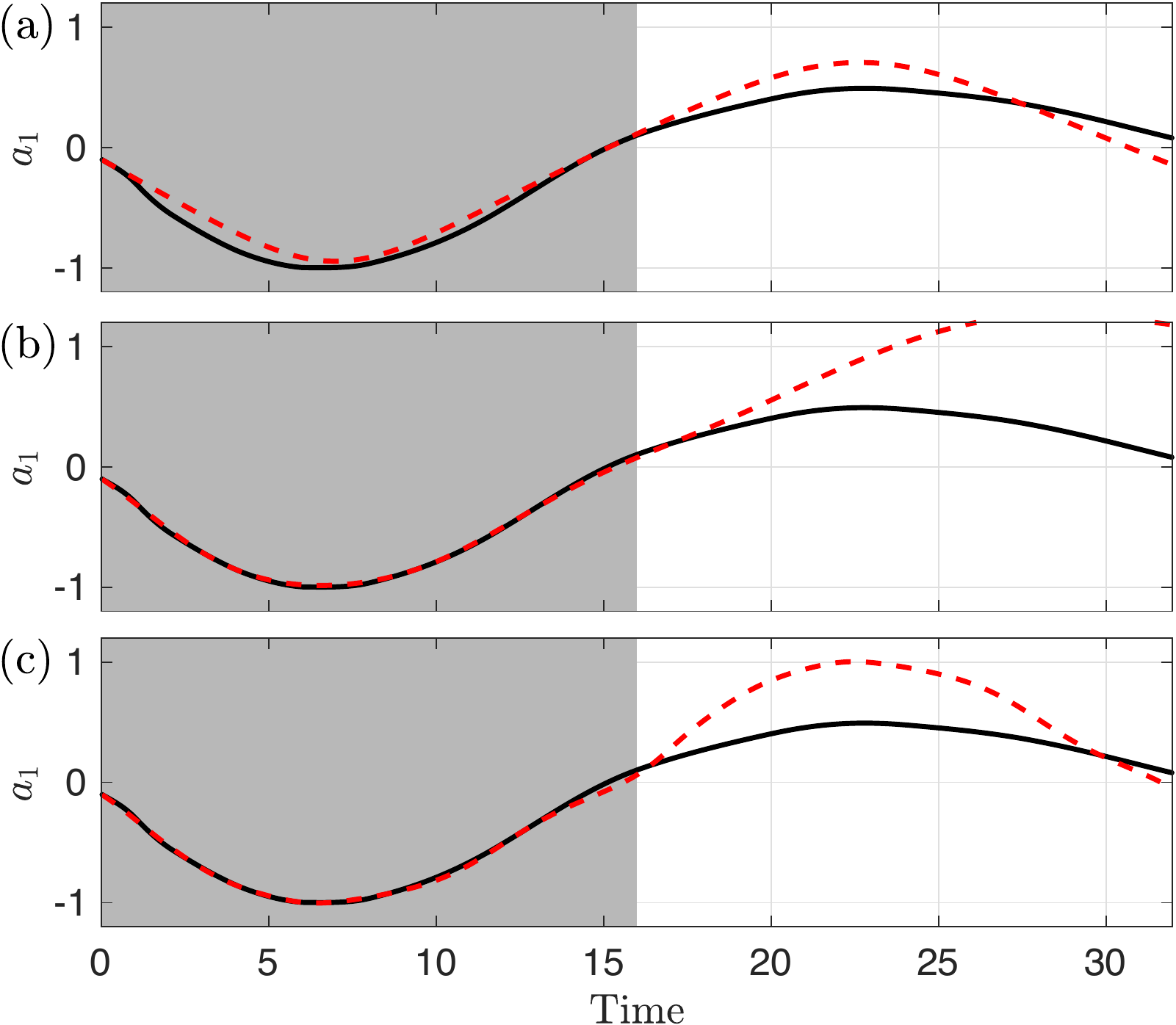}
\includegraphics[width=0.32\columnwidth]{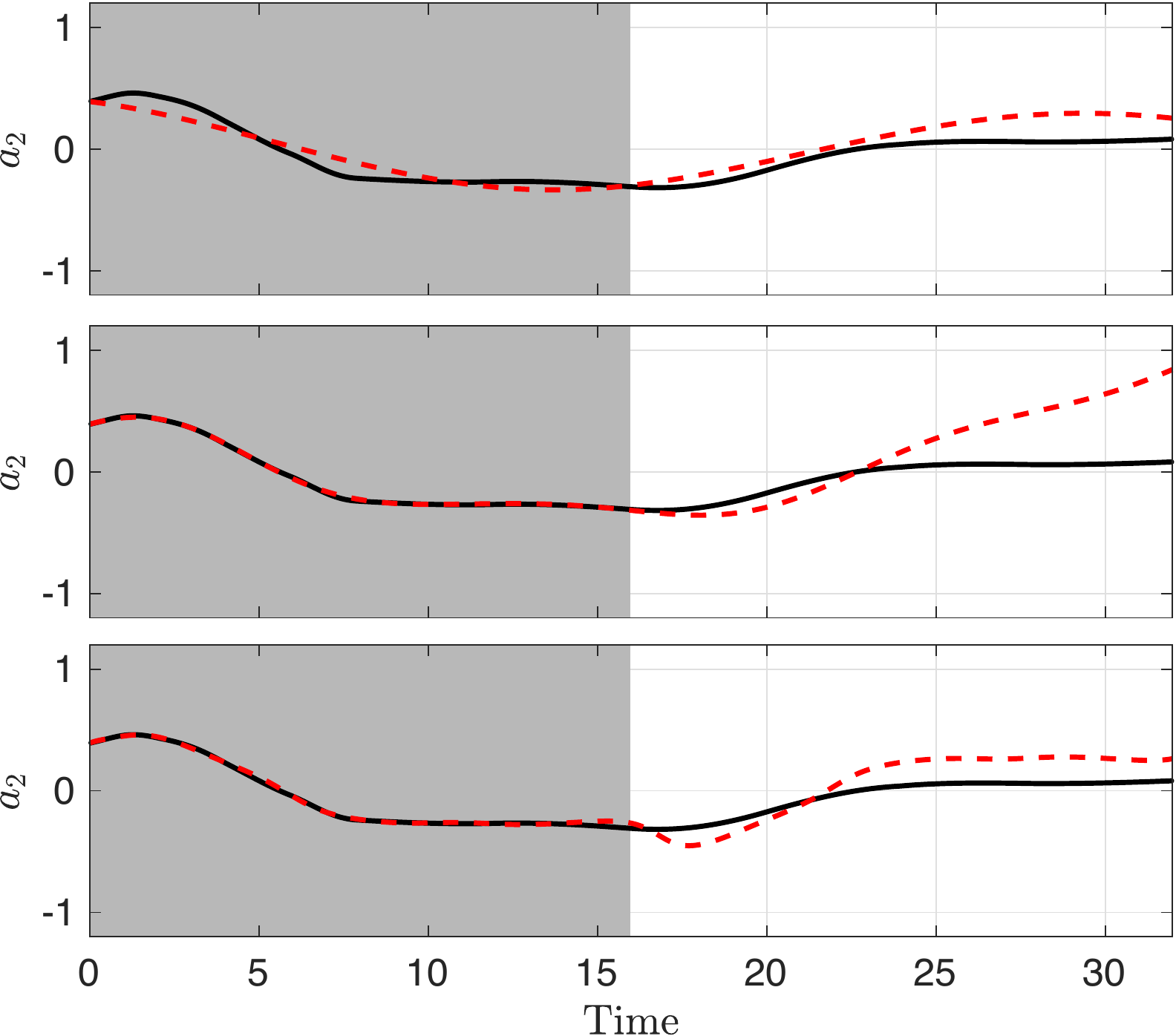}
\includegraphics[width=0.32\columnwidth]{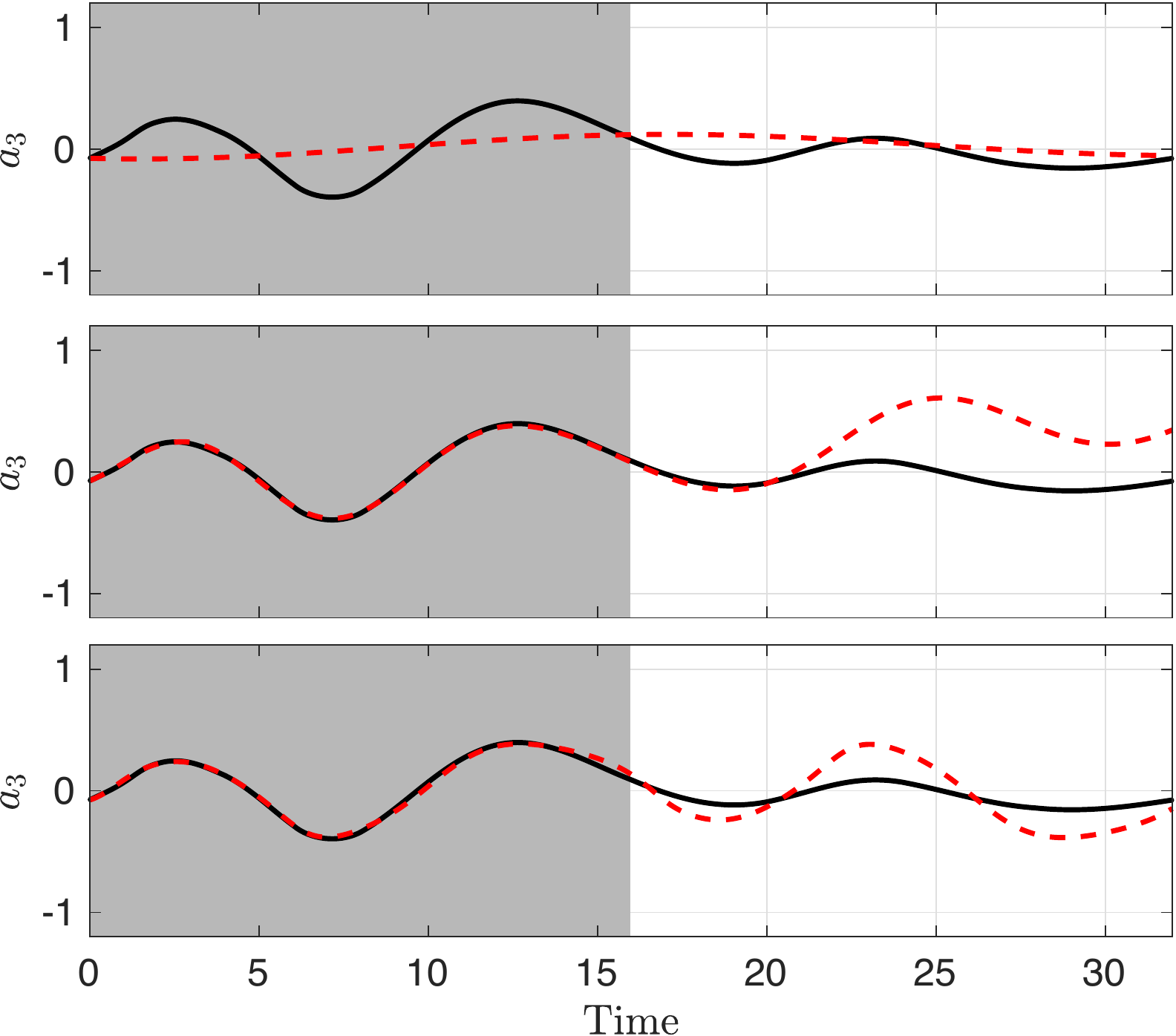}
\caption{Comparison of the time evolution of the posteriori prediction of the $3$ POD features generated from the different modes with the true data using $50\%$ data i.e. $800$ snapshots for training. The different plots correspond to:  (a) EDMD-TS1 , (b) EDMD-P2  and (c) FFNN with $N_f=3$.}
\label{f:F1_800}
\end{center}
\end{figure}

To assess the ability of the models to learn the underlying system dynamics, we split the dataset ($1600$ snapshots) equally into training (i.e. $800$ snapshots for training) and prediction regimes. Figure \ref{f:F1_800} compares the posteriori predictions of the three POD features for EDMD-TS1, EDMD-P2 and FFNN with $N_f=3$. For all these models, we clearly observe that reconstruction is better than true prediction performance indicating that dynamics is evolving rapidly and the data-driven models are finding it hard to forecast physics that it has not seen through training.  
Therefore, focusing on the predictions, the multilayer end-to-end learning FFNN outperforms the two MSM architectures considered here in terms of stability and accuracy. Particularly, the FFNN/MEM prediction using $50\%$ data (fig.\ref{f:F1_800}c ) is highly similar to that obtained from reconstruction of the entire dataset as shown in fig.~\ref{f:F3_1600}c. This shows that these models offer robust and stable performance even with limited data.  In summary, we see that MSM frameworks offer competitive reconstruction performance, but MEM learning models with different architectures (at least for the different examples considered in this study) offer stable and robust model performance for long time predictions using limited data. 

\section{Discussion and Summary}\label{s:conclusions}
Fluid flows represent multiscale PDE dynamical systems that often require low-dimensional data-driven representations and evolutionary models for a multitude of applications. 
In this article we explore the performance of multilayer sequential maps (MSMs) versus multilayer end-to-end maps (MEMs) in Markov models for learning and long-time prediction of nonlinear fluid flows using small amounts of training data. In particular, we assess the role of learning parameter dimension and nonlinear transfer functions on the ability of the architecture  to reconstruct and predict over long times without overfitting to the data. The sequential multilayer frameworks (MSMs) allow for both backward and forward mapping operations in symmetric architectures and can support the estimation of the Koopman operator for spectral analysis and linear control in addition to serving as data-driven models.  On the other hand, multilayer maps that incorporate end-to-end (MEMs) learning from data can support only forward maps within the framework of an asymmetric Markov model due to the use of gradient-based optimization algorithms employed to solve the nonlinear regression problem. Consequently, architectures like FFNN cannot learn the Koopman operator in base configuration although recent advancements~\cite{Shiva:18AIAA} can help bypass this limitation.

The major outcomes of the study are as follows. The success of both the MSM and MEM architectures is tied to the choice of nonlinearity in the mapping and the dimension of the learning parameter space embedded in the design of the multilayer architecture. We observe that for prediction of limit-cycle dynamics from limit-cycle data both the MSM and MEM-based models show reasonable success although MEM models such as FFNN control the growth of long-time prediction errors better than any of the MSM model considered. Further, MEM architectures generate the most accurate predictions for a given learning parameter ($\mathcal{LP}$) budget as long as the map incorporates appropriate nonlinear functions. In the absence of nonlinear functions in the map, the $\mathcal{LP}$ dimension did not impact the predictions. Further, any choice of nonlinear function will not produce good results. We observed that tansigmoid functions operate well with MEM architectures while polynomial nonlinearity fared well with MSMs. 

To assess the ability of these model architectures to generalize across diverse training data regimes, we considered two different case studies with different training regimes that differ in their proportion of limit-cycle to unstable wake growth dynamics. To mimic the availability of limited resolution data as is commonly the case, we chose to train these models using their low-dimensional representation with only three POD features.  With these constraints, we observed that for comparable number of learning parameters,  the FFNN (MEM) architectures outperform the corresponding MSM frameworks by a significant margin in terms of accuracy and robustness. 
To illustrate this, we show that the FFNN\cmnt{( 6-MEM-TS-1)} architecture with $N_f=1$ and $9$ learning parameters produce qualitatively accurate results as against the gross inaccuracy of MSM frameworks such as DMD\cmnt{ (4-MSM-$\mathcal{I}$-1)} and EDMD \cmnt{(6-MSM-P2-1 and 6-MSM-TS-1)}. With increase in $\mathcal{LP}$ dimension, both class of methods converge to the accurate predictions although the MSM reaches their slowly and results in significant overfitting as compared to MEM architectures. 

The downside of MEM-based models is the added computational cost and learning time which limits the dimension of the input feature space for practical applications. The use of iterative gradient-based search algorithms impact convergence with a tendency for being stuck in local minima. However, this is compensated by more efficient learning from data i.e. requires only a relatively modest increase in $\mathcal{LP}$ dimension for improved predictions. All these perceived advantages of MEM over MSM (and vice versa) are valid only in the limit of availability of sufficient data which injects a dose of reality regarding data-driven modeling approaches. We observed this when training a model for a different flow regime (TR-II) that contained little information about the limit-cycle dynamics where both class of methods found learning and prediction harder. Yet, the MEM architectures were able to generate qualitatively accurate predictions with as little as $135$ learning parameters whereas  the equivalent MSM architectures could not generate meaningful predictions. 
We also explored the performance of the various data-driven modeling approaches for an instability-driven, non-stationary, buoyant mixing flow which requires unlimited amounts of data to represent all the possible dynamics of the system. While we knew the challenge faced by data-driven methods for such problems, we analyzed how the various models fared in learning-based reconstruction and learning-based prediction of such flows. While both class of methods struggle to generate accurate predictions, the MSM-based models perform well in reconstruction whereas the MEM-based models offer better predictions and model generalization. 

In summary, the strategy of extending the $\mathcal{LP}$ space, learning the model parameters concurrently using a end-to-end maps and improved regularizations can help improve learning from data for robust and accurate predictions. However, as with machine learning in general, these outcomes are strongly tied to data sufficiency and quality. 


\begin{acknowledgement}
We acknowledge support from Oklahoma State University start-up grant and OSU HPCC for compute resources to generate the data used in this article. The authors thank Chen Lu, a former member of the Computational, Flow and Data science research group at OSU for providing the CFD data sets used in this article. \cmnt{BJ acknowledges discussions on data-driven modeling with Prof. Karthik Duraisamy at the University of Michigan.} 
\end{acknowledgement}

\section*{Author Contributions}
BJ conceptualized the work with input from SCP. SCP developed the data-driven modeling codes used in this article with input from BJ. BJ and SCP analyzed the results. BJ developed the manuscript with contributions from SCP.

%



\appendix
\section{Effect of Bias on Predictions}\label{s:Appendix1}

\begin{figure}[H]
\begin{center}
\includegraphics[width=0.33\columnwidth]{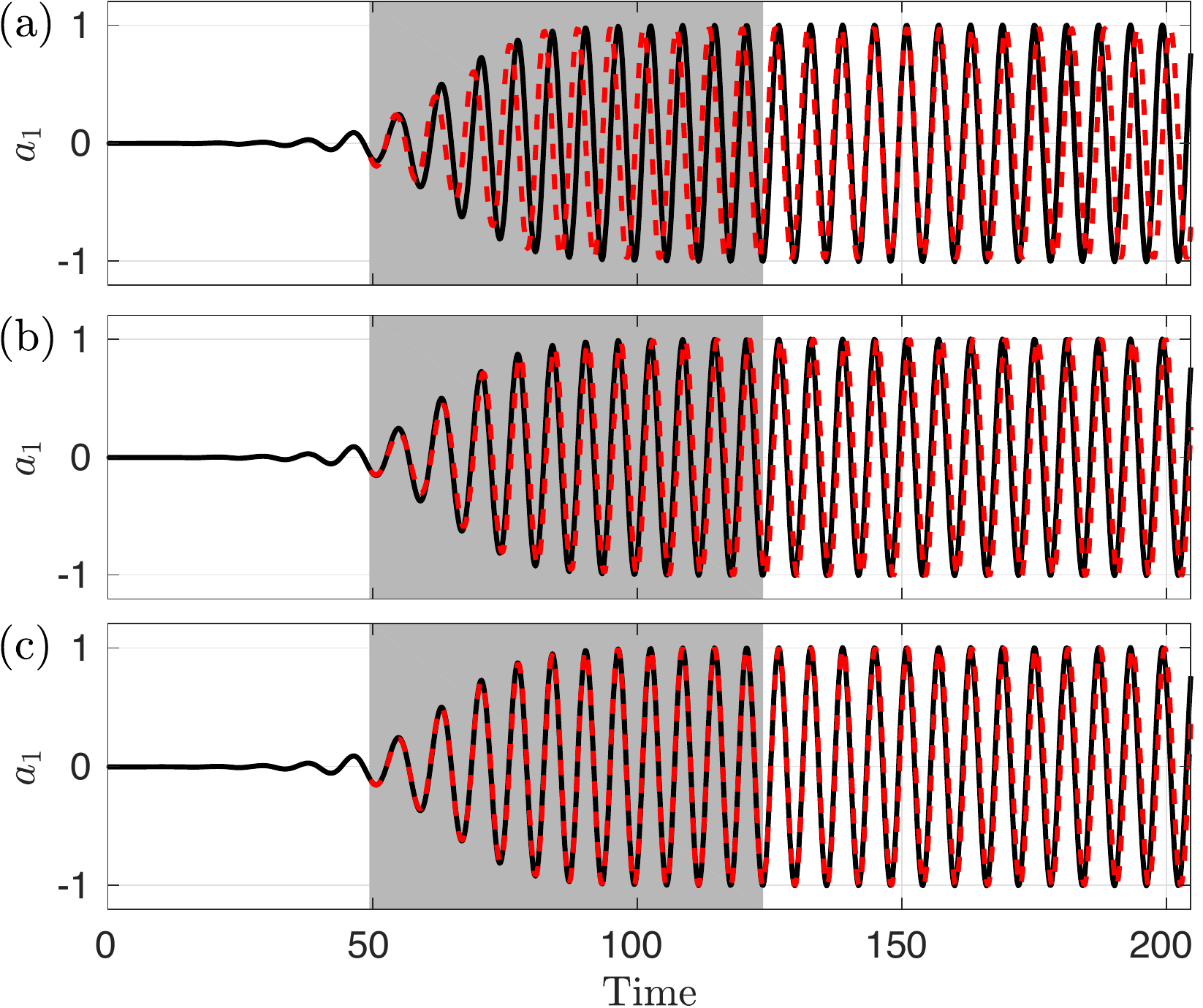}
\includegraphics[width=0.32\columnwidth]{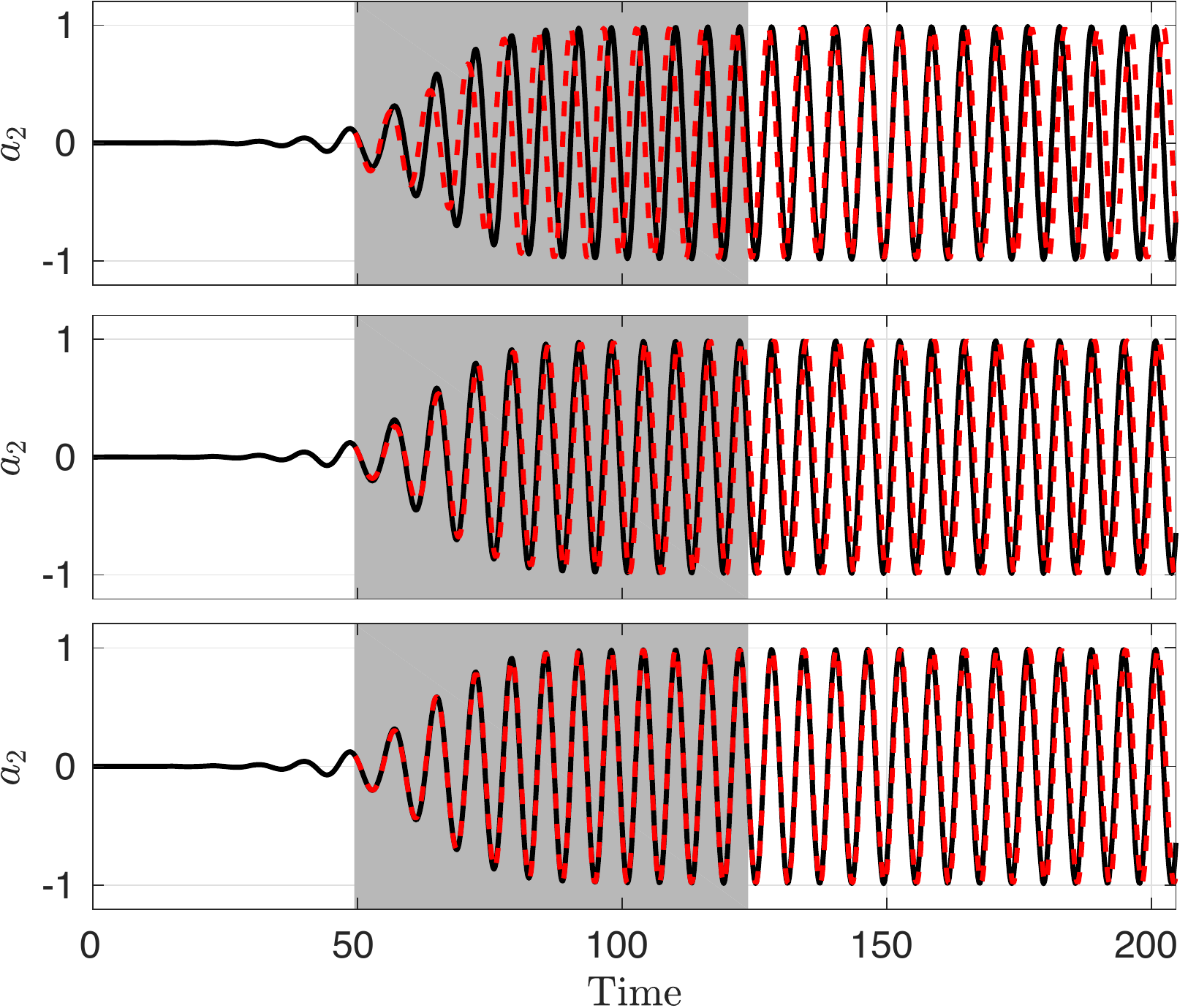}
\includegraphics[width=0.32\columnwidth]{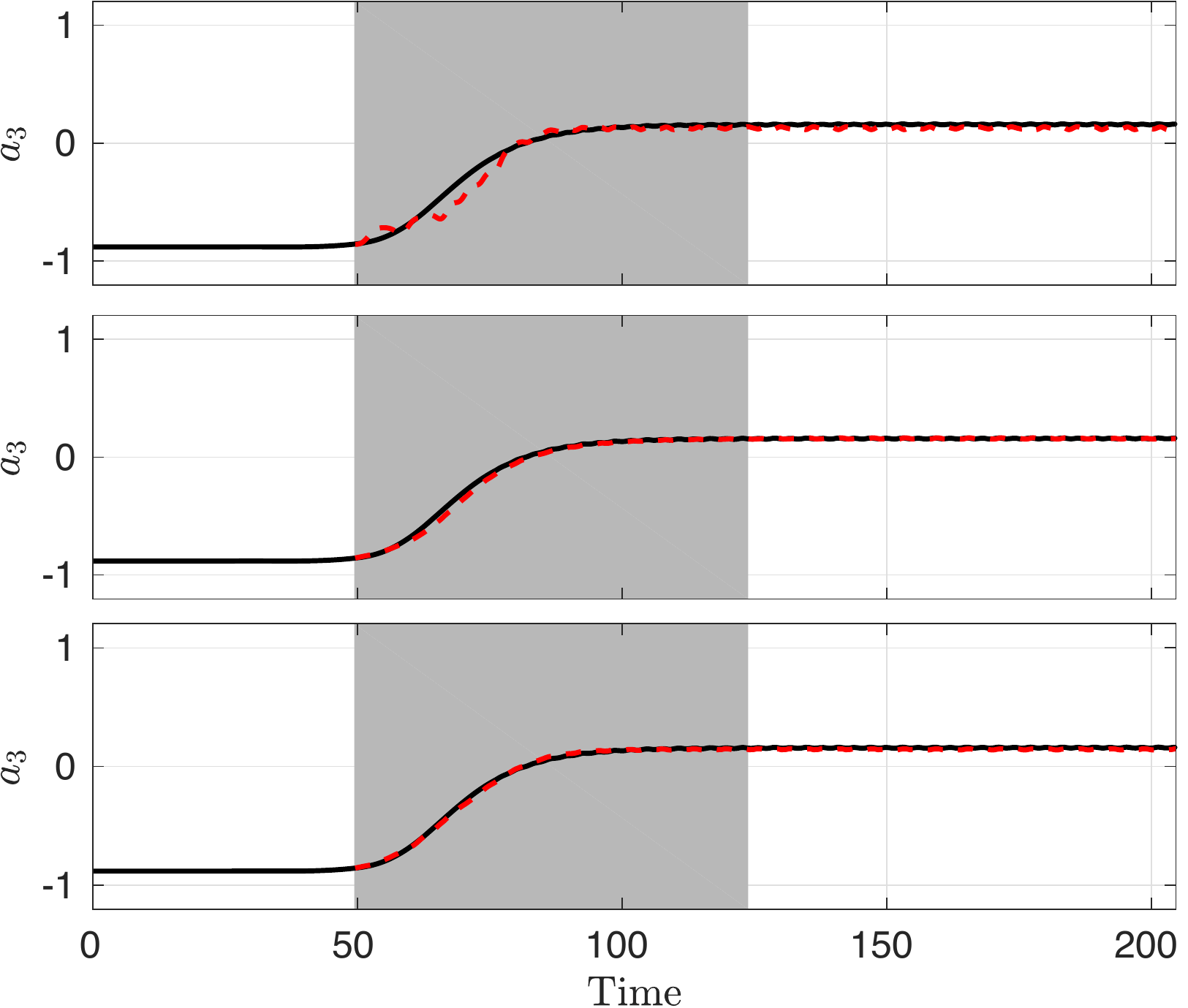}
\caption{Times series of  predicted $Re 100$ POD features obtained from (a) 6-MEM-TS1  (b) 6-MEM-TS3 and (c) 6-MEM-TS9 compared with their respective original coefficients for TR-I region.}
\label{f:WBias_820Re100}
\end{center}
\end{figure}

\begin{figure}[H]
\begin{center}
\includegraphics[width=0.33\columnwidth]{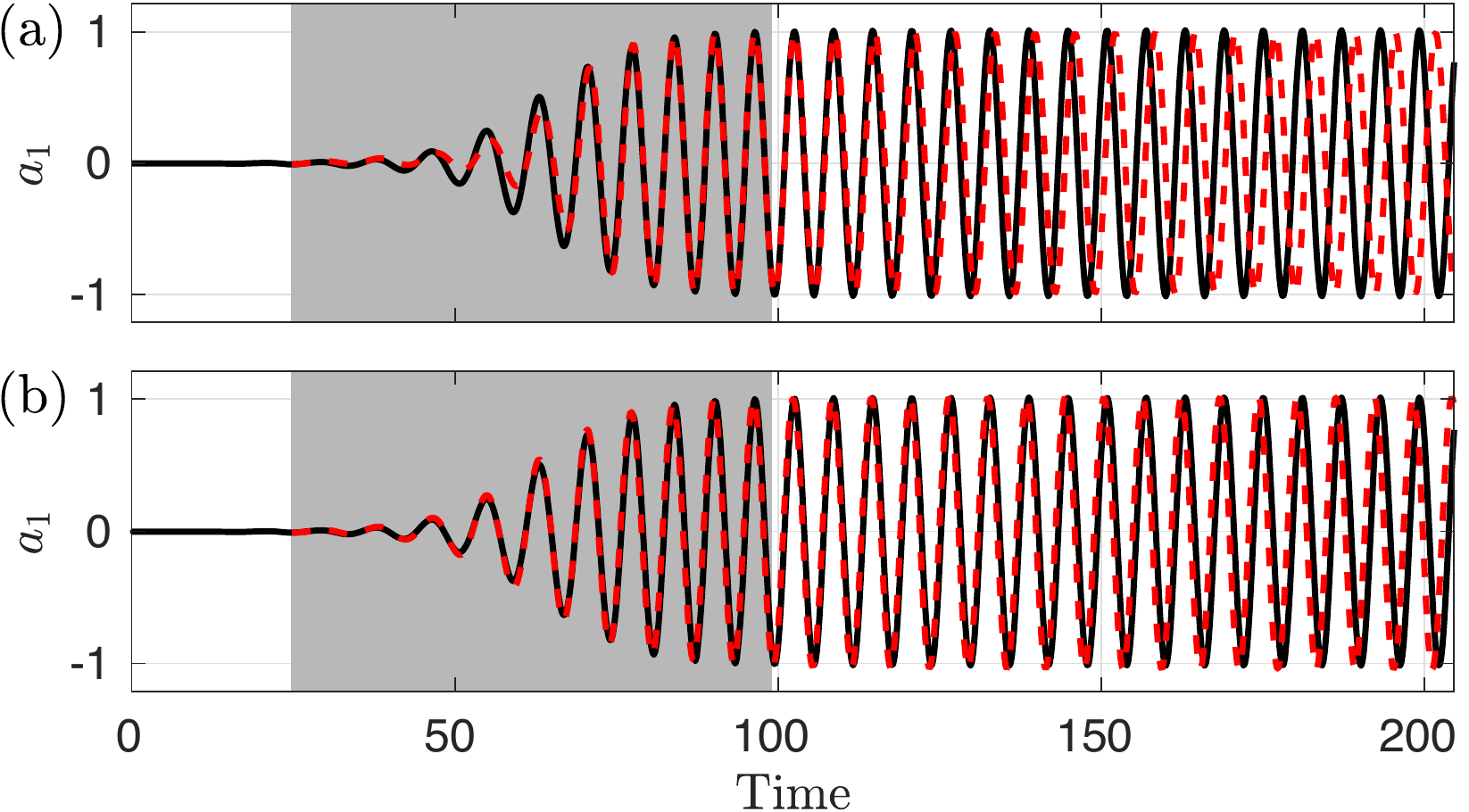}
\includegraphics[width=0.32\columnwidth]{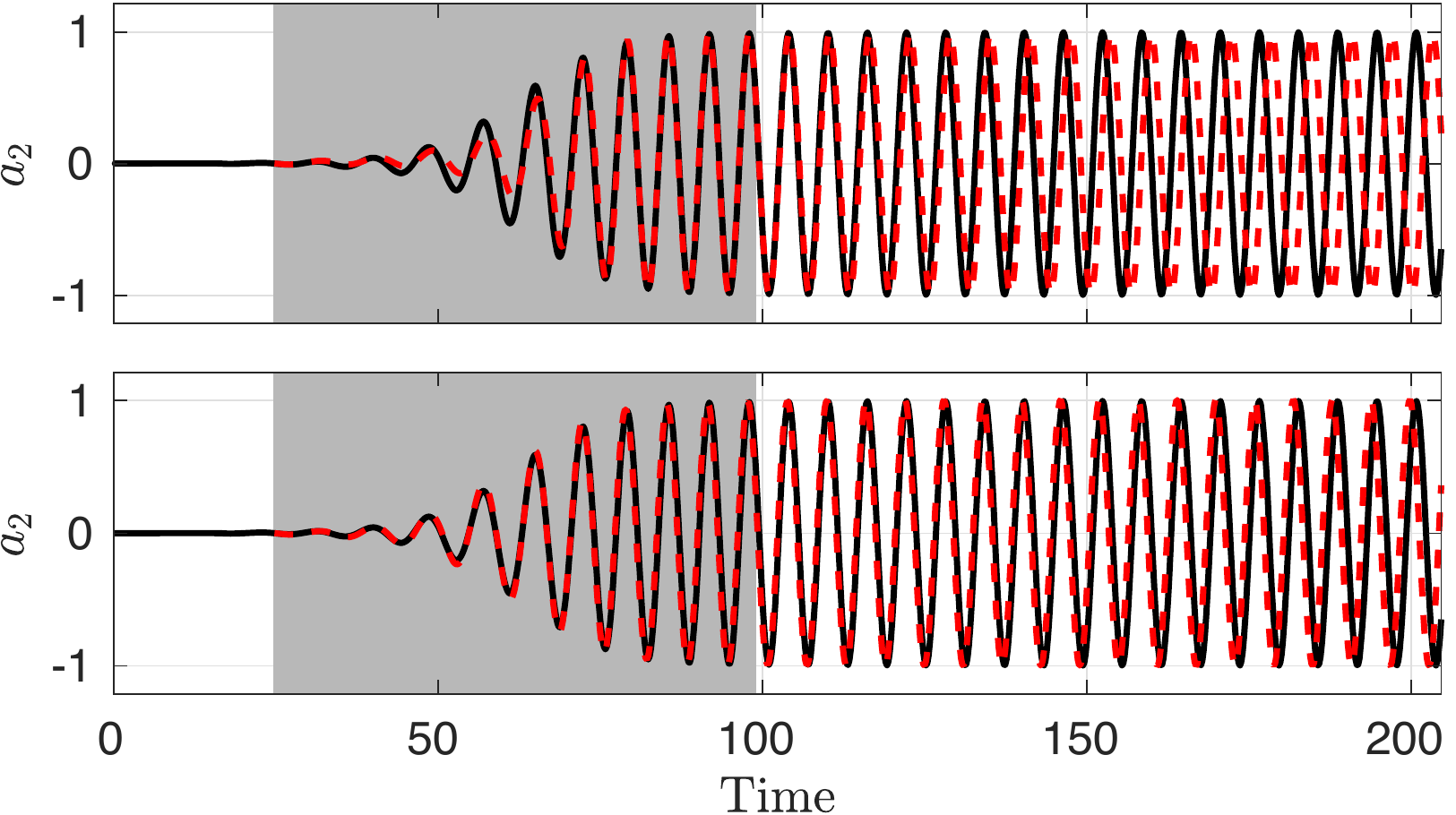}
\includegraphics[width=0.32\columnwidth]{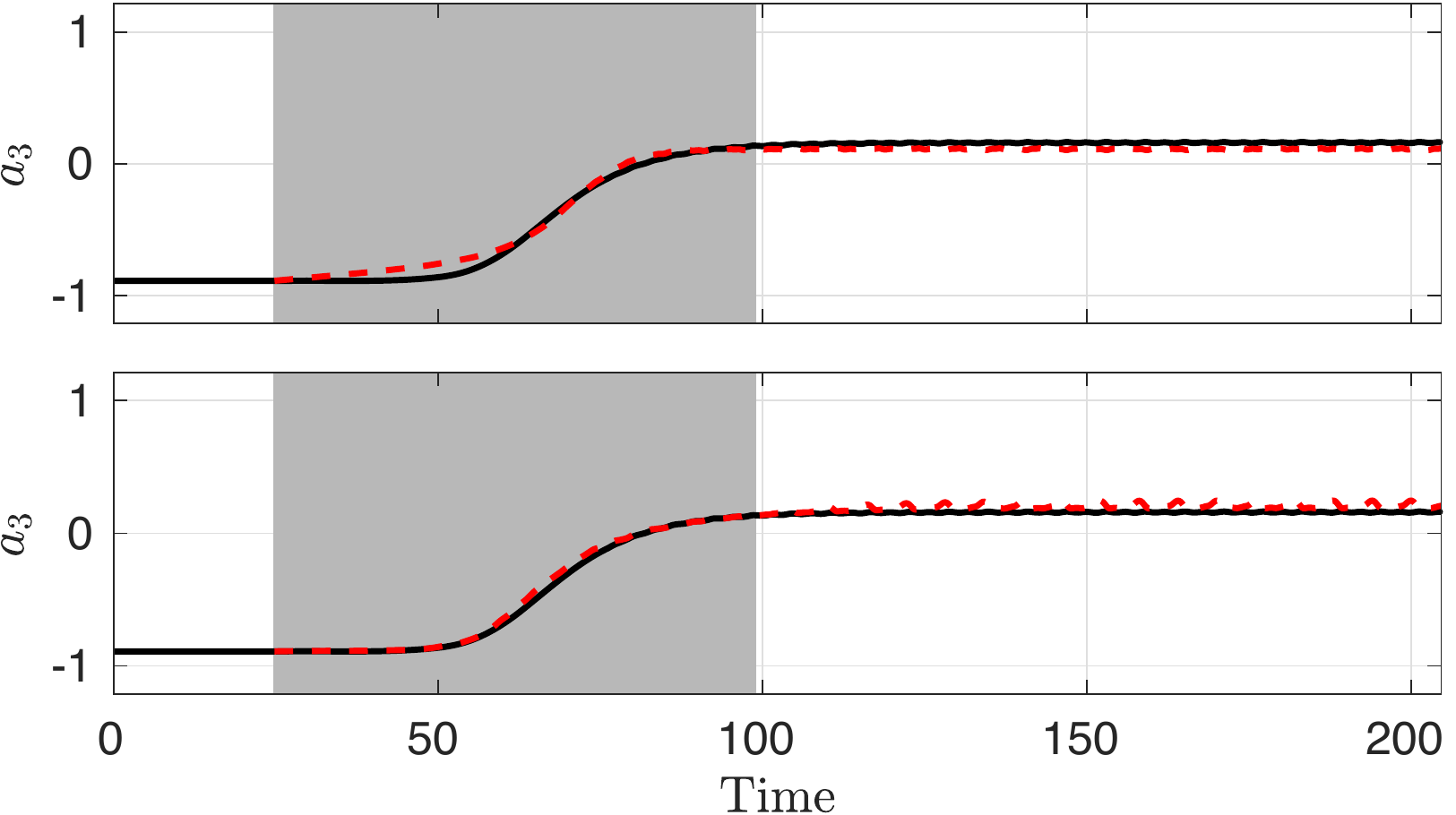}
\caption{Times series of  predicted $Re 100$ POD features obtained from (a) 6-MEM-TS3 and (b) 6-MEM-TS9 compared with their respective original coefficients for TR-II region.}
\label{f:WBias_416Re100}
\end{center}
\end{figure}

The results presented in the main sections of this article for MEM architectures were based on FFNNs devoid of the bias term. It is well known from machine learning literature~\cite{Hornik:89}  that the presence of a bias term helps with function approximation provided sufficient $\mathcal{LP}$s are used to capture the dynamics. In our studies, the FFNN had difficulty predicting the shift mode for the transient cylinder wake dynamics (section \ref{ss:transientdata}) whereas the modes with zero mean were predicted accurately.  Since the bias term helps in quantitative translation (shift) of the learned dynamics into higher or lower values, we expect its inclusion to improve predictions. In fig.~\ref{f:WBias_820Re100}, we show predictions of the features obtained from FFNNs with $N_f=1,3,9$ for the TR-I regime for $Re = 100$. In fig.~\ref{f:WBias_416Re100}, we include the predictions for the TR-II data. In both these cases, the shift mode (third POD feature) is accurately predicted with a bias term.

\bibliographystyle{ieeetr}%
\bibliography{ref}


%

\end{document}